\documentclass[fleqn,usenatbib]{mnras}

\usepackage{newtxtext,newtxmath}

\usepackage[T1]{fontenc}

\DeclareRobustCommand{\VAN}[3]{#2}
\let\VANthebibliography\thebibliography
\def\thebibliography{\DeclareRobustCommand{\VAN}[3]{##3}\VANthebibliography}

\usepackage{graphicx}	
\usepackage{amsmath}	
\usepackage{multicol}
\usepackage{subcaption}
\usepackage{multicol}
\usepackage{longtable}
\usepackage{comment}

\title[Gamma-ray and X-ray Emissions from Pulsars]{Observational Relationship between Spectral Properties of Gamma-ray and X-ray Emissions from Pulsars}

\author[Aravindaraj \& Chang]{
Ashwin Aravindaraj,$^{1}$\thanks{E-mail: ashwinaravindaraj@gmail.com}
and Hsiang-Kuang Chang,$^{1,2,3,4}$\thanks{E-mail: hkchang@mx.nthu.edu.tw}
\\
$^{1}$Institute of Astronomy, National Tsing Hua University, Hsinchu 300044, Taiwan  \\
$^{2}$Department of Physics, National Tsing Hua University, Hsinchu 300044, Taiwan\\
$^{3}$Institute of Space Engineering, National Tsing Hua University, Hsinchu 300044, Taiwan\\
$^{4}$Center for Theory-Computation-Data Science Research (CTCD), National Tsing Hua University, Hsinchu 300044, Taiwan\\
}

\date{Accepted XXX. Received YYY; in original form ZZZ}

\pubyear{2025}

\begin{document}
\label{firstpage}
\pagerange{\pageref{firstpage}--\pageref{lastpage}}
\maketitle

\begin{abstract}
Correlations between gamma-ray and X-ray spectral properties of pulsars are investigated in order to provide observational hints on 
physics involved in pulsars' high-energy emissions.
Using a sample of 43 pulsars detected in both X-ray and gamma-ray bands, we find that pulsars' gamma-ray luminosity, $L_\gamma$, clearly correlates with the luminosity of non-thermal X-ray emission, $L_{\rm p}$, and anti-correlates 
with non-thermal X-ray photon index. Other gamma-ray spectral parameters show weaker or negligible correlations. The found relation that $L_\gamma \propto L_{\rm p}^{0.49\pm 0.05}$ implies a certain connection between radiation mechanisms and energy distributions of radiating particles for these high-energy emissions.
Pulsars with and without detected thermal emissions seem to show different dependencies in those correlations, suggesting the possible existence of two different kinds of pulsars. The ones without detected thermal emissions may represent
a population of pulsars with low surface temperature.
The origin and energetics of high-energy emitting electron-positron pairs for this group of pulsars probably do not depend on their surface thermal emissions, while that of the other group do.
The low surface temperatures might be evidence for the working of some exotic processes of neutron-star cooling. 
Similar to $L_{\rm p}$, some tempting relationships are found among each gamma-ray spectral parameter, surface temperature and thermally radiating area radius. It again strengthens the connection between gamma-ray and X-ray emissions from pulsars.

\end{abstract}




\section{Introduction}
Pulsars are observed in electromagnetic radiation from radio to gamma rays. About four thousand are observed in the radio band (see the ATNF catalog (\citet{Manchester2005})) and 294 in gamma-rays
(see the Third Fermi LAT Catalog of Gamma-ray Pulsars \citep{Smith2023}; hereafter 3PC). Pulsars' radio emissions are considered to be from a certain kind of coherent mechanism, but emissions at higher energies are not.
Although there is no demand on devising a coherent mechanism for these high-energy emissions, the origin and mechanisms of pulsar high-energy emissions are still in debate.

Models for high energy emissions from pulsars have evolved from the earlier polar-gap (\citet{1982ApJ...252..337D}), slot-gap (\citet{1983ApJ...266..215A}), outer-gap \citep{Cheng1986}, 
and striped-wind (\citet{1990ApJ...349..538C, 1994ApJ...431..397M, 2002A&A...388L..29K}) models to more recent, more involved ones, which include, for example, 
the striped-wind model with particle-in-cell simulations (\citet{2024A&A...687A.169P, 2025A&A...695A..93C}), the extended-slot-gap and equatorial-current-sheet model (\citet{2021ApJ...923..194H, 2022ApJ...925..184B}), the synchro-curvature model (\citet{2019MNRAS.489.5494T, 2022MNRAS.516.2475I}) and the non-stationary outer-gap model (\citet{2017ApJ...834....4T}).

A comprehensive picture to understand all the properties of high-energy emissions across different energy bands, from optical to gamma rays, has not yet appeared in current models. In observational aspect, population studies on cross-band correlations are very few. Most are on the correlation between spectral properties of a certain band and timing parameters (e.g. \citet{possenti2002,Li2008,HK2023}) or on spectral properties in a certain band for a certain group of pulsars, such as non-thermal X-ray emissions from gamma-ray pulsars \citep{Coti2020}. We therefore conduct an effort to investigate possible correlations between spectral properties of GeV gamma-ray emission and X rays from pulsars. These correlations may inspire and constrain modelling works. 

Fermi LAT has significantly increased the number of gamma-ray pulsars to 294 (3PC). Gamma-ray spectra of these pulsars are described in great details in 3PC  
with a particular cut-off power-law model (Eq.(15) in \citet{Smith2023}).
In this paper, we report the results of examining correlations between spectral properties of gamma-ray and X-ray emissions from pulsars.

Besides those consistent with earlier literature, we found that the
gamma-ray luminosity clearly correlates with the luminosity of non-thermal X-ray emission and anti-correlates with its photon index. 
Pulsars with and without detected thermal emissions seem to show different dependencies in those relations, suggesting possible existence of two different kinds of pulsars. 

In Section 2 we explain the data selection criteria. Correlation analysis results are presented in Section 3. The dependence of the gamma-ray luminosity 
and other spectral parameters
on timing parameters is shown in Section 3.1 as a check with earlier literature. 

Then the relationships between gamma-ray and non-thermal X-ray emissions are reported in Section 3.2. That between gamma-ray and thermal X-ray emissions are presented in Section 3.3.
Correlations among gamma-ray spectral parameters are shown in Section 3.4.
Discussions on these results are in Section 4.

\section{Sample Selection}
To collect samples for this study, we cross-matched the list of pulsars with X-ray emissions reported in \citet{HK2023} with 3PC and found 43 pulsars with both GeV gamma-ray and X-ray emissions. We note that about 200 more candidates of pulsars with X-ray emissions have been reported based on position coincidence and flux ratios in different wave bands (\citet{Mayer&becker2024,Xu2025}). Future confirmation of these candidates will further increase the sample size.
Among these 43 pulsars, PSR J2021+4026 is observed to switch between two gamma-ray emission states, with flux in 0.1 -- 100 GeV being  $(6.86\pm0.13) \times 10^{-10}\:\rm{erg\:cm^{-2}\:s^{-1}}$ (\citet{2013ApJ...777L...2A}) and $(8.33\pm0.08) \times 10^{-10}\:\rm{erg\:cm^{-2}\:s^{-1}}$ (\citet{2017ApJ...842...53Z}), respectively. 
Its X-ray flux (0.2 keV -- 12 keV) is found to be $3.5^{+1.3}_{-0.7}\times 10^{-14}\:\rm{erg\:cm^{-2}\:s^{-1}}$ (\citet{2013ApJ...770L...9L}) and $4.7^{+1.4}_{-1.0}\times 10^{-14}\:\rm{erg\:cm^{-2}\:s^{-1}}$ (\citet{Wang2018})  correspondingly and there is a 6\% change in the time derivative of its period between the two states (see Figure 1 in \citet{wang2024}). This pulsar in different states is treated as two pulsars in this study. Hereafter we therefore refer to the number of pulsars in our sample as 44.   

Among the 44 pulsars in this study, the X-ray spectra of 22 pulsars are well fitted by a power law. Another 18 can be well described by a two-component model of a power law plus a blackbody. For the other four, that is, PSR J0633+1746, PSR J0659+1414, PSR J1057-5226, and PSR J1740+1000, a three-component model consisting of a power law plus two blackbodies is required. To have a uniform comparison base, a flux-weighted average of the temperature and the emitting-region radius of the two blackbody components, as reported in \citet{HK2023}, is adopted to describe their thermal emission. 

To calculate  luminosity from flux, we consider the luminosity ($L = 4\pi d^2 f_{\Omega}F$) with a beaming fraction of $f_{\Omega} = 1$ \citep{2009ApJ...695.1289W} and use distance estimates from various independent measurements, along with their associated uncertainties listed in the ATNF catalog \citep{Manchester2005}, whenever available. If there are no independent measurements, we adopt the dispersion-measure–based distance estimate (also provided in the ATNF catalog) and assign, following \citet{Li2008} and \citet{HK2023}, a 40\% uncertainty to the distance. Due to the large uncertainty in distances, we use the Monte Carlo error propagation method to derive luminosity errors.
The X-ray luminosity of the power-law component, $L_{\rm p}$, reported in this study is for the energy range of 0.5 -- 8 keV, and the gamma-ray one, $L_\gamma$, is from 0.1 to 100 GeV. All the gamma-ray fluxes  are taken from 3PC.

The timing variables and distances taken from the ATNF catalog are shown in Table \ref{Timing_table}.
The gamma-ray spectral parameters listed in Table~\ref{Spectral_table} are taken directly from 3PC, where $\Gamma_{100}$ denotes the photon index, $E_{\rm p}$ the peak energy, and $d_{\rm p}$ the curvature of the gamma-ray spectrum around the peak energy; readers are referred to 3PC for details.
The X-ray spectral properties of these 44 pulsars are also listed in Table \ref{Spectral_table}.

\begin{table*}
    \caption{Timing variables of the 44 pulsars. $P$ is the pulsar period and $\dot{P}$ is its first time derivative. Other timing variables, the spin-down power $\dot{E}$, characteristic age $\tau$, surface dipole magnetic field strength $B_{\rm s}$, and magnetic field strength at the light cylinder $B_{\rm lc}$, are all functions of $P$ and $\dot{P}$. With the conventional values of the moment of inertia as $10^{45}$ g cm$^2$ and neutron star radius as 10 km for
the pulsar, these variables are (all in gaussian units)
$\dot{E}=3.9\times 10^{46}P^{-3}\dot{P}$, 
$\tau=0.5P\dot{P}^{-1}$,
$B_{\rm s}=3.2\times 10^{19}(P\dot{P})^{0.5}$, and
$B_{\rm lc}=2.9\times 10 ^8 P^{-2.5}\dot{P}^{0.5}$.}
    \label{Timing_table}
    \begin{tabular}{lccccccr}
         \hline
         Pulsars & Distances (kpc)  & $P$ (s) & $\dot{P}$ (s/s) & $\dot{E}$ (erg/sec) &$\tau$ (years) & $B_{\rm s}$ (gauss) & $B_{\rm lc}$ (gauss)\\
         \hline
         J0007+7303 &$1.4\pm0.3$&0.3159&$3.56 \times 10^{-13}$&$4.45 \times 10^{35}$ &$1.39 \times 10^{4}$&$1.08 \times 10^{13}$&$3.08 \times 10^{3}$\\

         J0205+6449 &$3.2\pm0.2$&0.0657&$1.92 \times 10^{-13}$&$2.67 \times 10^{37}$ &$5.37 \times 10^{3}$&$3.61 \times 10^{12}$&$1.15 \times 10^{5}$\\

         J0357+3205 &$0.6\pm0.3$&0.44&$1.31 \times 10^{-14}$&$5.9 \times 10^{33}$ &$5.4 \times 10^{5}$&$2.43 \times 10^{12}$&$2.53 \times 10^{2}$\\

         J0534+2200 &$2^{+0.4}_{-0.3}$&0.03&$4.2 \times 10^{-13}$&$4.35 \times 10^{38}$ &$1.26 \times 10^{3}$&$3.79 \times 10^{12}$&$9.01 \times 10^{5}$\\

         J0540-6919 &$49.7\pm1.1$&0.05&$4.79 \times 10^{-13}$&$1.46 \times 10^{38}$ &$1.67 \times 10^{3}$&$4.98 \times 10^{12}$&$3.48 \times 10^{5}$\\

         J0633+0632 &$1.35\pm0.65$&0.297&$7.96 \times 10^{-14}$&$1.19 \times 10^{35}$ &$5.92 \times 10^{4}$&$4.92 \times 10^{12}$&$1.7 \times 10^{3}$\\

         J0633+1746 &$0.2\pm0.1$&0.237&$1.1 \times 10^{-14}$&$3.25 \times 10^{34}$ &$3.42 \times 10^{5}$&$1.63 \times 10^{12}$&$1.11 \times 10^{3}$\\

         J0659+1414 &$0.29\pm0.03$&0.385&$5.5 \times 10^{-14}$&$3.8 \times 10^{34}$ &$1.11 \times 10^{5}$&$4.65 \times 10^{12}$&$7.4 \times 10^{2}$\\

         J0835-4510 &$0.28\pm0.02$&0.089&$1.22 \times 10^{-13}$&$6.76 \times 10^{36}$ &$1.13 \times 10^{4}$&$3.38 \times 10^{12}$&$4.24 \times 10^{4}$\\

         J0922+0638 &$1.1^{+0.2}_{-0.1}$&0.43&$1.37 \times 10^{-14}$&$6.77 \times 10^{33}$ &$4.97 \times 10^{5}$&$2.46 \times 10^{12}$&$2.79 \times 10^{2}$\\

         J1016-5857 &$3.16\pm1.26$&0.107&$8.04 \times 10^{-14}$&$2.56 \times 10^{36}$ &$2.1 \times 10^{4}$&$2.98 \times 10^{12}$&$2.18 \times 10^{4}$\\

         J1023-5746 &$4.02^{+2.08}_{-1.12}$&0.11&$3.8 \times 10^{-13}$&$1.08 \times 10^{37}$ &$4.6 \times 10^{3}$&$6.62 \times 10^{12}$&$4.31 \times 10^{4}$\\

         J1048-5832 &$2.9^{+1.2}_{-0.7}$&0.124&$9.55 \times 10^{-14}$&$1.99 \times 10^{36}$ &$2.04 \times 10^{4}$&$3.49 \times 10^{12}$&$1.67 \times 10^{4}$\\

         J1057-5226 &$0.09\pm0.04$&0.197&$5.8 \times 10^{-15}$&$3.01 \times 10^{34}$ &$5.35 \times 10^{5}$&$1.09 \times 10^{12}$&$1.28 \times 10^{3}$\\

         J1112-6103 &$4.46\pm1.79$&0.065&$3.15 \times 10^{-14}$&$4.54 \times 10^{36}$ &$3.27 \times 10^{4}$&$1.45 \times 10^{12}$&$4.78 \times 10^{4}$\\

         J1124-5916 &$4.8^{+1.2}_{-0.7}$&0.135&$7.52 \times 10^{-13}$&$1.19 \times 10^{37}$ &$2.85 \times 10^{3}$&$1.02 \times 10^{13}$&$3.72 \times 10^{4}$\\

         J1357-6429 &$3.1\pm1.24$&0.166&$3.54 \times 10^{-13}$&$3.05 \times 10^{36}$ &$7.31 \times 10^{3}$&$7.83 \times 10^{12}$&$1.53 \times 10^{4}$\\

         J1418-6058 &$1.6\pm0.7$&0.11&$1.71 \times 10^{-13}$&$4.99 \times 10^{36}$ &$1.03 \times 10^{4}$&$4.38 \times 10^{12}$&$2.95 \times 10^{4}$\\

         J1420-6048 &$5.63\pm2.25$&0.0682&$8.24 \times 10^{-14}$&$1.03 \times 10^{37}$ &$1.3 \times 10^{4}$&$2.41 \times 10^{12}$&$6.85 \times 10^{4}$\\

         J1459-6053 &$2.27^{+1.38}_{-0.632}$&0.103&$2.53 \times 10^{-14}$&$9.08 \times 10^{35}$ &$6.47 \times 10^{4}$&$1.63 \times 10^{12}$&$1.35 \times 10^{4}$\\

         J1509-5850 &$3.37\pm1.35$&0.089&$9.2 \times 10^{-15}$&$5.16 \times 10^{35}$ &$1.54 \times 10^{5}$&$9.14 \times 10^{11}$&$1.18 \times 10^{4}$\\

         J1709-4429 &$2.6^{+0.5}_{-0.6}$&0.102&$9.48 \times 10^{-14}$&$3.48 \times 10^{36}$ &$1.75 \times 10^{4}$&$3.12 \times 10^{12}$&$2.65 \times 10^{4}$\\

         J1718-3825 &$3.49\pm1.4$&0.075&$1.32 \times 10^{-14}$&$1.25 \times 10^{36}$ &$8.95 \times 10^{4}$&$1.01 \times 10^{12}$&$2.18 \times 10^{4}$\\

         J1732-3131 &$0.64\pm0.26$&0.196&$2.8 \times 10^{-14}$&$1.45 \times 10^{35}$ &$1.11 \times 10^{5}$&$2.38 \times 10^{12}$&$2.84 \times 10^{3}$\\

         J1740+1000 &$1.23\pm0.49$&0.154&$2.13 \times 10^{-14}$&$2.30 \times 10^{35}$ &$1.14 \times 10^{5}$&$1.84 \times 10^{12}$&$4.54 \times 10^{3}$\\

         J1741-2054 &$0.3^{+0.7}_{-0.1}$&0.413&$1.7 \times 10^{-14}$&$9.48 \times 10^{33}$ &$3.86 \times 10^{5}$&$2.68 \times 10^{12}$&$3.43 \times 10^{2}$\\

         J1747-2958 &$2.52\pm 1$&0.098&$6.13 \times 10^{-14}$&$2.51 \times 10^{36}$ &$2.55 \times 10^{4}$&$2.49 \times 10^{12}$&$2.34 \times 10^{4}$\\

         J1801-2451 &$3.8\pm1.52$&0.125&$8.95 \times 10^{-14}$&$1.81 \times 10^{36}$ &$1.55 \times 10^{4}$&$4.04 \times 10^{12}$&$1.57 \times 10^{4}$\\

         J1809-2332 &$1.7\pm1$&0.147&$3.44 \times 10^{-14}$&$4.29 \times 10^{35}$ &$6.76 \times 10^{4}$&$2.27 \times 10^{12}$&$6.51 \times 10^{3}$\\

         J1813-1246 &$2.64\pm0.14$&0.048&$1.76 \times 10^{-14}$&$6.24 \times 10^{36}$ &$4.34 \times 10^{4}$&$9.3 \times 10^{11}$&$7.58 \times 10^{4}$\\

         J1826-1256 &$3.5\pm0.1$&0.11&$1.21 \times 10^{-13}$&$3.56 \times 10^{36}$ &$1.44 \times 10^{4}$&$3.7 \times 10^{12}$&$2.5 \times 10^{4}$\\

         J1833-1034 &$4.1\pm0.3$&0.062&$2.02 \times 10^{-13}$&$3.36 \times 10^{37}$ &$4.85 \times 10^{3}$&$3.58 \times 10^{12}$&$1.37 \times 10^{5}$\\

         J1836+5925& $0.5\pm0.3$&0.173&$1.5 \times 10^{-15}$&$1.14 \times 10^{34}$ &$1.83 \times 10^{6}$&$5.16 \times 10^{11}$&$8.98 \times 10^{2}$\\

         J1838-0537 &$4.02^{+2.47}_{-1.13}$&0.146&$4.51 \times 10^{-13}$&$5.75 \times 10^{36}$ &$4.89 \times 10^{3}$&$8.39 \times 10^{12}$&$2.4 \times 10^{4}$\\

         J1907+0602 &$2.6\pm1.04$&0.106&$8.65 \times 10^{-14}$&$2.81 \times 10^{36}$ &$1.95 \times 10^{4}$&$3.08 \times 10^{12}$&$2.30 \times 10^{4}$\\

         J1952+3252 &$3\pm 2$&0.039&$5.8 \times 10^{-15}$&$3.72 \times 10^{36}$ &$1.07 \times 10^{5}$&$4.86 \times 10^{11}$&$7.12 \times 10^{4}$\\

         J1957+5033 &$0.55\pm0.45$&0.374&$7.08 \times 10^{-15}$&$5.31 \times 10^{33}$ &$8.39 \times 10^{5}$&$1.65 \times 10^{12}$&$2.84 \times 10^{2}$\\

         J2021+3651 &$1.8^{+1.7}_{-1.4}$&0.104&$9.48 \times 10^{-14}$&$3.35 \times 10^{36}$ &$1.72 \times 10^{4}$&$3.19 \times 10^{12}$&$2.58 \times 10^{4}$\\

         $\mathrm{J2021+4026_{LGF}}$ &$2.15\pm0.45$&0.2653&$5.77\times 10^{-14}$&$1.22 \times 10^{35}$ &$7.28 \times 10^{4}$&$3.97 \times 10^{12}$&$1.93 \times 10^{3}$\\

         $\mathrm{J2021+4026_{HGF}}$ &$2.15\pm0.45$&0.2653&$5.45\times 10^{-14}$&$1.14 \times 10^{35}$ &$7.71 \times 10^{4}$&$3.85 \times 10^{12}$&$1.87 \times 10^{3}$\\

         J2022+3842 &$10^{+4}_{-2}$&0.048&$8.62 \times 10^{-14}$&$2.97 \times 10^{37}$ &$8.94 \times 10^{3}$&$2.07 \times 10^{12}$&$1.64\times 10^{5}$\\

         J2043+2740 &$1.48\pm0.592$&0.096&$1.2 \times 10^{-15}$&$5.5 \times 10^{34}$ &$1.2 \times 10^{6}$&$3.54 \times 10^{11}$&$3.51 \times 10^{3}$\\

         J2055+2539 &$0.62\pm0.15$&0.32&$4.1 \times 10^{-15}$&$4.96 \times 10^{33}$ &$1.24 \times 10^{6}$&$1.16 \times 10^{12}$&$3.22 \times 10^{2}$\\

         J2229+6114 &$3\pm1.2$&0.052&$7.53 \times 10^{-14}$&$2.16 \times 10^{37}$ &$1.05 \times 10^{4}$&$2.03 \times 10^{12}$&$1.31 \times 10^{5}$\\
         \hline
    \end{tabular}
\end{table*}

\begin{table*}
    \caption{Spectral properties of the 44 pulsars.
    Columns 2-5 list gamma-ray emission properties. $\Gamma_{100}$ denotes the photon index, $E_{\rm p}$ the peak energy, and $d_{\rm p}$ the curvature of the gamma-ray spectrum around the peak energy.
    Columns 6-10 are of X-rays.
    $\Gamma_{\rm p}$ is the power-law photon index of the X-ray spectrum. $kT$ is the surface temperature and $R$ is the emission radius from the X-ray spectral fitting. 
    Those pulsars without $kT$ and $R$ information are designated as Group 1 pulsars and the others are Group 2 ones. X-ray spectral information is from the following references listed in the last column:
    (1) \citet{HK2023} (2) \citet{Marelli2013} (3) \citet{Li2008} (4) \citet{Ge2019}   (5)\citet{Danilenko2015} (6) \citet{Posselt2017} (7) \citet{Luca2005} (8) \citet{Pavlov2001} (9) \citet{Rigoselli2018} (10) \citet{Klingler2022} (11) \citet{Kargaltsev2007} (12) \citet{Posselt2015} (13) \citet{Chang2012} (14) \citet{Kim2020} (15) \citet{Pancrazi2012} (16) \citet{Klingler2016} (17) \citet{Romani2005} (18) \citet{Rigoselli2022} (19) \citet{Karpova2014} (20) \citet{Klingler2018} (21) \citet{kaspi2001} (22) \citet{Karpova2019} (23) \citet{Lin2014} (24) \citet{Coti2020} (25) \citet{Li2005} (26) \citet{Zyuzin2021} (27) \citet{Etten2008} (28) \citet{2013ApJ...770L...9L} (29) \citet{2013ApJ...777L...2A} (30) \citet{Wang2018} (31) \citet{2017ApJ...842...53Z}
    (32) \citet{Becker2004} (33) \citet{Marelli2016} }
    \label{Spectral_table}
    
    \begin{tabular}{lcccccccccr}
         \hline
         Pulsars & $L_{\gamma}$ (erg/sec) & $\Gamma_{100}$ & $E_{\rm p}$ (GeV) & $d_{\rm p}$ & $L_{\rm p}$ (erg/sec) & $\Gamma_{\rm p}$ & $kT$ (eV) & $R$ (km) & References\\
         \hline
         J0007+7303 & ${1.01^{+0.48}_{-0.39} \times 10^{35}} $ & $1.2 \pm 0.04$ &$2.44\pm0.03$&
         $0.36\pm0.01$
         &${8.81^{+4.3}_{-3.41} \times 10^{30}}$ &$1.71^{+0.3}_{0.28}$&--&-- & 1\\
         \\
         J0205+6449 & ${7.85^{+1.04}_{-0.97} \times 10^{34}} $ & $1.6\pm0.22$ &$0.26\pm0.1$&
         $0.45\pm0.08$ &
         ${1.06^{+0.14}_{-0.13} \times 10^{33}}$ &$1.47^{+0.03}_{-0.03}$&$167^{+31}_{-19}$&$1.3^{+2.02}_{-1.06}$ & 1\\
         \\
         J0357+3205 & ${2.58^{+3.22}_{-1.94} \times 10^{33}} $ & $0.73\pm0.23$ &$0.87\pm0.03$& $0.95\pm0.05$
         &${1.75^{+2.19}_{-1.3} \times 10^{30}}$ &$2.28^{+0.17}_{-0.16}$&$94^{+12}_{-9}$&$0.54^{+0.81}_{-0.49}$ &2 \\
         \\
         J0534+2200 & ${7.17^{+2.7}_{-2.3} \times 10^{35}} $ & $2\pm0.03$ &$0.04\pm0.03$&
         $0.23\pm0.01$
         &${1.32^{+0.61}_{-0.46} \times 10^{36}}$ &$1.63^{+0.09}_{-0.09}$&--&-- & 3 \\
         \\
         J0540-6919 & ${7.94^{+0.57}_{-0.56} \times 10^{36}} $ & - & 
         $0.02\pm0.08$&
         $0.26\pm0.05$
         &${3^{+0.2}_{-0.19} \times 10^{36}}$ &$0.78^{+0.09}_{-0.09}$&--&-- &3,4 \\
         \\
         J0633+0632 & ${2.10^{+2.47}_{-1.51} \times 10^{34}} $ & $1.1\pm0.17$ & $1.43\pm0.06$&
         $0.6\pm0.03$
         &${1.29^{+2.35}_{-0.92} \times 10^{31}}$ &$1.6^{+0.6}_{-0.6}$&$105^{+23}_{-18}$&$3.24^{+11.68}_{-2.86}$ &5\\
         \\
         J0633+1746 & ${2.01^{+2.48}_{-1.49} \times 10^{34}} $ &$1.1\pm0.01$ & $1.63\pm0.01$ & $0.66\pm0.0001$
         &${1.64^{+2.04}_{-1.22}\times 10^{30}}$ &$1.47^{+0.06}_{-0.07}$&$72^{+4}_{-4}$&$2.24^{+2.19}_{-2.07}$ &6\\
         \\
         J0659+1414 & ${2.72^{+0.6}_{-0.54} \times 10^{32}} $  &$1.5\pm0.27$ & $0.16\pm0.03$ &
         $0.86\pm0.09$
         &${1.51^{+0.64}_{-0.54} \times 10^{30}}$ &$2.30^{+0.68}_{-0.57}$&$66^{+3}_{-6}$&$0.41^{+0.36}_{-0.25}$ &7\\
         \\
         J0835-4510 & ${8.72^{+1.28}_{-1.19} \times 10^{34}} $ & $1.4\pm0.01$  &$1.27\pm0.01$ &
         $0.52\pm0.0001$ &
         ${5.27^{+1.13}_{-1} \times 10^{32}}$ &$2.7^{+0.4}_{-0.4}$&$128^{+3}_{-3}$&$2.35^{+1.57}_{-1.44}$ & 8 \\
         \\
         J0922+0638 & ${3.18^{+1.15}_{-0.92} \times 10^{32}} $  & $-$ & $-$ & $0.12\pm0.16$
         &${1.53^{+0.95}_{-0.75} \times 10^{30}}$ &$2.20^{+0.6}_{-0.6}$&$110^{+20}_{-30}$&$0.21^{+0.34}_{-0.16}$ & 9 \\
         \\
         J1016-5857 & ${8.36^{+8.07}_{-5.3} \times 10^{34}} $  &$1.5\pm0.58$& $0.89\pm0.3$&
         $0.63\pm0.1$
         &${7.2^{+6.84}_{-4.61} \times 10^{32}}$ &$1.08^{+0.08}_{-0.08}$&--&-- &10 \\
         \\
         J1023-5746 & ${2.9^{+3.16}_{-2} \times 10^{35}} $ & $-$ & $0.62\pm0.13$ & 
         $0.67\pm0.06$
         &${8.46^{+9.23}_{-5.83} \times 10^{33}}$ &$0.84^{+0.42}_{-0.42}$&--&-- &1\\
         \\
         J1048-5832 & ${1.91^{+1.45}_{-1.06} \times 10^{35}} $  & $1.2\pm0.15$ & $0.97\pm0.07$&
         $0.41\pm0.02$
         &${2.42^{+1.91}_{-1.31} \times 10^{31}}$ &$1.5^{+0.3}_{-0.3}$&--&--&11 \\
         \\
         J1057-5226 & ${3.1^{+2.97}_{-1.97} \times 10^{32}} $ & $1\pm0.06$ & $1.21\pm0.01$&
         $0.88\pm0.02$
         &${8.4^{+1.04}_{-0.54} \times 10^{28}}$ &$1.9^{+0.2}_{-0.2}$&$85^{+6}_{-6}$&$0.75^{+0.65}_{-0.56}$ &12 \\
         \\
         J1112-6103 & ${5.26^{+5.15}_{-3.31} \times 10^{34}} $ & $-$   &$0.55\pm0.54$ & 
         $0.55\pm0.18$&
         ${9.85^{+9.83}_{-6.11} \times 10^{31}}$ &$1.09^{+0.65}_{-0.51}$&--&--&1 \\
         \\
         J1124-5916 & ${1.68^{+0.73}_{-0.6} \times 10^{35}} $ & $1.5\pm0.22$ & $0.36\pm0.09$ &
         $0.41\pm0.04$
         &${1.13^{+0.65}_{-0.46} \times 10^{33}}$ &$0.94^{+0.07}_{-0.08}$&$460^{+70}_{-80}$&$0.34^{+0.34}_{-0.22}$ &1\\
         \\
         J1357-6429 & ${3.34^{+3.21}_{-2.11} \times 10^{34}} $  & $-$& $0.35\pm0.12$ &
         $0.7\pm0.12$
         &${6.2^{+7.25}_{-3.94} \times 10^{31}}$ &$1.72^{+0.55}_{-0.63}$&$140^{+60}_{-40}$&$2.52^{+18.69}_{-2.54}$&13 \\
         \\
         J1418-6058 & ${9.20^{+9.72}_{-6.28} \times 10^{34}} $  &$1.3\pm0.2$ &$1\pm0.13$&
         $0.76\pm0.05$&
         ${3.98^{+4.39}_{-2.65} \times 10^{31}}$ &$1.5^{+0.4}_{-0.4}$&--&-- & 14 \\
         \\
         J1420-6048 & ${4.94^{+4.74}_{-3.12} \times 10^{35}} $  &$-$
         &$0.73\pm0.18$& $0.78\pm0.1$ &
         ${3.75^{+3.6}_{-2.39} \times 10^{32}}$ &$0.63^{+0.33}_{-0.34}$&--&-- & 1\\
         \\
         J1459-6053 & ${7.41^{+7.95}_{-5.06} \times 10^{34}} $ & $1.6\pm0.16$ & $0.37\pm0.07$ & $0.44\pm0.03$
         &${4.38^{+4.72}_{-2.98} \times 10^{31}}$ &$1.08^{+0.18}_{-0.17}$&--&--&15 \\
         \\
         J1509-5850 & ${1.63^{+1.55}_{-1.05} \times 10^{35}} $ & $1.5\pm0.18$ & $1.22\pm0.13$ & $0.53\pm0.04$
         &${7.13^{+6.86}_{-4.56} \times 10^{31}}$ &$1.9^{+0.12}_{-0.12}$&--&--&16 \\
         \\
         J1709-4429 & ${1.13^{+5.24}_{-4.25} \times 10^{36}} $ & $1.3\pm0.04$ &$1.43\pm0.02$& $0.37\pm0.01$ &
         ${1.82^{+0.86}_{-0.68} \times 10^{32}}$ &$1.62^{+0.2}_{-0.2}$&$172^{+15}_{-14}$&$2.08^{+2.92}_{-1.66}$ &17 \\
         \\
         J1718-3825 & ${1.46^{+1.4}_{-0.93} \times 10^{35}} $  &$1.3\pm0.4$& $0.56\pm0.14$
         &$0.68\pm0.07$
         &${6.48^{+6.26}_{-4.09} \times 10^{31}}$ &$1.62^{+0.33}_{-0.3}$&--&-- & 1 \\
         \\
         J1732-3131 & ${8.86^{+8.43}_{-5.64} \times 10^{33}} $  &$0.37\pm0.31$& $2.09\pm0.05$&
         $1.04\pm0.05$&
         ${3.07^{+3.01}_{-1.94} \times 10^{30}}$ &$1.14^{+0.35}_{-0.22}$&--&-- &1\\
         \\
         J1740+1000 & ${6.46^{+6.52}_{-4.04} \times 10^{32}} $ & $-$&
         $0.33\pm0.14$ & $1.72\pm0.98$
         &${4.92^{+4.94}_{-3.04} \times 10^{30}}$ &$1.80^{+0.17}_{-0.17}$&$83^{+5}_{-5}$&$4.84^{+4.5}_{-3.65}$ &18\\
         \\
         J1741-2054 & ${1.29^{+5.39}_{-1.6} \times 10^{33}} $ & $0.96\pm0.15$ & 
         $0.84\pm0.02$ &
         $1.4\pm0.05$
         &${4.16^{+}_{-2.74} \times 10^{30}}$ &$2.66^{+0.06}_{-0.06}$&$60^{+2}_{-2}$&$5.1^{+12.57}_{-3.7}$ & 19\\
         \\
         J1747-2958 & ${1.22^{+1.16}_{-0.77} \times 10^{35}} $ & $0.62\pm0.54$ & $0.53\pm0.12$ & $0.68\pm0.06$
         &${1.48^{+1.42}_{-0.96} \times 10^{33}}$ &$1.55^{+0.04}_{-0.04}$&--&-- &20\\
         \\
         J1801-2451 & ${5.50^{+5.38}_{-3.51} \times 10^{34}} $ & $-$ &
         $0.7\pm0.47$& $0.46\pm0.12$&
         ${1.53^{+1.47}_{-0.98} \times 10^{33}}$ &$1.6^{+0.6}_{-0.5}$&--&-- &21\\
         \hline
    \end{tabular}
\end{table*}
\begin{table*}
    \contcaption{}
    \begin{tabular}{lccccccccccr}
         \hline
         Pulsars&$L_{\gamma}$  (erg/sec)& $\Gamma_{100}$  &$E_{\rm p}$ (GeV)& $d_{\rm p}$ &$L_{\rm p}$ (erg/sec)& $\Gamma_{\rm p}$&$kT$ (eV) &$R$ (km)&References\\
         \hline
         J1809-2332 & ${1.45^{+2.18}_{-1.19} \times 10^{35}} $ & $1.1\pm0.12$ & $1.29\pm0.05$ & $0.5\pm0.02$&${5.3^{+8.01}_{-4.28} \times 10^{31}}$ &$1.36^{+0.16}_{-0.15}$&$178^{+23}_{-20}$&$0.83^{+1.28}_{-0.79}$ & 1\\
         \\
         J1813-1246 & ${2.08^{+0.22}_{-0.21} \times 10^{35}} $ & $1.5\pm0.12$ & $0.48\pm0.05$ &
         $0.45\pm0.02$ &
         ${6.77^{+0.71}_{-0.68} \times 10^{32}}$ &$0.85^{+0.03}_{-0.03}$&--&-- &1\\
         \\
         J1826-1256 & ${6.01^{+0.4}_{-0.38} \times 10^{35}} $ & $1.6\pm0.1$ & $0.87\pm0.07$ & $0.63\pm0.03$
         &${1.18^{+0.17}_{-0.16} \times 10^{32}}$ &$0.92^{+0.25}_{-0.24}$&--&--&22 \\
         \\
         J1833-1034 & ${1.8^{+0.3}_{-0.27} \times 10^{35}} $ & $1.1\pm0.76$ & $0.31\pm0.34$ & $0.49\pm0.15$&${3.77^{+0.94}_{-0.82} \times 10^{33}}$ &$1.5^{+0.33}_{-0.31}$&--&-- &1\\
         \\
         J1836+5925 & ${1.85^{+2.9}_{-1.52} \times 10^{34}} $  &$1.1\pm0.03$ & $1.48\pm0.01$ & $0.58\pm0.01$
         &${8.79^{+1.41}_{-6.86} \times 10^{29}}$ &$2.1^{+0.3}_{-0.3}$&$45^{+10}_{-8}$&$3.93^{+5.04}_{-1.71}$&23 \\
         \\
         J1838-0537 & ${2.32^{+2.55}_{-1.6} \times 10^{35}} $ & $0.92\pm0.66$ & $1.28\pm0.35$ & $0.66\pm0.09$
         &${1.07^{+1.92}_{-0.79} \times 10^{32}}$ &$1.2^{+1}_{-1}$&--&-- &24\\
         \\
         J1907+0602 & ${2.42^{+2.32}_{-1.54} \times 10^{35}} $ & $1.2\pm0.16$ & $0.95\pm0.06$ & $0.48\pm0.02$
         &${5.69^{+6}_{-3.52} \times 10^{31}}$ &$0.76^{+1.07}_{-0.34}$&--&-- & 1\\
         \\
         J1952+3252 & ${1.62^{+2.83}_{-1.38} \times 10^{35}} $ & $1.3\pm0.12$ & $0.99\pm0.05$ & $0.42\pm0.02$
         &${2.91^{+5.12}_{-2.47} \times 10^{33}}$ &$1.63^{+0.05}_{-0.05}$&$130^{+20}_{-20}$&$3.3^{+4.26}_{-2.04}$ &25\\
         \\
         J1957+5033 & ${9.44^{+21.3}_{-8.63} \times 10^{32}} $ & $0.5\pm0.4$ & $0.76\pm0.03 $ & $1.03\pm0.07$
         &${1.05^{+2.4}_{-0.94} \times 10^{30}}$ &$1.76^{+0.11}_{-0.11}$&$54^{+5}_{-7}$&$2.22^{+8.21}_{-1.91}$ & 26 \\
         \\
         J2021+3651 & ${1.9^{+4.59}_{-1.78} \times 10^{35}} $ & $0.99\pm0.1$ & $1.06\pm0.03$ & $0.5\pm0.01$
         &${1.86^{+4.49}_{-1.72} \times 10^{31}}$ &$1.73^{+1.15}_{-1.02}$&$159^{+17}_{-21}$&$1.25^{+1.99}_{-0.68}$&27 \\
         \\
         $\mathrm{J2021+4026_{LGF}}$ & ${3.79^{+1.74}_{-1.43}\times 10^{35}}$ & $1.662\pm0.011$ & $0.827\pm0.002$ &
         $-$
         &${1.23^{+0.85}_{-0.58}\times10^{31}}$ & $1^{+0.7}_{-0.8}$ &$210^{+30}_{-30}$ &$0.272^{+0.591}_{-0.255}$ & 28,29
         \\
         \\
         $\mathrm{J2021+4026_{HGF}}$ & ${4.6^{+2.13}_{-1.72}\times 10^{35}}$ & $1.639\pm0.015$& $0.927\pm0.003$ & $-$&
         ${1.77^{+1.07}_{-0.76}\times10^{31}}$ & $1.3^{+0.9}_{-0.9}$ &$210^{+40}_{-40}$ &$0.255^{+0.622}_{-0.245}$ & 30,31
         \\
         \\
         J2022+3842 & ${3.24^{+2.33}_{-1.62} \times 10^{35}} $ & $-$ & $0.2\pm0.22$& $0.61\pm0.15$
         &${7.69^{+5.45}_{-3.91} \times 10^{33}}$ &$1.23^{+0.22}_{-0.22}$&$222^{+76}_{-37}$&$5.23^{+16.5}_{-4.56}$ &1\\
         \\
         J2043+2740 & ${2.36^{+2.27}_{-1.52} \times 10^{33}} $ & $-$ &$0.83\pm0.1$& $1.06\pm0.18$&
         ${2.85^{+3.4}_{-1.82} \times 10^{30}}$ &$3.1^{+1.1}_{-0.6}$&--&--& 32 \\
         \\
         J2055+2539 & ${2.28^{+1.29}_{-0.99} \times 10^{33}} $ & $0.69\pm0.17$&$1.21\pm0.03$ & $0.89\pm0.04$ &
         ${1.08^{+0.62}_{-0.47} \times 10^{30}}$ &$2.36^{+0.14}_{-0.14}$&--&--&33 \\
         \\
         J2229+6114 & ${2.58^{+2.48}_{-1.64} \times 10^{35}} $ & $1.5\pm0.07$ & $0.74\pm0.05$ & $0.29\pm0.01$
         &${8.29^{+7.92}_{-5.28} \times 10^{32}}$ &$1.39^{+0.11}_{-0.12}$&$126^{+19}_{-13}$&$5.17^{+10.13}_{-3.99}$ &1\\
         \hline  
    \end{tabular}
\end{table*}

\section{Correlation Analysis}
In this section, we report results of studying the correlation between GeV gamma-ray and keV X-ray spectral properties of the 44 pulsars in the hope to shed some light on our understanding of pulsar emission mechanisms. The Spearmann rank-order coefficient, $r_{\rm s}$, and its corresponding p-value are employed to indicate the significance of a possible correlation between two physical quantities.  The best linear fit for these parameters is calculated with the orthogonal distance regression (ODR) method when both variables have quoted errors. Otherwise, the weighted least squares method is used for the fitting. 

\subsection{Correlation between gamma-ray spectral properties and timing parameters}
We examine the correlation between gamma-ray spectral properties and the timing parameters for the 44 pulsars in our sample to verify consistency with previous studies. Specifically, we examine their relations with period ($P$), period derivative ($\dot{P}$), spin-down power ($\dot{E}$), characteristic age ($\tau$), surface magnetic field ($B_{\rm s}$), and magnetic field at the light cylinder ($B_{\rm lc}$). The definitions of these parameters are in the caption of Table~\ref{Timing_table}.

\begin{figure*}
\centering

\begin{subfigure}{0.32\textwidth}
  \includegraphics[width=\textwidth]{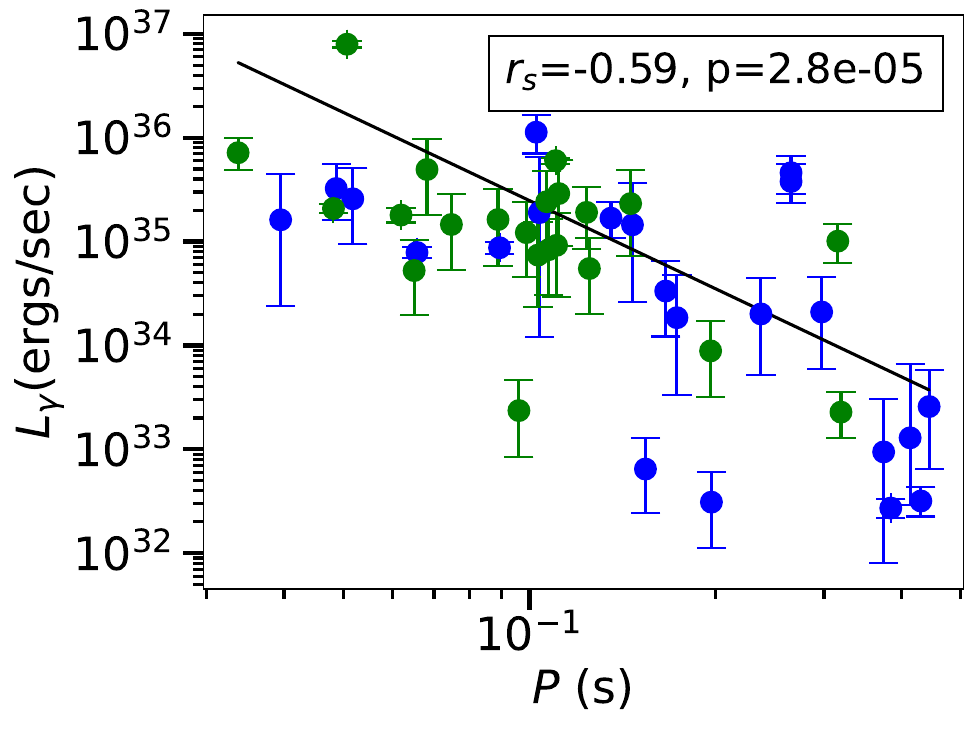}
\end{subfigure}
\hfill
\begin{subfigure}{0.32\textwidth}
  \includegraphics[width=\textwidth]{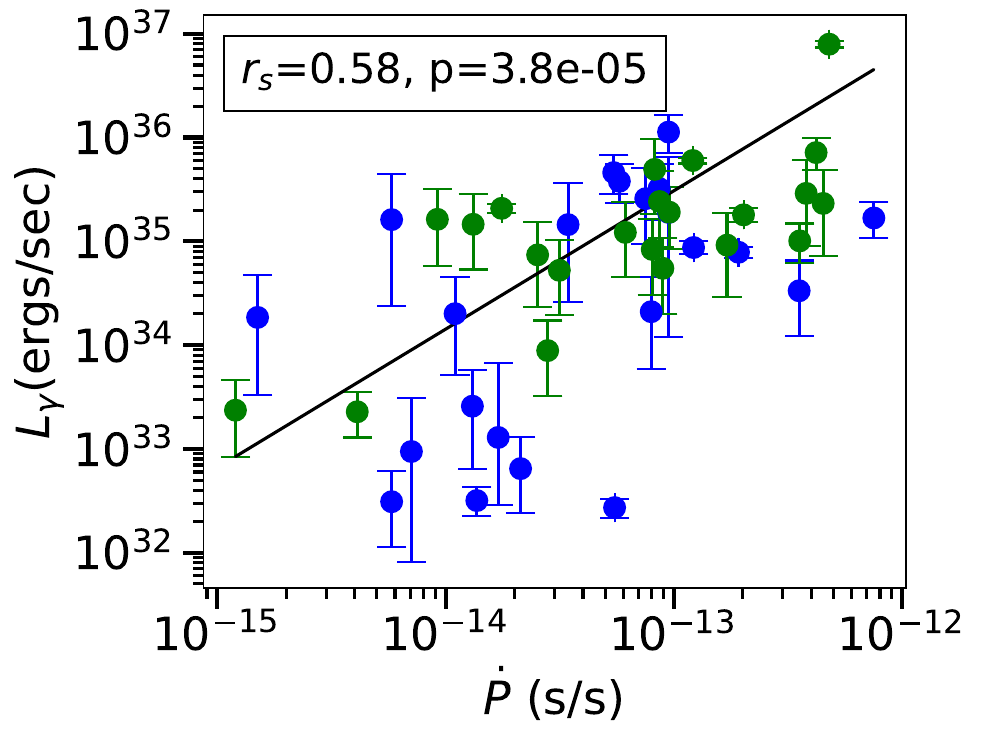}
\end{subfigure}
\hfill
\begin{subfigure}{0.32\textwidth}
  \includegraphics[width=\textwidth]{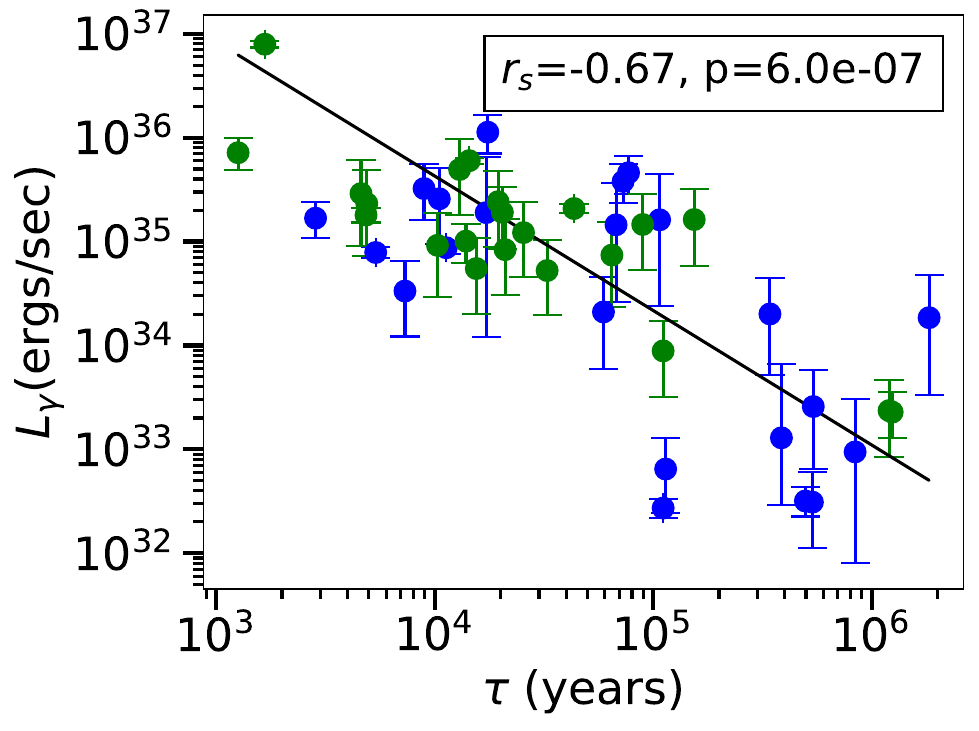}
\end{subfigure}

\vspace{3mm}

\begin{subfigure}{0.32\textwidth}
  \includegraphics[width=\textwidth]{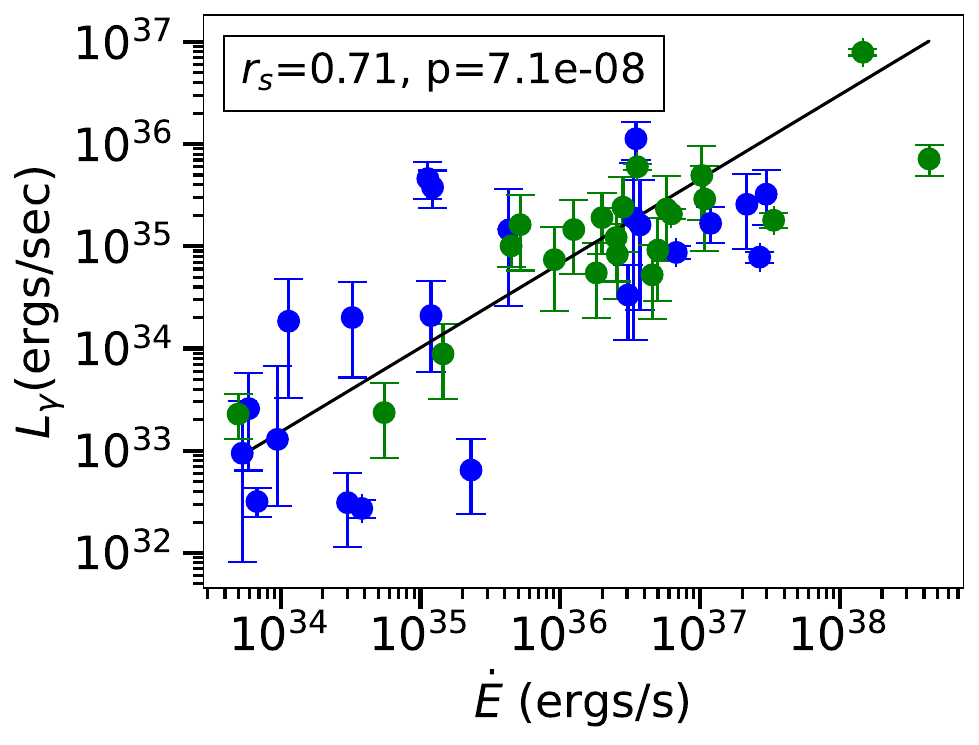}
\end{subfigure}
\hfill
\begin{subfigure}{0.32\textwidth}
  \includegraphics[width=\textwidth]{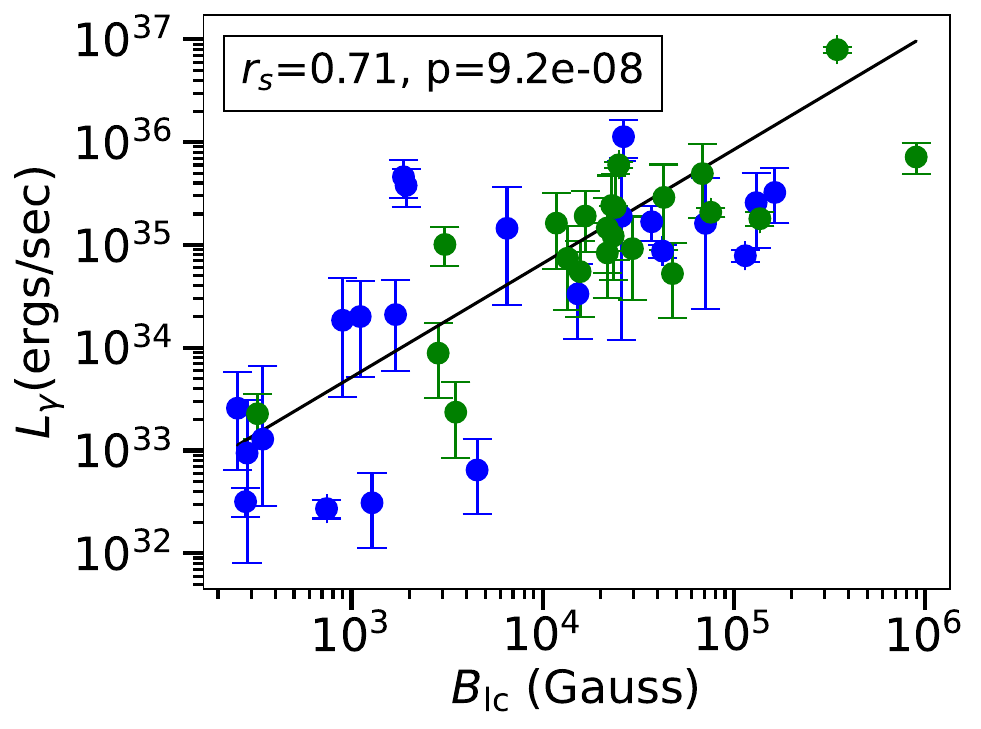}
\end{subfigure}
\hfill
\begin{subfigure}{0.32\textwidth}
  \includegraphics[width=\textwidth]{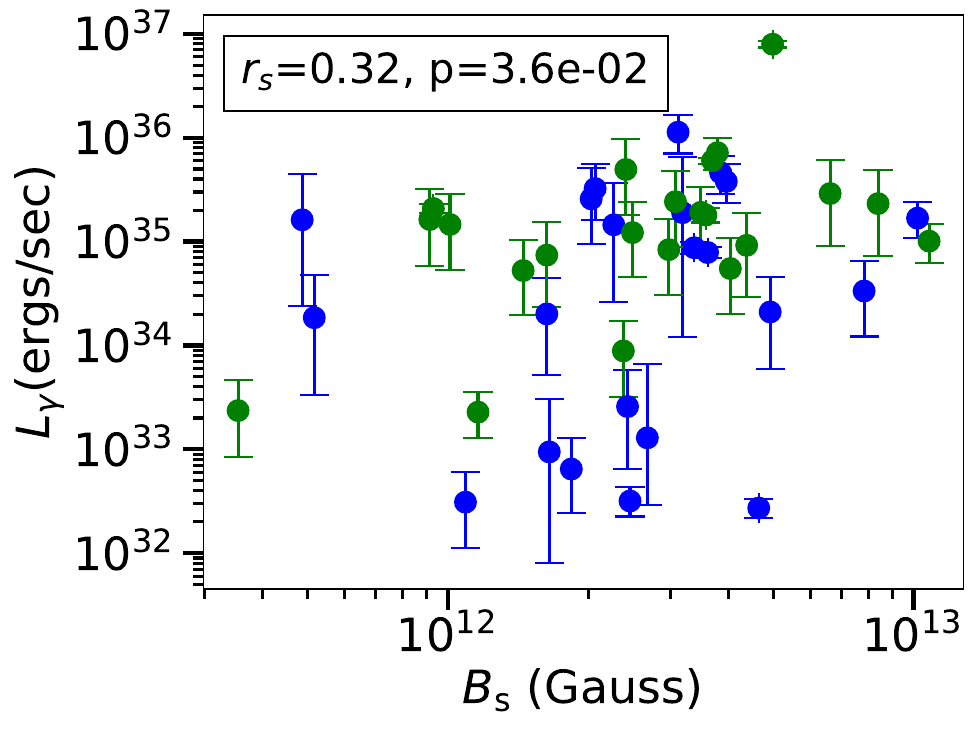}
\end{subfigure}

\caption{Gamma-ray luminosity ($L_{\gamma}$) versus timing parameters ($P$, $\dot{P}$, $\tau$, $\dot{E}$, $B_{\rm lc}$, $B_{\rm s}$).
  $r_{\rm s}$ in the legend is the Spearman rank-order correlation coefficient, followed by its corresponding p-value. Green solid data points are for Group-1 pulsars (without detected thermal X-rays) and blue solid ones are for Group 2 (with detected thermal X-rays). These two groups do not distinguish themselves in these distributions.
  The straight line is the best linear fit.}
\label{fig:Lgamma_vs_timing}
\end{figure*}

One can see that
the gamma-ray luminosity is clearly correlated with timing variables $P$, $\dot{P}$, $\tau$ and $\dot{E}$ (Fig. \ref{fig:Lgamma_vs_timing}). The best fit for these relations are
\begin{equation}
    \centering
    L_{\gamma}\propto \: P^{-2.81 \pm 0.49} \: \:(\chi^{2}_{\nu} = 57.6) 
\end{equation}
\begin{equation}
    \centering
    L_{\gamma}\propto \: \dot{P}^{1.33 \pm 0.24} \: \:(\chi^{2}_{\nu} = 59.6) 
\end{equation}
\begin{equation}
    \centering
    L_{\gamma}\propto \: \dot{E}^{0.82 \pm 0.1} \: \:(\chi^{2}_{\nu} = 36.6) 
    \label{gamma_edot}
\end{equation}
\begin{equation}
    \centering
    L_{\gamma}\propto \: \tau^{-1.29 \pm 0.16} \: \:(\chi^{2}_{\nu} = 40.4) 
\end{equation}

The relation between the gamma-ray luminosity and the spindown power (the middle-right panel in  Fig. \ref{fig:Lgamma_vs_timing} and Eq.(\ref{gamma_edot})) is consistent with those shown in Figure 9 of \citet{2013ApJS..208...17A} and in Figure 23 of \citet{Smith2023}.

We note that all the above reported $\chi^2_\nu$ and those in the following are only suggestive for various reasons, 
such as unknown beaming factors, viewing geometry and distance uncertainties.
There is usually a large scatter in the correlation
among spectral and timing parameters, as discussed in \citet{possenti2002}, \citet{Li2008} and \citet{HK2023}. Furthermore, in those earlier works, the distance uncertainty is all set to be 40\%. Typical $\chi^2_\nu$ obtained for those best linear fitting is around 3 or 4. In this work, we adopt the uncertainties associated with the reported distance estimates, most of which are smaller than 40\%. The best-fit $\chi^2_\nu$ is therefore much larger than those in earlier literature.

We also found that the gamma-ray luminosity is strongly correlated with $B_{\rm lc}$ with a Spearmann coefficient of 0.71 but it does not seem to be correlated with the surface magnetic field ($B_{\rm s}$) as shown in Fig. \ref{fig:Lgamma_vs_timing}. The relationship of the gamma-ray luminosity with $B_{\rm lc}$ follows as
\begin{equation}
    \centering
    L_{\gamma}\propto \: B_{\rm lc}^{1.11 \pm 0.14} \: \:(\chi^{2}_{\nu} =40.7 ) \,\,\, .
\end{equation}

Since the above timing parameters are all functions of $P$ and $\dot{P}$, a generic description for the gamma-ray luminosity as a function of $P$ and $\dot{P}$ might also be helpful. The best fit of such a function for our sample pulsars is 
\begin{equation}
    \centering
    L_{\gamma}\propto \: P^{-2.17 \pm 0.41}\dot{P}^{1.01\pm 0.20} \:(\chi^{2}_{\nu} = 36.5)
\end{equation}
This best fit does not improve the fitting in terms of $\chi^2_\nu$. However, it shows that $\dot{E}$, which is proportional to $P^{-3}\dot{P}$, is the most relevant quantity to the gamma-ray luminosity among all the timing parameters. 

The relationship between other gamma-ray spectral parameters and the timing properties of pulsars are shown in Fig.\ref{fig:gamma_100_timing}-\ref{fig:dp_timing}. 
Since their correlations are all weak or negligible,
they are put in the Appendix to allow a better reading  fluence in the main text. 
\textbf{Overall}, among the gamma-ray spectral properties of the 44 pulsars with X-ray emissions, only $L_\gamma$ shows clear correlations with $P$, $\dot{P}$, $\dot{E}$, $\tau$ and $B_{\rm lc}$.

\subsection{Correlation of gamma-ray spectral properties with non-thermal X-ray emissions}
\begin{figure}
  \centering
  \begin{subfigure}{0.48\columnwidth}
    \includegraphics[width=\textwidth]{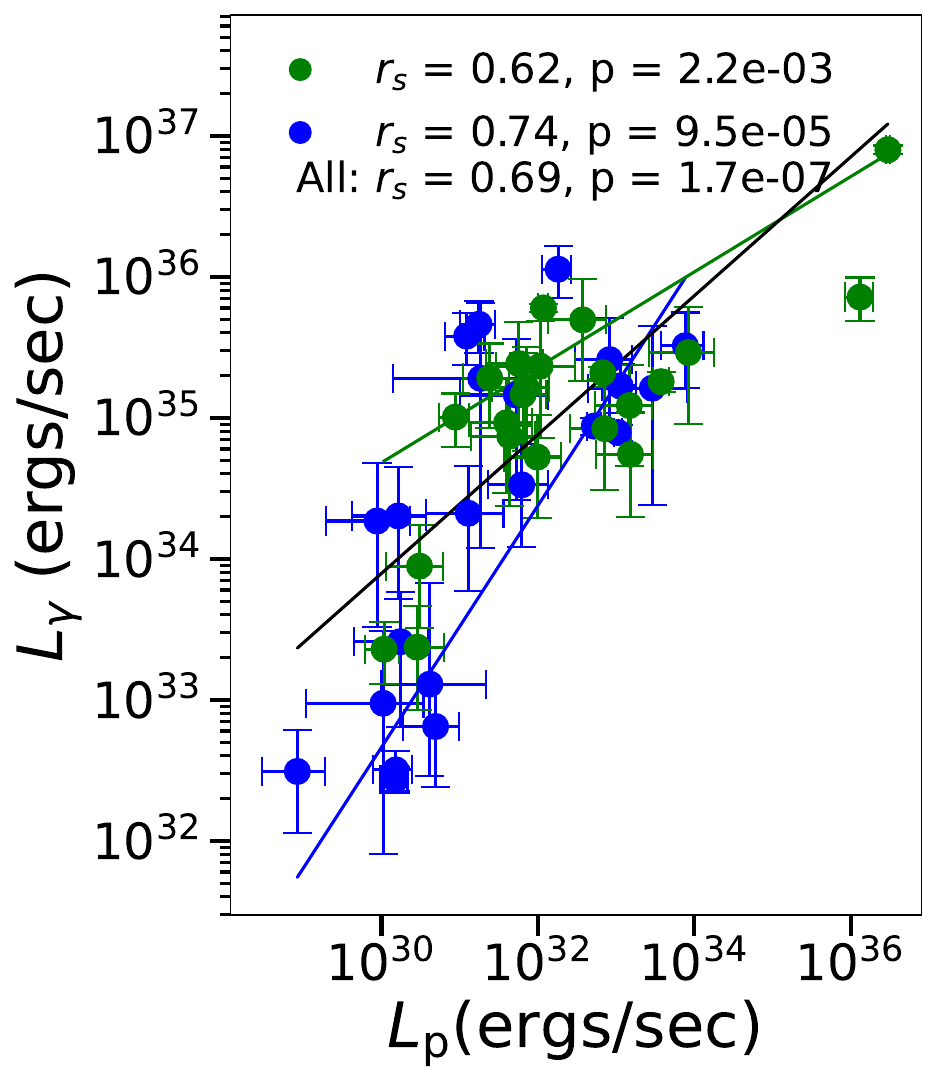}
  \end{subfigure}
  \hfill
  \begin{subfigure}{0.48\columnwidth}
    \includegraphics[width=\textwidth]{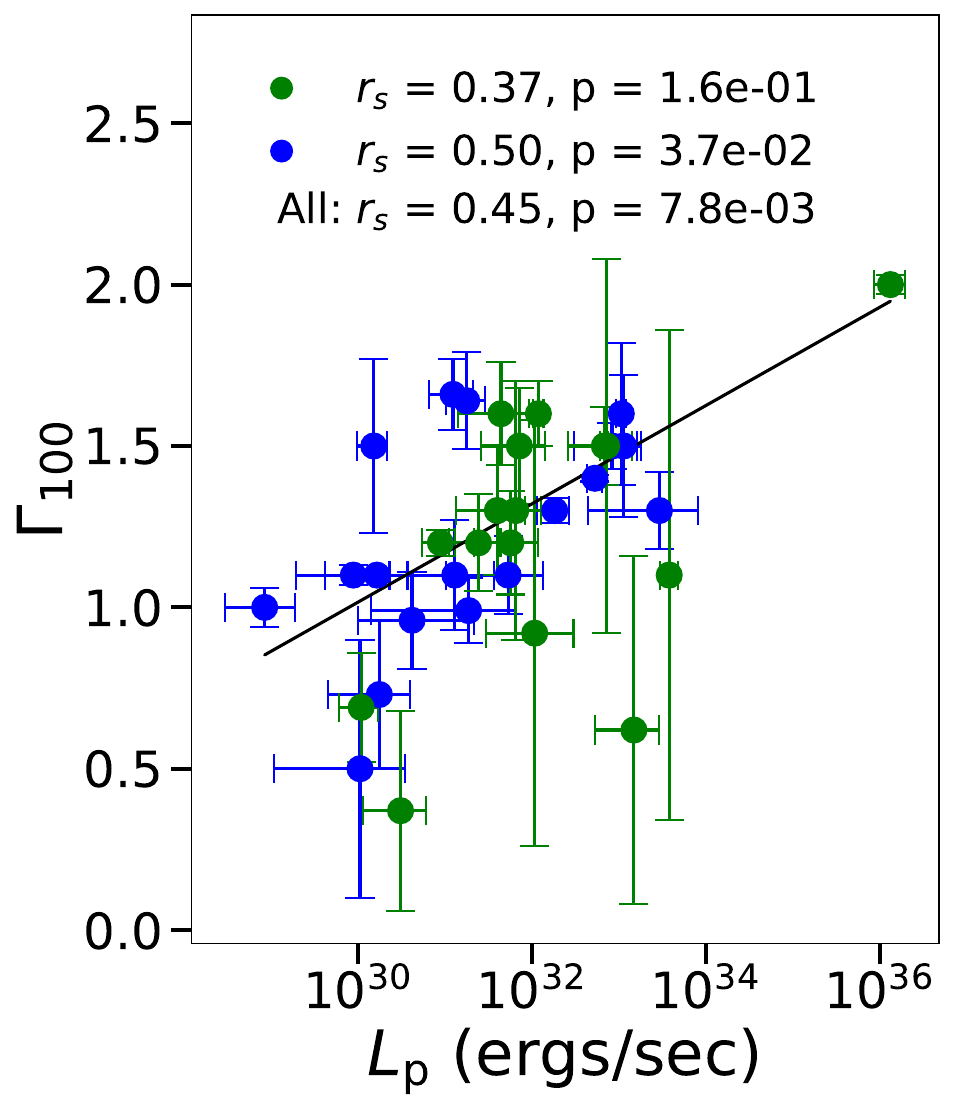}
  \end{subfigure}

  \vspace{0.3cm} 

  \begin{subfigure}{0.48\columnwidth}
    \includegraphics[width=\textwidth]{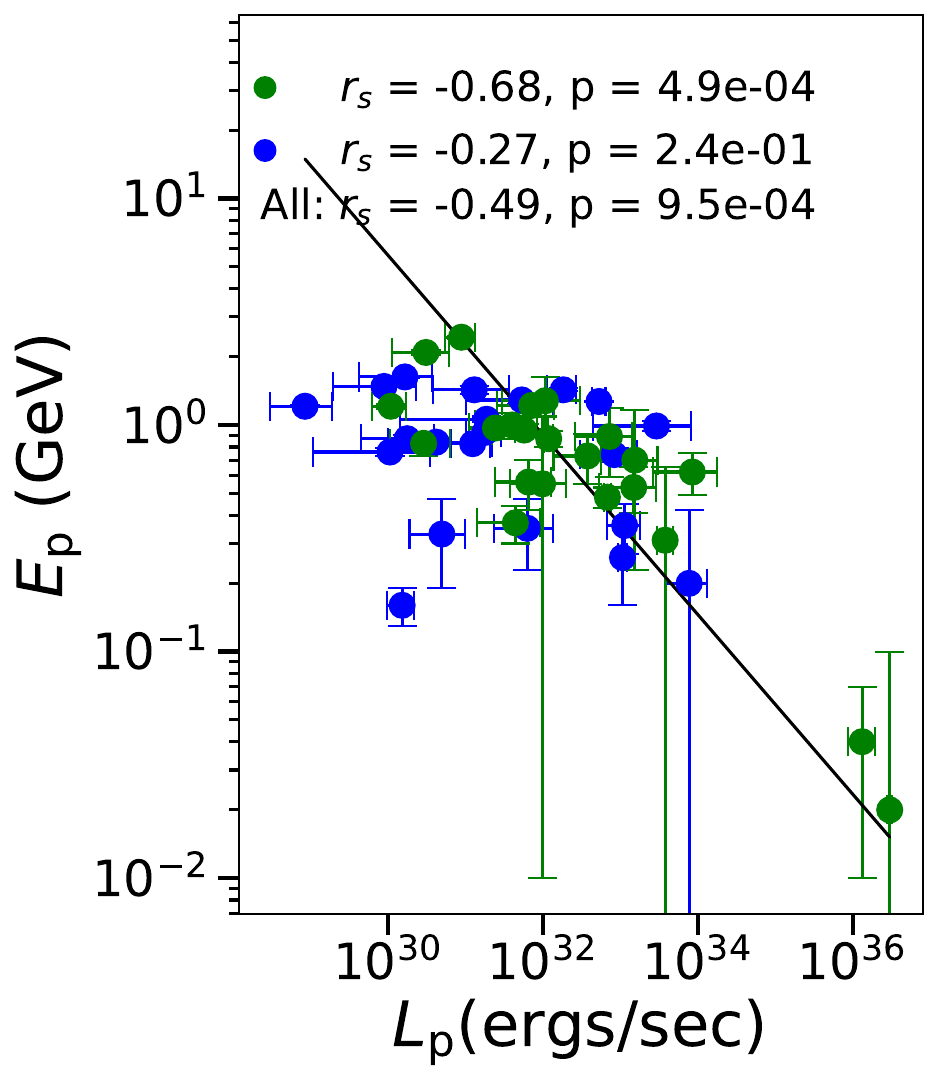}
  \end{subfigure}
  \hfill
  \begin{subfigure}{0.48\columnwidth}
    \includegraphics[width=\textwidth]{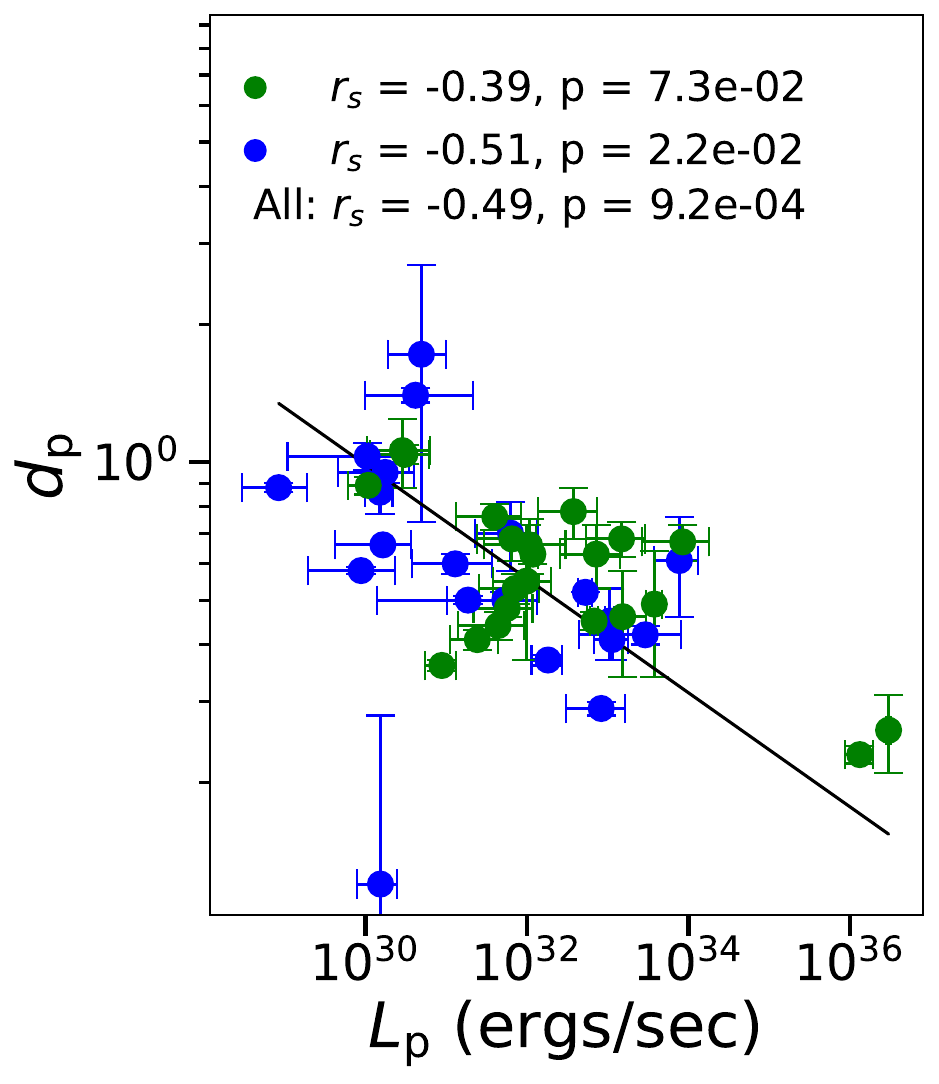}
  \end{subfigure}

  \caption{Gamma-ray spectral parameters versus non-thermal X-ray luminosity ($L_{\rm p}$). $r_{\rm s}$ in the legend is the Spearman rank-order correlation coefficient,followed by its corresponding p-value. Green solid data points are for Group-1 pulsars (without detected thermal X-rays) and blue solid ones are for Group 2 (with detected thermal X-rays). The black line indicates the best-fit for all data points, green line indicates the best-fit for Group 1 data and blue line indicates the best-fit for Group 2 data.}
  \label{fig:Gamma_spectral_vs_Lp}
\end{figure}

\begin{figure}
  \centering
  \begin{subfigure}{0.48\columnwidth}
    \includegraphics[width=\textwidth]{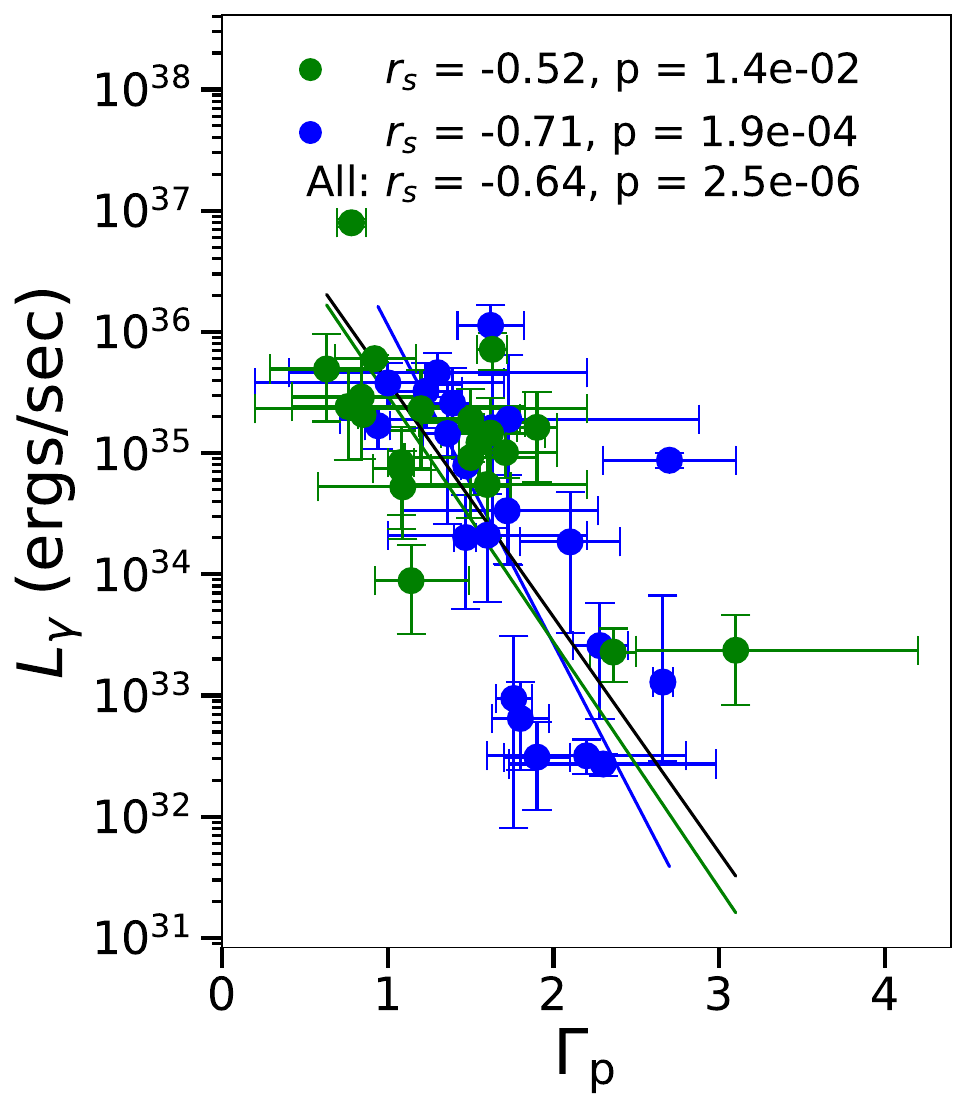}
  \end{subfigure}
  \hfill
  \begin{subfigure}{0.48\columnwidth}
    \includegraphics[width=\textwidth]{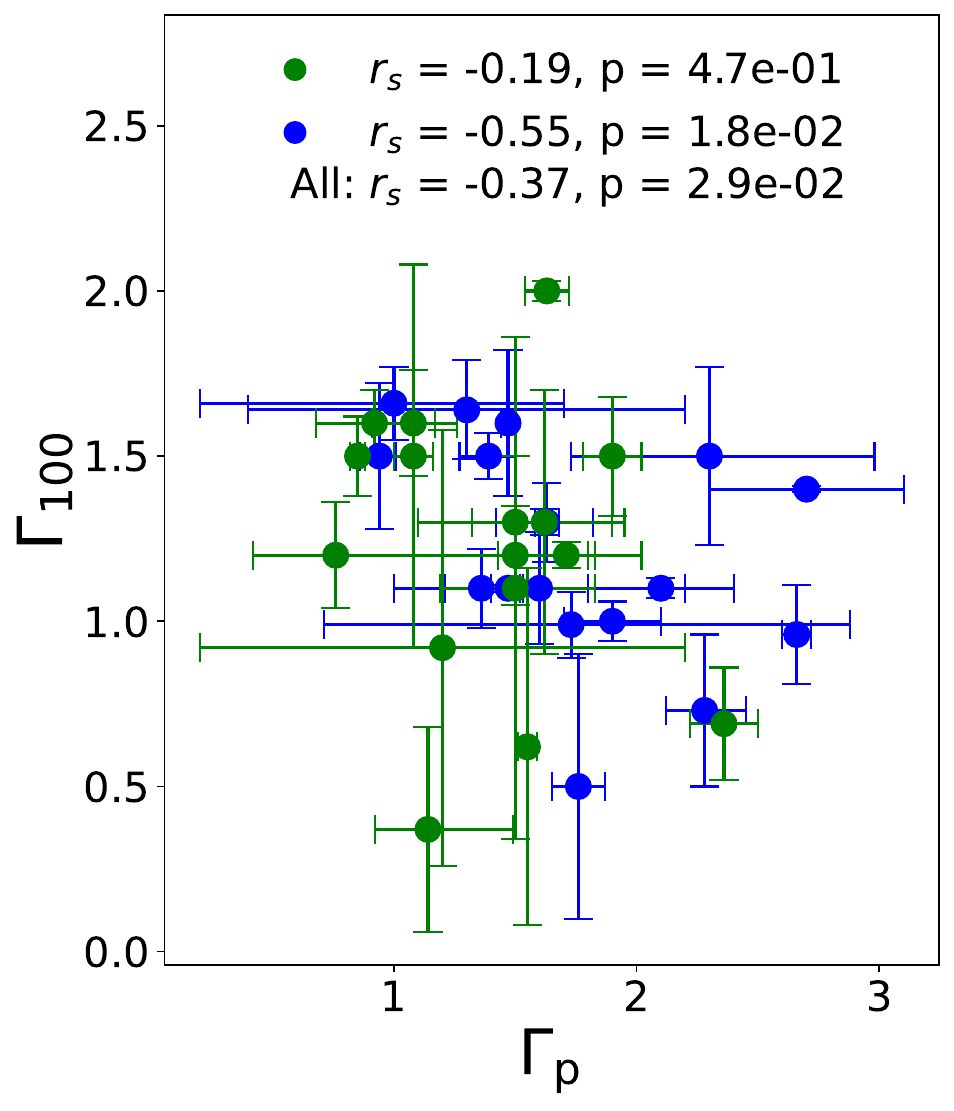}
  \end{subfigure}

  \vspace{0.3cm} 

  \begin{subfigure}{0.48\columnwidth}
    \includegraphics[width=\textwidth]{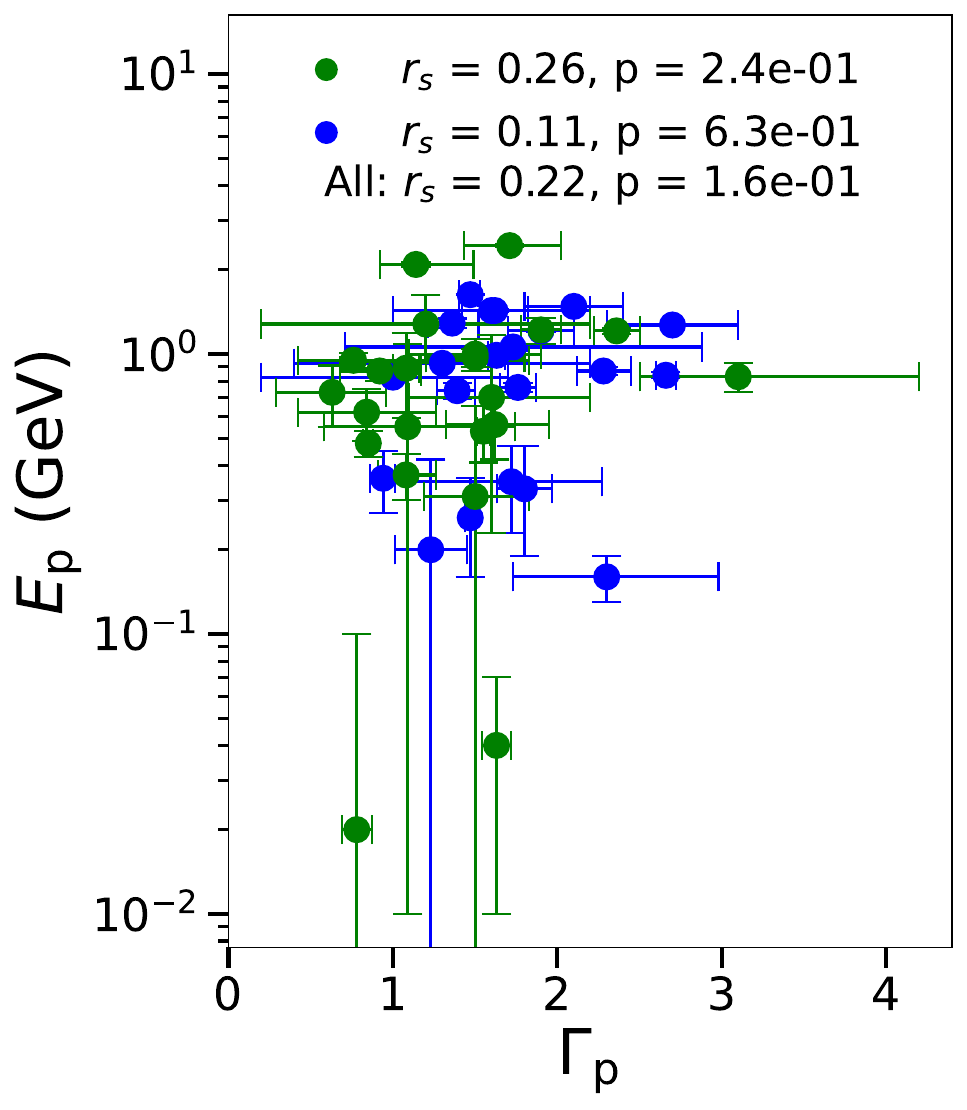}
  \end{subfigure}
  \hfill
  \begin{subfigure}{0.48\columnwidth}
    \includegraphics[width=\textwidth]{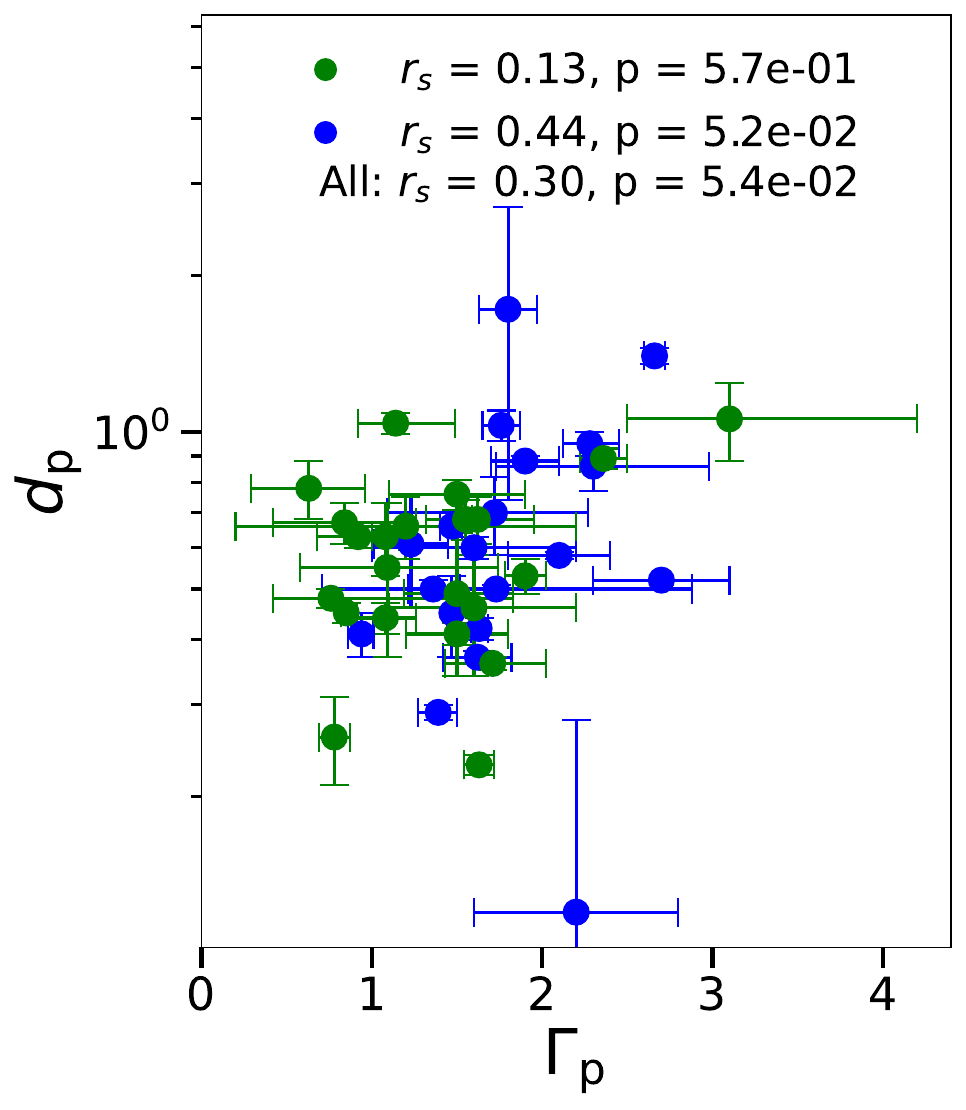}
  \end{subfigure}

  \caption{Gamma-ray spectral parameters vs non-thermal X-ray photon index ($\Gamma_{\rm p}$). Legends are the same as previous figures.}
  \label{fig:Gamma_spectral_vs_gammap}
\end{figure}

We check the correlation between gamma-ray spectral parameters ($L_{\gamma}, \Gamma_{100},d_{\rm p},E_{\rm p}$) and the non-thermal X-ray spectral parameters ($L_{\rm p}, \Gamma_{\rm p}$). They are shown in Fig. \ref{fig:Gamma_spectral_vs_Lp} $\&$ \ref{fig:Gamma_spectral_vs_gammap}.

For the relation between the gamma-ray luminosity and non-thermal X-ray emission, represented by the power-law component in the X-ray spectrum, we found that 
the gamma-ray luminosity and the non-thermal X-ray luminosity are strongly correlated as shown in Fig. \ref{fig:Gamma_spectral_vs_Lp}. A relationship between X-ray and gamma-ray luminosities has been reported as $L_{\rm{x-ray}} \propto L_{\gamma}^{1.22 \pm 0.21}$ \citep{Malov2019}. This relationship was obtained 
with the least-square method which only takes into account the error in the ordinate (y-direction). 
It gives an inconsistent relation when one
interchanges the two luminosities from the ordinate to abscissa and vice versa.  We use the orthogonal distance regression method which takes into account errors in both x- and y-directions to find the relationship from our sample. The best-fit relationship is given as 
\begin{equation}
   L_{\rm{p}} \propto L_{\gamma}^{2.03 \pm 0.19}\,\,\,
    (\chi^{2}_{\nu} = 23.1) \,\,\, . 
   \label{eq:lplg}
\end{equation}
The best fit for its inverse relationship, that is, putting $L_\gamma$ in the ordinate and $L_{\rm p}$ in the abscissa, is given as 
\begin{equation}
    L_{\gamma} \propto L_{\rm{p}}^{0.49 \pm 0.05}\,\,\,
    (\chi^{2}_{\nu} = 23.1) \,\,\, ,
    \label{L_gamma_L_xray}
\end{equation} 
which is consistent with Eq.(\ref{eq:lplg}).

As one can see from Fig. \ref{fig:Gamma_spectral_vs_Lp}, pulsars without and with detected thermal emission seem to show different dependence between $L_\gamma$ and $L_{\rm p}$.
We designate these two groups of pulsars as Group 1 (without detected thermal emission) and Group 2 (with detected thermal emission). The best-fit relationships between $L_\gamma$ and $L_{\rm p}$ for these two groups are 
\begin{equation}
    L_{\gamma} \propto L_{\rm{p}}^{0.34 \pm 0.04}\,\,\,
    (\chi^{2}_{\nu} = 17.4) \,\,\, (\mathrm{Group\: 1}) ,
    \label{L_gamma_L_xray_group1}
\end{equation}
and
\begin{equation}
    L_{\gamma} \propto L_{\rm{p}}^{0.86 \pm 0.14}\,\,\,
    (\chi^{2}_{\nu} = 9.8) \,\,\, (\mathrm{Group\: 2}) .
    \label{L_gamma_L_xray_group2}
\end{equation}
It is important to note that while the non-thermal X-ray luminosity is correlated with the gamma-ray luminosity, the gamma-ray energy flux shows no correlation with radio flux density, as shown in Fig. 6 of the 3PC.

For the rest of the gamma-ray spectral parameters ($E_{\rm p}$, $\Gamma_{100}$ and $d_{\rm p}$), the two groups separately do not show a strong correlation, except for $E_{\rm p}$ versus $L_{\rm p}$ in Group 1 pulsars (the lower left panel in Fig.\ref{fig:Gamma_spectral_vs_Lp}) The relationship for these gamma-ray spectral parameters with $L_{\rm p}$ is given below for readers' reference:
\begin{equation}
    \Gamma_{100}=(0.2\pm0.03) \log L_{\rm p} + (-5.04\pm1.03) \:\: (\chi^2_{\nu}=8.2)\,\,\, ,
\end{equation}
\begin{equation}
    E_{\rm p} \propto L_{\rm p}^{-0.40\pm0.12}\:\: (\chi^2_{\nu}=15.4)\,\,\, ,
\end{equation}
and
\begin{equation}
    d_{\rm p} \propto L_{\rm p}^{-0.12\pm0.02}\:\: (\chi^2_{\nu}=10.4)\,\,\, .
\end{equation}

The correlation of gamma-ray spectral parameters with non-thermal X-ray photon index ($\Gamma_{\rm p}$) is shown in Fig. \ref{fig:Gamma_spectral_vs_gammap}, in which one can see that only gamma-ray luminosity has an strong anti-correlation with the X-ray photon index with a Spearman coefficient of -0.64, whose best fit to describe the correlation is
 \begin{equation}
    \centering
    \log{L_{\gamma}}= \: ({-1.94 \pm 0.28})\Gamma_{\rm p} + (37.53\pm0.35)\: \:(\chi^{2}_{\nu} = 5.2) \,\,\, .
\end{equation}
It means that the luminosity of gamma rays increases when the non-thermal X-rays becomes harder. 
A similar anti-correlation between $L_{\rm p}$ and $\Gamma_{\rm p}$ was reported in \citet{HK2023}. Given the positive correlation between $L_\gamma$ and $L_{\rm p}$, it is not surprising to see the anti-correlation between $L_\gamma$ and $\Gamma_{\rm p}$. However, what these relations imply is yet to be revealed.
It is also interesting to check whether this relation differs between Group-1 and -2 pulsars. If fitted separately, the best fits are
\begin{equation}
    \centering
    \log{L_{\gamma}}= \: ({-2.03 \pm 0.49})\Gamma_{\rm p} + (37.5\pm0.5)\: \:(\chi^{2}_{\nu} = 7.5) \,\,\, (\rm{Group\: 1})\, ,
\end{equation}
and
\begin{equation}
    \centering
    \log{L_{\gamma}}= \: ({-2.65 \pm 0.57})\Gamma_{\rm p} + (38.72\pm0.85)\: \:(\chi^{2}_{\nu} = 2.6) \,\,\, (\rm{Group\: 2})\, ,
\end{equation}
which do not show a statistically significant difference.

To explore the dependence of $L_{\gamma}$ on $L_{\rm p}$ and $\Gamma_{\rm p}$ jointly, we adopt a function form like the following and their best fits are given as 
\begin{equation}
    L_\gamma\propto L_{\rm p}^{(0.33\pm 0.04)} e^{(-1.03\pm0.15)\Gamma_{\rm p}} \:\: (\chi^{2}_{\nu} = 2.9) \,\,\, ,
    \label{eqn:Lgamma_Lp_gammap}
\end{equation}
\begin{equation}
    L_\gamma\propto L_{\rm p}^{(0.34\pm 0.04)} e^{(-0.84\pm0.18)\Gamma_{\rm p}} \:\: (\chi^{2}_{\nu} = 2.6) \,\,\, (\rm{Group\:1})
\end{equation}
and
\begin{equation}
    L_\gamma\propto L_{\rm p}^{(0.21\pm 0.14)} e^{(-1.83\pm0.48)\Gamma_{\rm p}} \:\: (\chi^{2}_{\nu} = 2.6) \,\,\, (\rm{Group\:2})\, .
\end{equation}
In all the above relationships, that is, $L_\gamma(L_{\rm p})$, $L_\gamma(\Gamma_{\rm p})$ and
$L_\gamma(L_{\rm p},\Gamma_{\rm p})$, 
Group-2 pulsars seem to have a more sensitive dependence than Group 1, although it is only more apparent to separate these two group in $L_\gamma(L_{\rm p})$. 
Other gamma-ray spectral parameters as a function of $L_{\rm p}$ and $\Gamma_{\rm p}$ jointly are listed below:
\begin{equation}
\Gamma_{100}\propto L_{\rm p}^{(0.15\pm 0.01)} e^{(-0.27\pm0.08)\Gamma_{\rm p}} \:\: (\chi^{2}_{\nu} = 1.6) \,\,\, ,
\label{eqn:Gamma100_Lp_gammap}
\end{equation}
\begin{equation}
\Gamma_{100}\propto L_{\rm p}^{(0.17\pm 0.02)} e^{(-0.26\pm0.12)\Gamma_{\rm p}} \:\: (\chi^{2}_{\nu} = 1.4) \,\,\, (\rm{Group\:1})
\end{equation}
and
\begin{equation}
\Gamma_{100}\propto L_{\rm p}^{(0.12\pm 0.03)} e^{(-0.32\pm0.15)\Gamma_{\rm p}} \:\: (\chi^{2}_{\nu} = 1.8) \,\,\, (\rm{Group\:2})\, .
\end{equation}
\begin{equation}
    E_{\rm p}\propto L_{\rm p}^{(-0.42\pm 0.1)} e^{(-0.63\pm0.19)\Gamma_{\rm p}} \:\: (\chi^{2}_{\nu} = 4.3) \,\,\, ,  \label{eqn:Ep_Lp_gammap}
\end{equation}
\begin{equation}
    E_{\rm p}\propto L_{\rm p}^{(-0.37\pm 0.07)} e^{(-0.35\pm0.13)\Gamma_{\rm p}} \:\: (\chi^{2}_{\nu} = 2.3) \,\,\, (\rm{Group\:1})
\end{equation}
and
\begin{equation}
    E_{\rm p}\propto L_{\rm p}^{(-0.26\pm 0.13)} e^{(0.47\pm0.26)\Gamma_{\rm p}} \:\: (\chi^{2}_{\nu} = 13.2) \,\,\, (\rm{Group\:2})\, .
\end{equation}
\begin{equation}
    d_{\rm p}\propto L_{\rm p}^{(-0.07\pm 0.01)} e^{(0.24\pm0.05)\Gamma_{\rm p}} \:\: (\chi^{2}_{\nu} = 6.1) \,\,\, ,  \label{eqn:dp_Lp_gammap}
\end{equation}
\begin{equation}
    d_{\rm p}\propto L_{\rm p}^{(-0.11\pm 0.02)} e^{(-0.17\pm0.06)\Gamma_{\rm p}} \:\: (\chi^{2}_{\nu} = 7.1) \,\,\, (\rm{Group\:1})
\end{equation}
and
\begin{equation}
    d_{\rm p}\propto L_{\rm p}^{(-0.12\pm 0.02)} e^{(0.15\pm0.07)\Gamma_{\rm p}} \:\: (\chi^{2}_{\nu} = 3.9) \,\,\, (\rm{Group\:2})\, .
\end{equation}

\subsection{Correlation between gamma-ray spectral parameters and thermal X-ray spectral parameters}

\begin{figure}
  \centering
  \begin{subfigure}{0.48\columnwidth}
    \includegraphics[width=\textwidth]{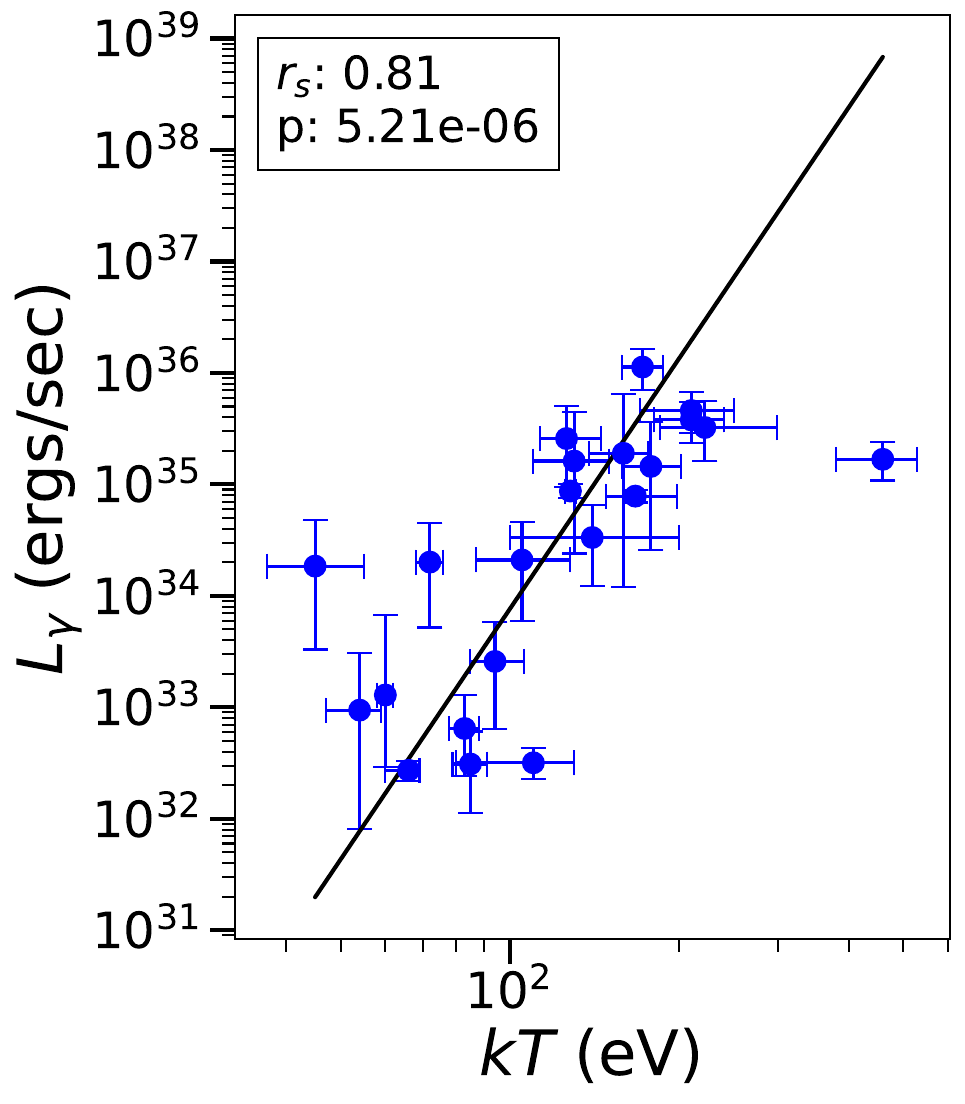}
  \end{subfigure}
  \hfill
  \begin{subfigure}{0.48\columnwidth}
    \includegraphics[width=\textwidth]{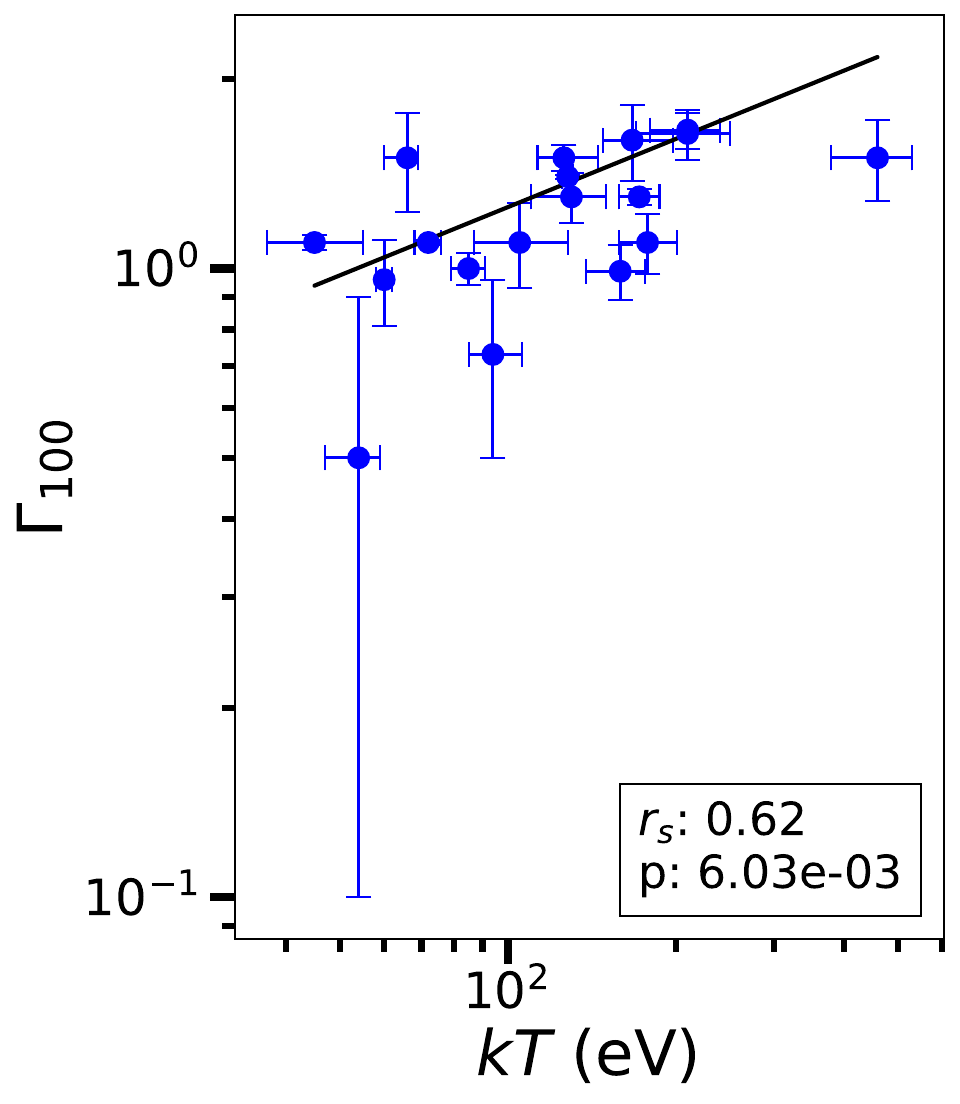}
  \end{subfigure}

  \vspace{0.3cm} 

  \begin{subfigure}{0.48\columnwidth}
    \includegraphics[width=\textwidth]{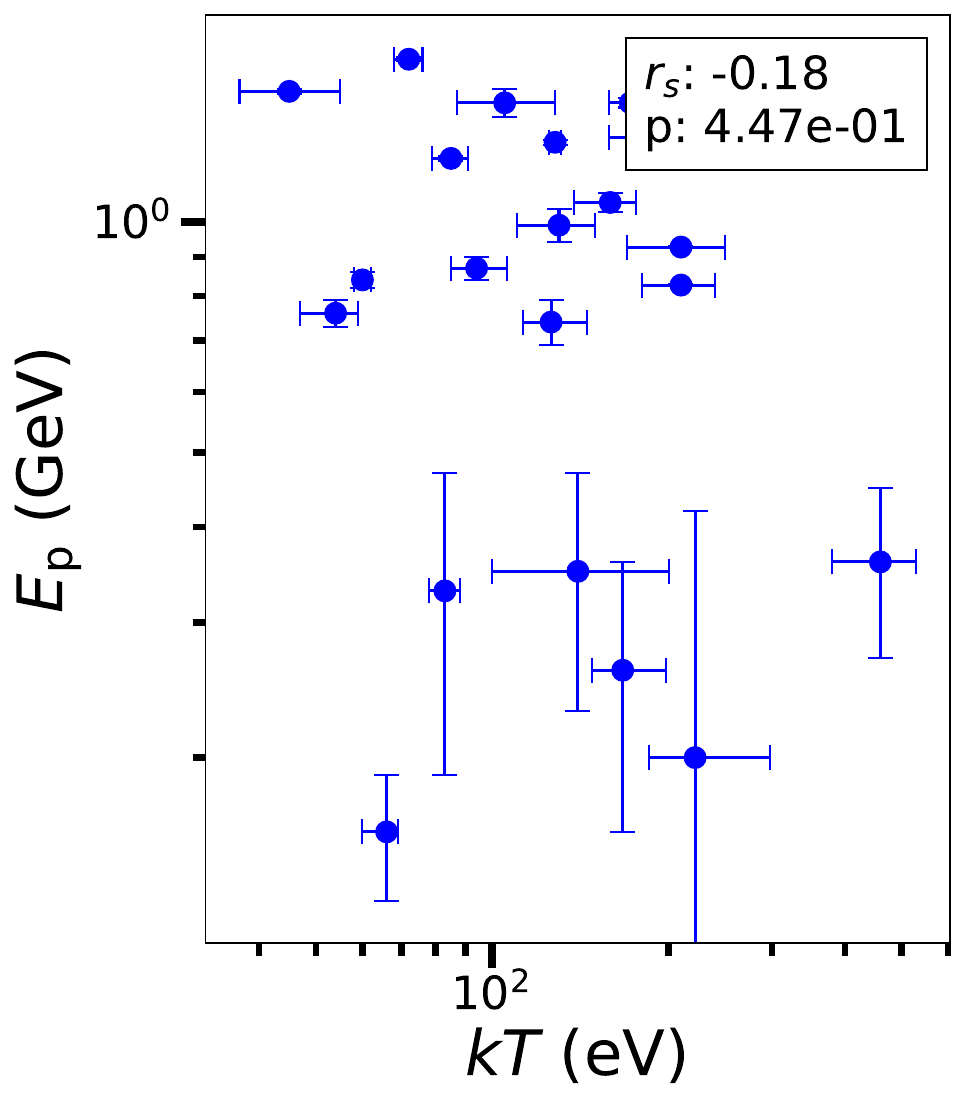}
  \end{subfigure}
  \hfill
  \begin{subfigure}{0.48\columnwidth}
    \includegraphics[width=\textwidth]{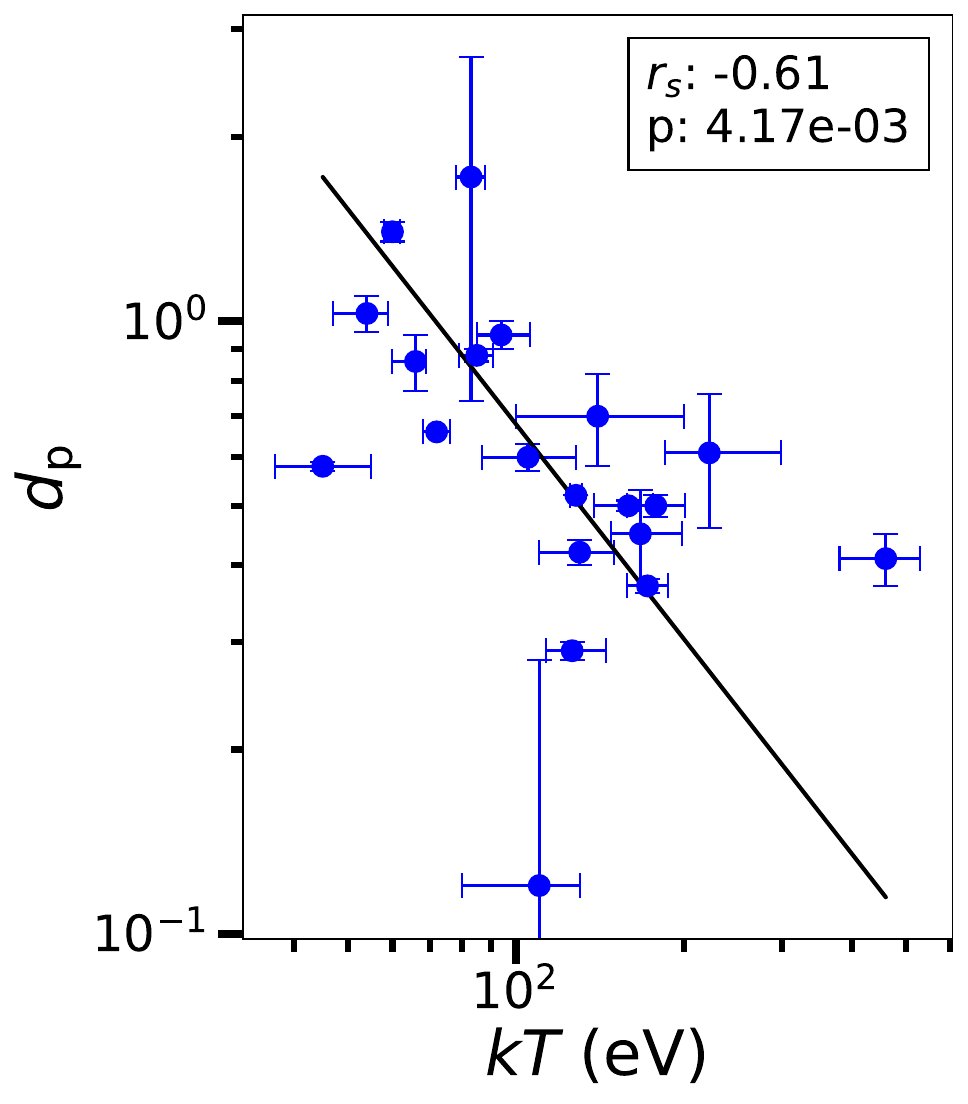}
  \end{subfigure}

  \caption{Gamma-ray spectral parameters versus blackbody temperature ($kT$)}
  \label{fig:gamma_spectral_kt}
\end{figure}

\begin{figure}
  \centering
  \begin{subfigure}{0.48\columnwidth}
    \includegraphics[width=\textwidth]{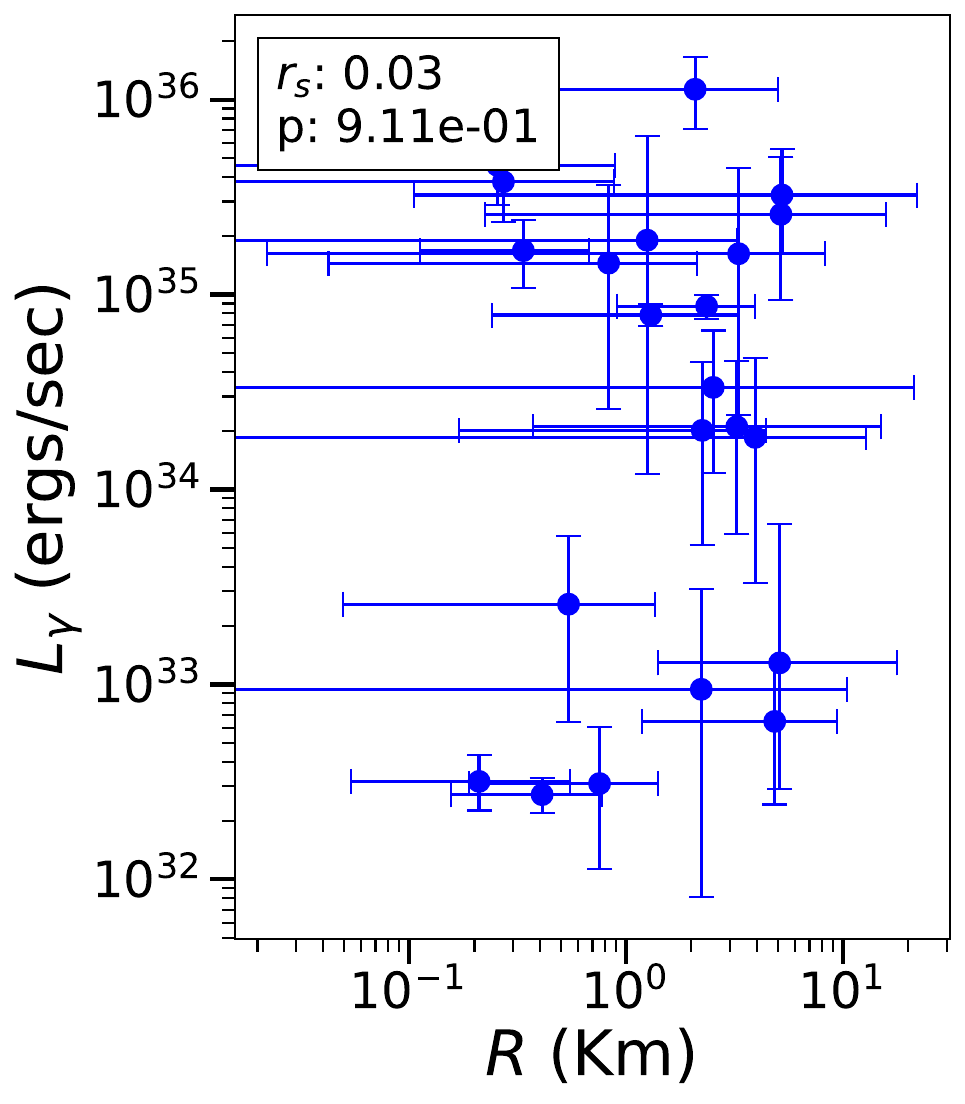}
  \end{subfigure}
  \hfill
  \begin{subfigure}{0.48\columnwidth}
    \includegraphics[width=\textwidth]{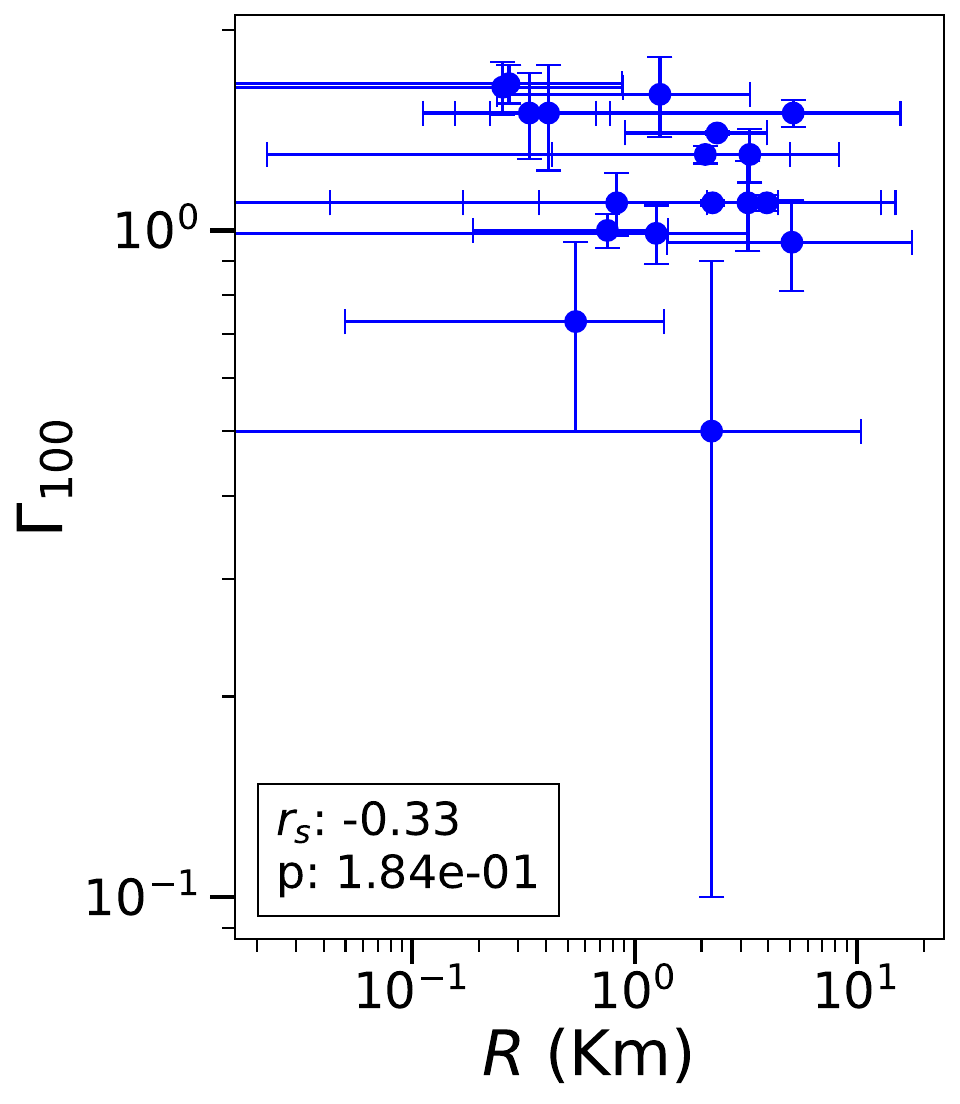}
  \end{subfigure}

  \vspace{0.3cm} 

  \begin{subfigure}{0.48\columnwidth}
    \includegraphics[width=\textwidth]{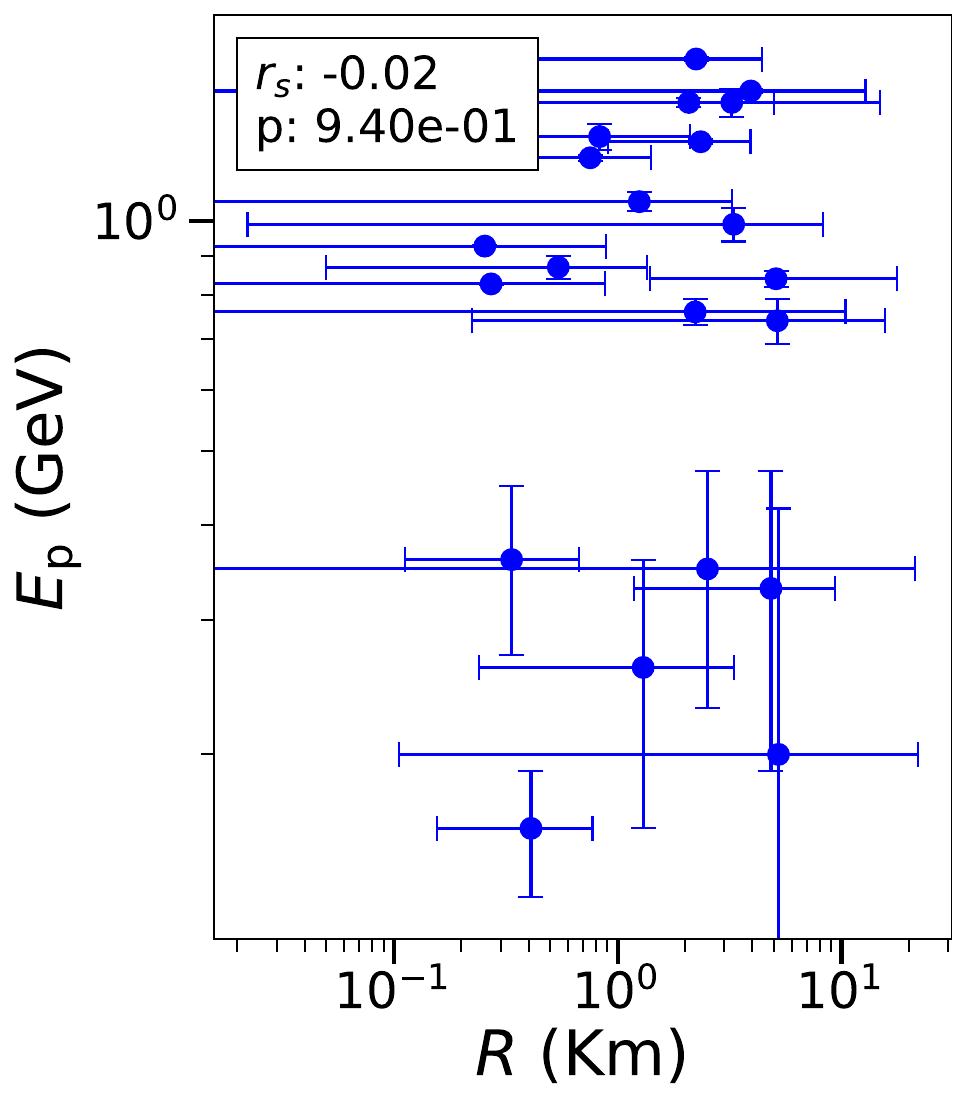}
  \end{subfigure}
  \hfill
  \begin{subfigure}{0.48\columnwidth}
    \includegraphics[width=\textwidth]{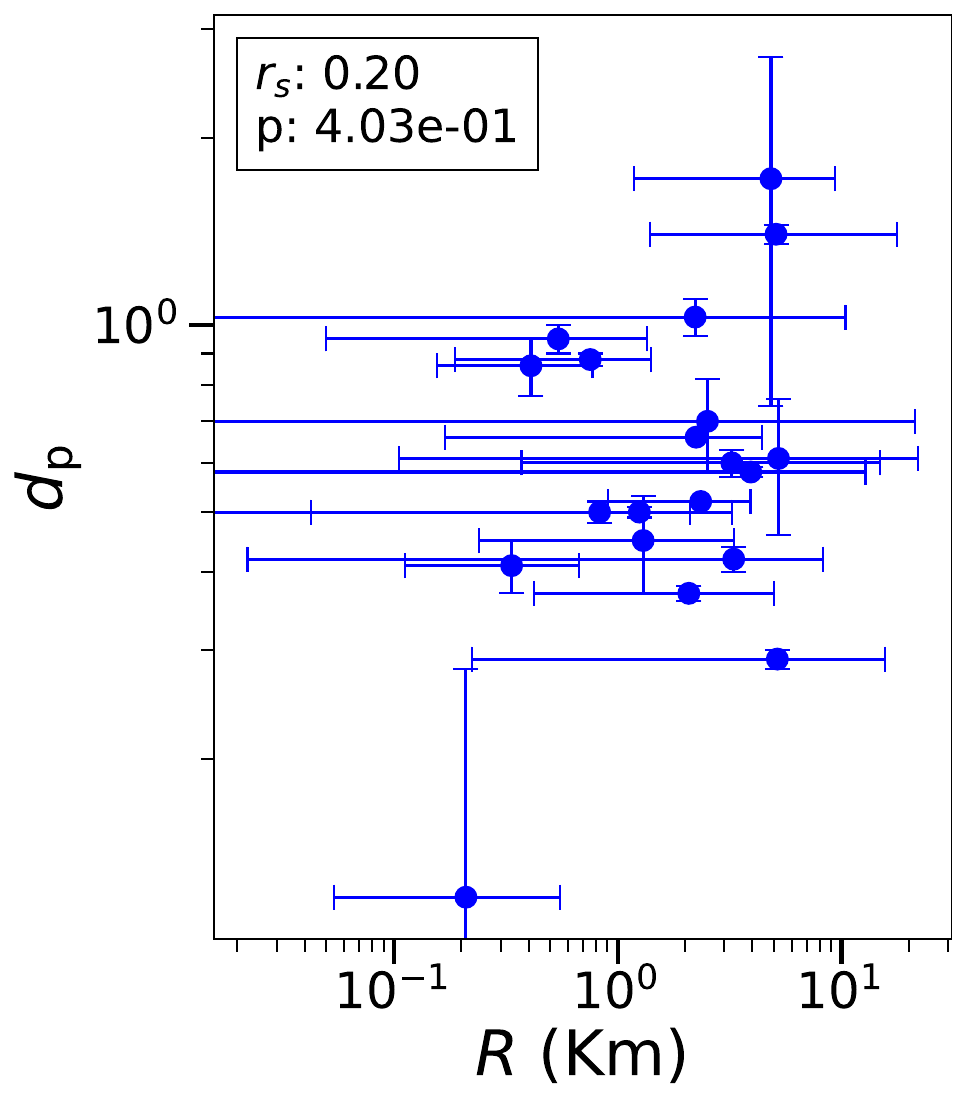}
  \end{subfigure}

  \caption{Gamma-ray spectral parameters versus blackbody radius ($R$)}
  \label{fig:gamma_spectral_R}
\end{figure}

Among our samples, 22 pulsars exhibit thermal emission. The correlations between the gamma-ray spectral parameters ($L_{\gamma}$, $\Gamma_{100}$, $d_{\rm p}$, $E_{\rm p}$) and the thermal X-ray spectral parameters ($kT$, $R$) are shown in Figs.~\ref{fig:gamma_spectral_kt} and \ref{fig:gamma_spectral_R}.

As discussed earlier, we expect to see a relation between gamma ray luminosity and thermal emission properties ($kT$, $R$), since thermal photons from the stellar surface may play a certain role in pair production in pulsars',
such as providing photons for two-photon pair production in the outer-gap model and cooling the pairs created in the polar gap region with inverse Compton scattering (e.g. \citet{Cheng1986, 1995A&A...301..456C}).
Thermal emissions of 22 pulsars in this sample are detected and their correlation with the gamma-ray luminosity is shown in Fig. \ref{fig:gamma_spectral_kt} $\&$ \ref{fig:gamma_spectral_R}, where one can see that the gamma-ray spectral properties are not correlated with the radius of the emitting region $R$ but there is a strong correlation between the gamma-ray luminosity and the temperature of the thermal emission, indicated by a Spearmann coefficient of 0.81 (p-value of about $10^{-6}$). $E_{\rm p}$ shows no correlation, while $d_{\rm p}$ and $\Gamma_{100}$ shows a correlation with weak significance and their best fit is described by
\begin{equation}
    \centering
    L_{\gamma}\propto \: kT^{7.24 \pm 1.17} \: \:(\chi^{2}_{\nu} = 4.8) \,\,\, .
    \label{gamma_kt}
\end{equation}

\begin{equation}
     \Gamma_{100}= (1.12\pm0.21)\log kT + (-0.98\pm0.42)
     \:\: (\chi^2_{\nu}=4.7)
\end{equation}

\begin{equation}
    d_{\rm p} \propto kT^{-1.2\pm0.2} \:\: (\chi^2_{\nu}=8.3)
\end{equation}

\begin{figure*}
\centering
\begin{subfigure}{0.24\textwidth}
  \includegraphics[width=\textwidth]{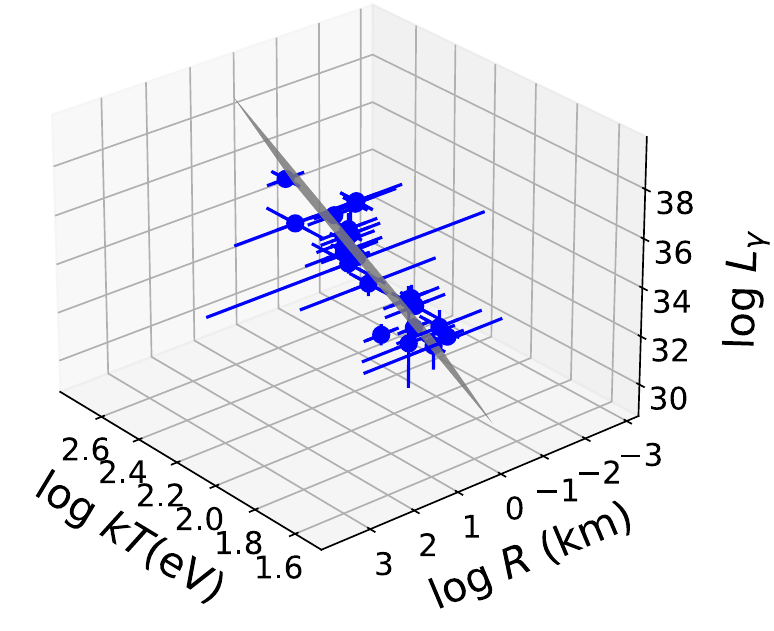}
\end{subfigure}
\hfill
\begin{subfigure}{0.24\textwidth}
  \includegraphics[scale=0.33]{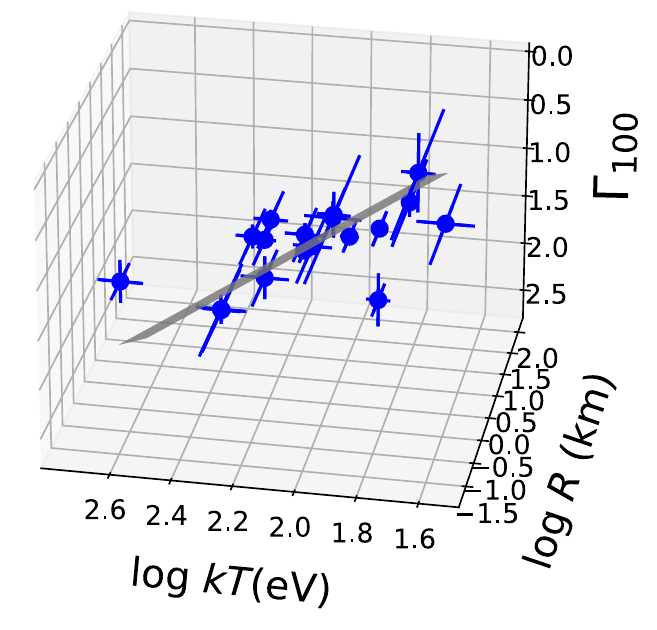}
\end{subfigure}
\hfill
\begin{subfigure}{0.24\textwidth}
  \includegraphics[scale=0.32]{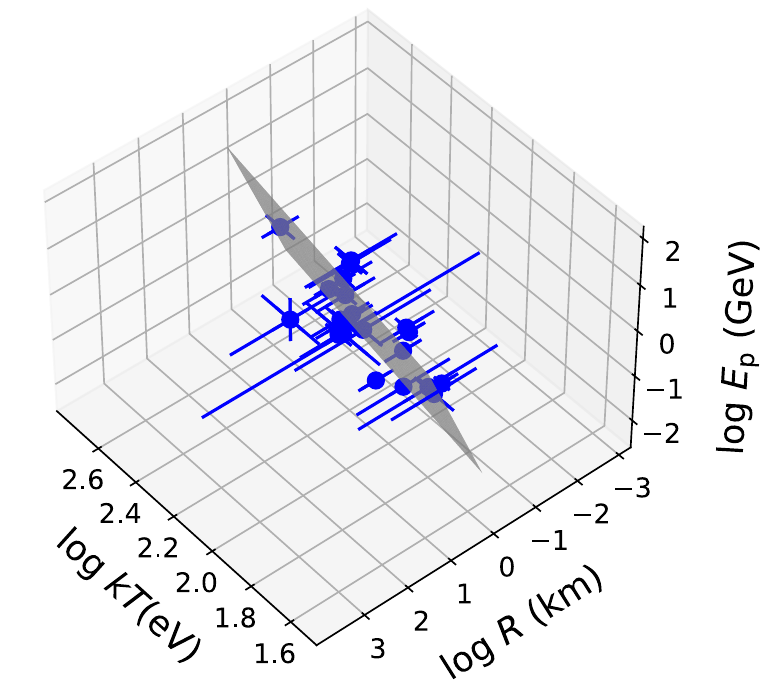}
\end{subfigure}
\hfill
\begin{subfigure}{0.24\textwidth}
  \includegraphics[scale=0.25]{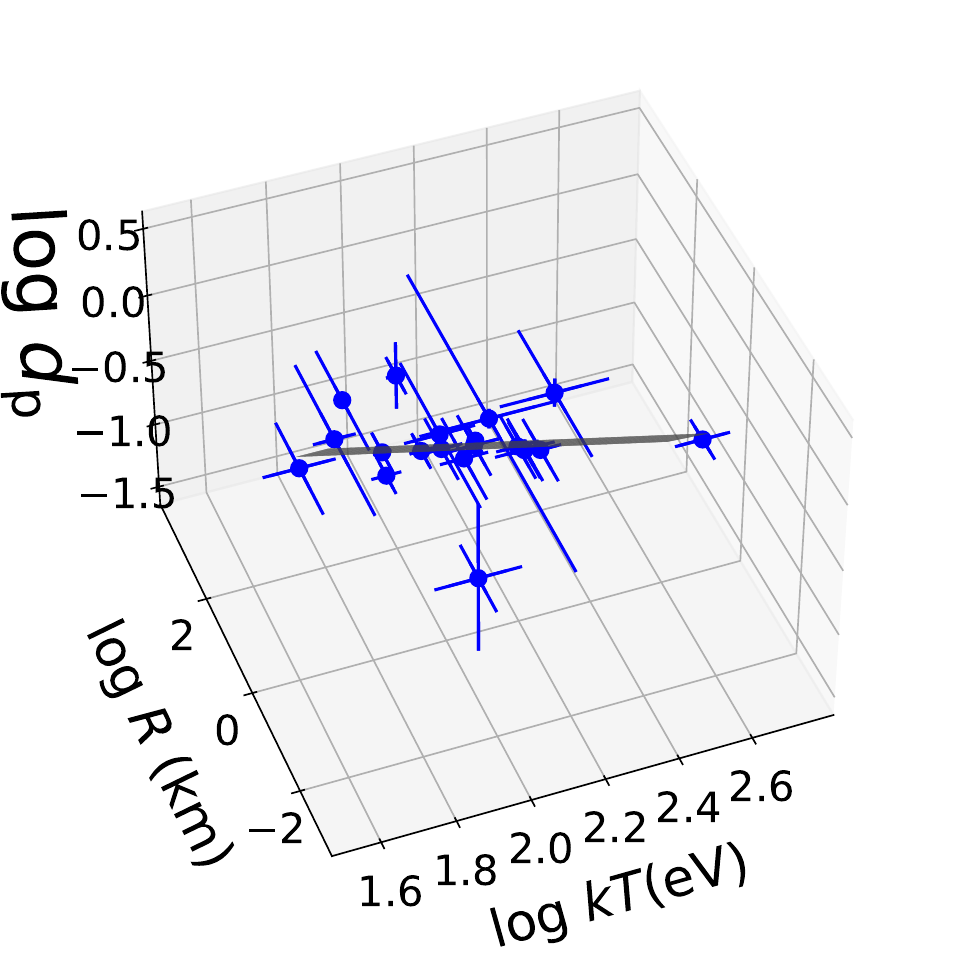}
\end{subfigure}

\medskip

\begin{subfigure}{0.24\textwidth}
  \includegraphics[scale=0.32]{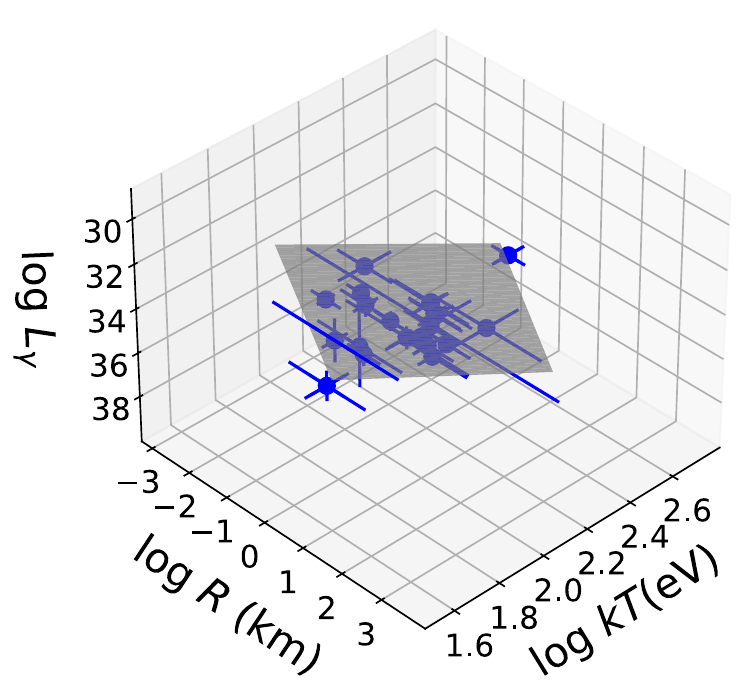}
\end{subfigure}
\hfill
\begin{subfigure}{0.24\textwidth}
  \includegraphics[scale=0.32]{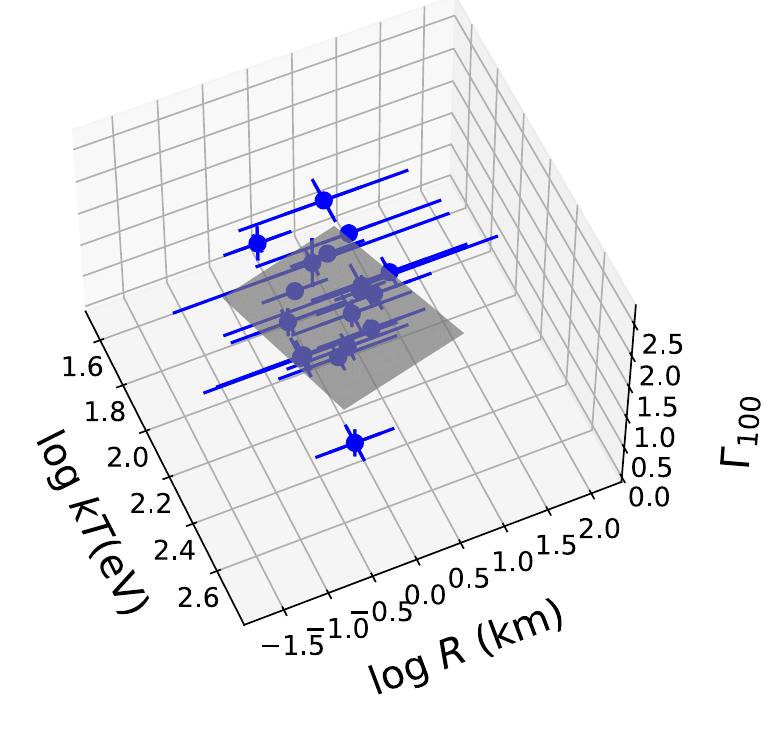}
\end{subfigure}
\hfill
\begin{subfigure}{0.24\textwidth}
  \includegraphics[scale=0.33]{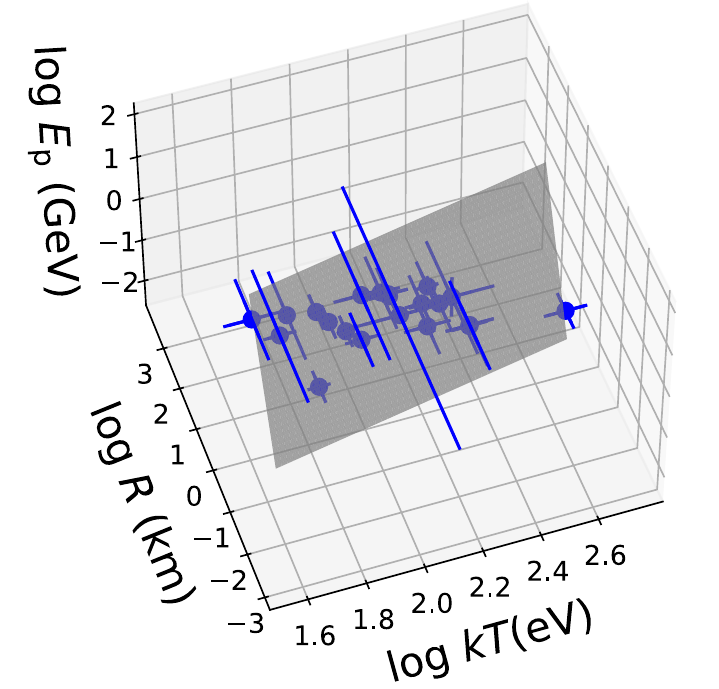}
\end{subfigure}
\hfill
\begin{subfigure}{0.24\textwidth}
  \includegraphics[scale=0.33]{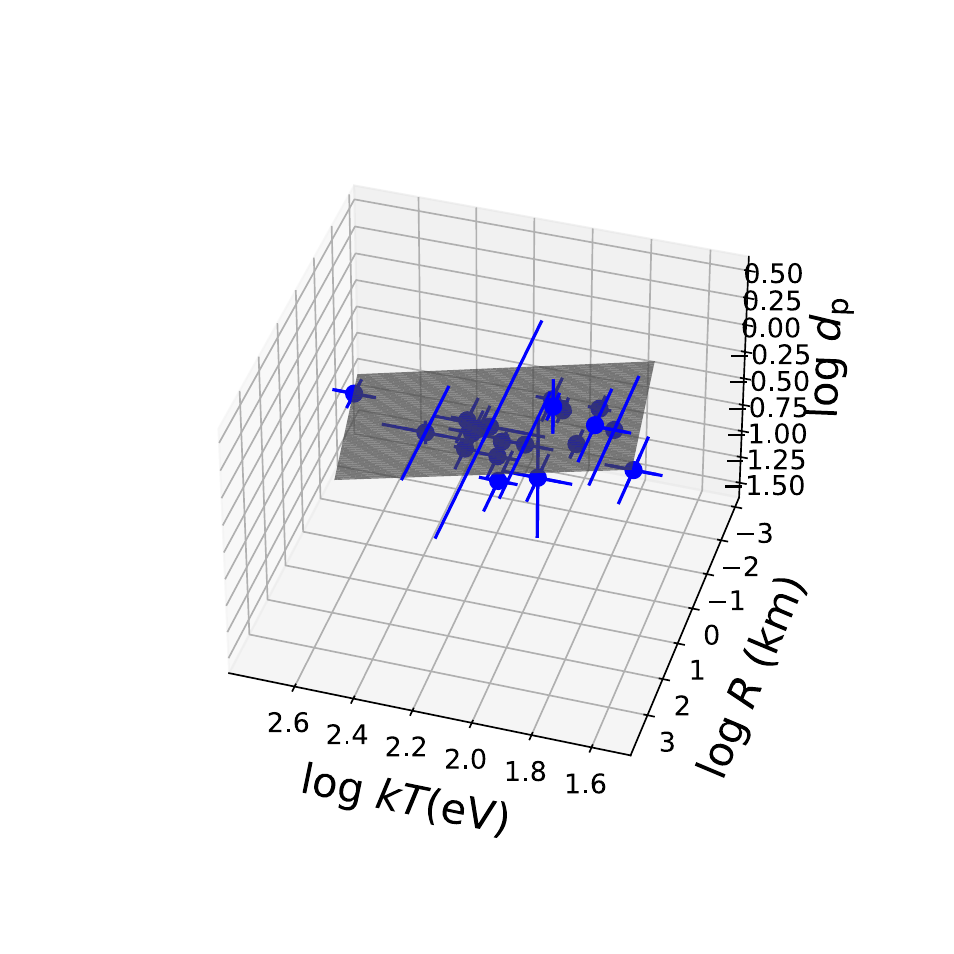}
\end{subfigure}

\caption{Fundamental planes of gamma-ray spectral parameters ($L_{\gamma}$, $\Gamma_{100}$, $E_{\rm p}$, $d_{\rm p}$) in the space of thermal parameters ($kT$, $R$). The first row shows the edge-on view of the plane and the second row shows the top-on view of the plane.}
\label{fig:all_thermal_planes}
\end{figure*}

We further examine the relation between the gamma-ray spectral parameters and both thermal parameters, $kT$ and $R$, jointly in a 3-dimensional space, since gamma-ray emissions very likely depend not only on the surface temperature but also on the flux level of thermal photons. The best fit to describe their relationship is found to be the following:
\begin{equation}
    \centering
    L_{\gamma}\propto \: kT^{4.79 \pm 0.90}\: R^{2.55 \pm 0.58} \: \:(\chi^{2}_{\nu} = 0.54) \,\,\, .
    \label{gamma_kt_r}
\end{equation}
This fitting has a $\chi^2_\nu$ of 0.54, showing improvement over Eq.(\ref{gamma_kt}) by adding one more parameter ($R$).
The significance of this improvement can be checked with the F-test, which indicates a very significant improvement at a random probability much lower than $10^{-5}$.
It suggests a fundamental plane in the space of 
\{$L_{\gamma}$, $kT$, $R$\}, as shown in Fig. \ref{fig:all_thermal_planes}. We note, however, the statistical statement here is only suggestive, as discussed earlier.

In fact, we also found that
the nonthermal X-ray luminosity of these Group-2 pulsars follows a relationship of $L_{\rm p}\propto kT^{5.97\pm1.08}R^{3.28\pm0.77}$ ($\chi^2_\nu=0.42$). 
This relationship is consistent with Eq.(11) in \citet{HK2023}. The sample used in this study is a subset of that used in \citet{HK2023}. 
It is also compatible with Eqs.(\ref{gamma_kt_r}) and (\ref{L_gamma_L_xray_group2}). 

Since the gamma-ray spectral parameters ($E_{\rm p}$, $d_{\rm p}$, $\Gamma_{100}$) do not exhibit a strong correlation with surface temperature, we further investigate their relationships with the two thermal parameters jointly. The resulting correlation is illustrated in Fig. \ref{fig:all_thermal_planes} Their corresponding best-fit descriptions are shown in the following.

\begin{equation}
    \Gamma_{100}=\log ((kT)^{1.96\pm0.64} R^{0.39\pm0.30}) +(-2.89\pm1.36) 
   (\chi^2_{\nu}=0.67)
\end{equation}
\begin{equation}
    E_{\rm p} \propto kT^{1.28\pm1.12}R^{2.26\pm1.34} \:\: (\chi^2_{\nu}=0.64)
\end{equation}

\begin{equation}
    d_{\rm p} \propto kT^{-0.56\pm0.16}R^{-0.37\pm0.11} \:\: (\chi^2_{\nu}=0.76)
\end{equation}

As shown by the above equations, the gamma-ray spectral parameters ($L_{\gamma}$, $\Gamma_{100}$, $E_{\rm p}$, $d_{\rm p}$) exhibit improved fits, indicated by lower reduced chi-square values, when the blackbody radius ($R$) is included in addition to the blackbody temperature ($kT$).

\subsection{Correlations among gamma-ray spectral parameters}

\begin{figure*}
    \centering
    \begin{subfigure}{0.32\textwidth}
        \includegraphics[width=\linewidth]{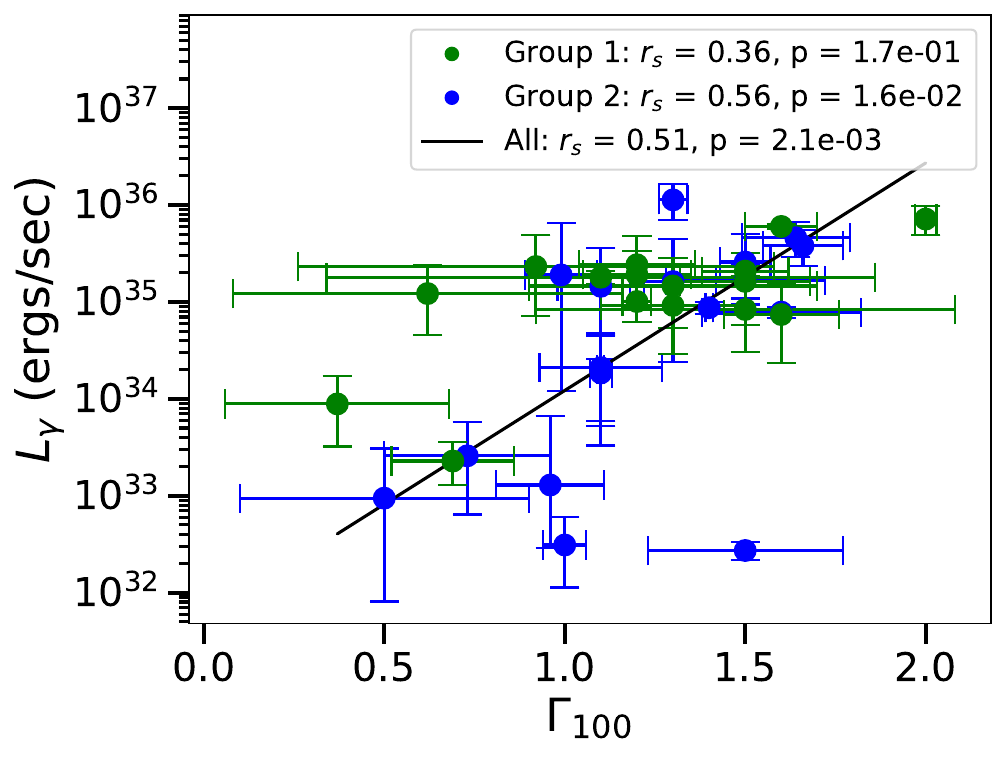}
    \end{subfigure}
    \hfill
    \begin{subfigure}{0.32\textwidth}
        \includegraphics[width=\linewidth]{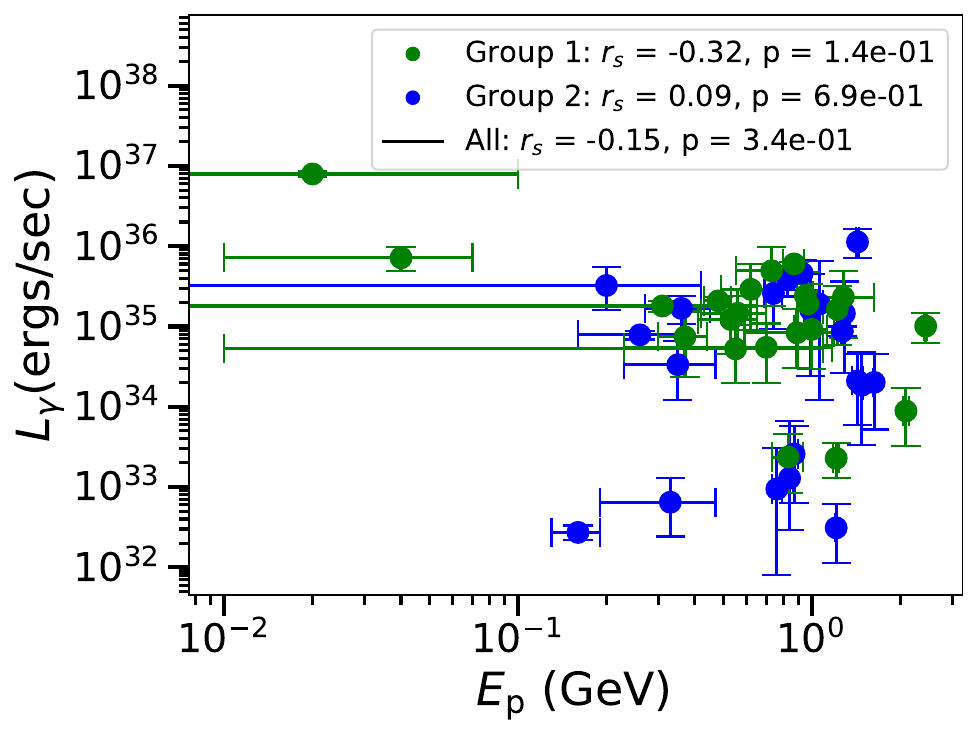}
    \end{subfigure}
    \hfill
    \begin{subfigure}{0.32\textwidth}
        \includegraphics[width=\linewidth]{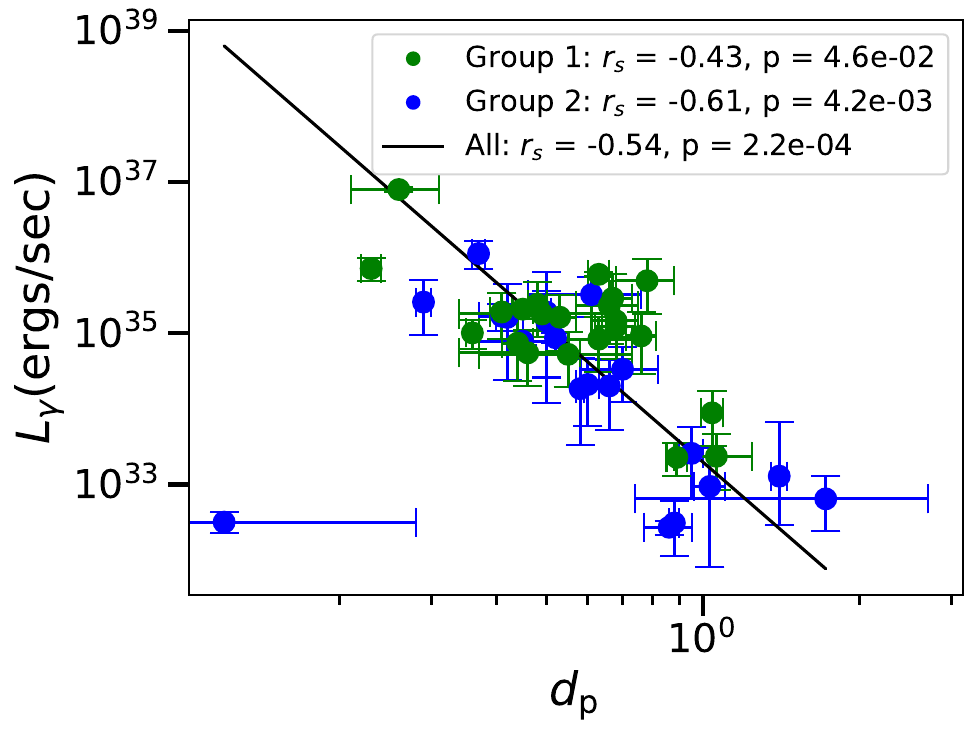}
    \end{subfigure}

    \vspace{1em}

    \begin{subfigure}{0.32\textwidth}
        \includegraphics[width=\linewidth]{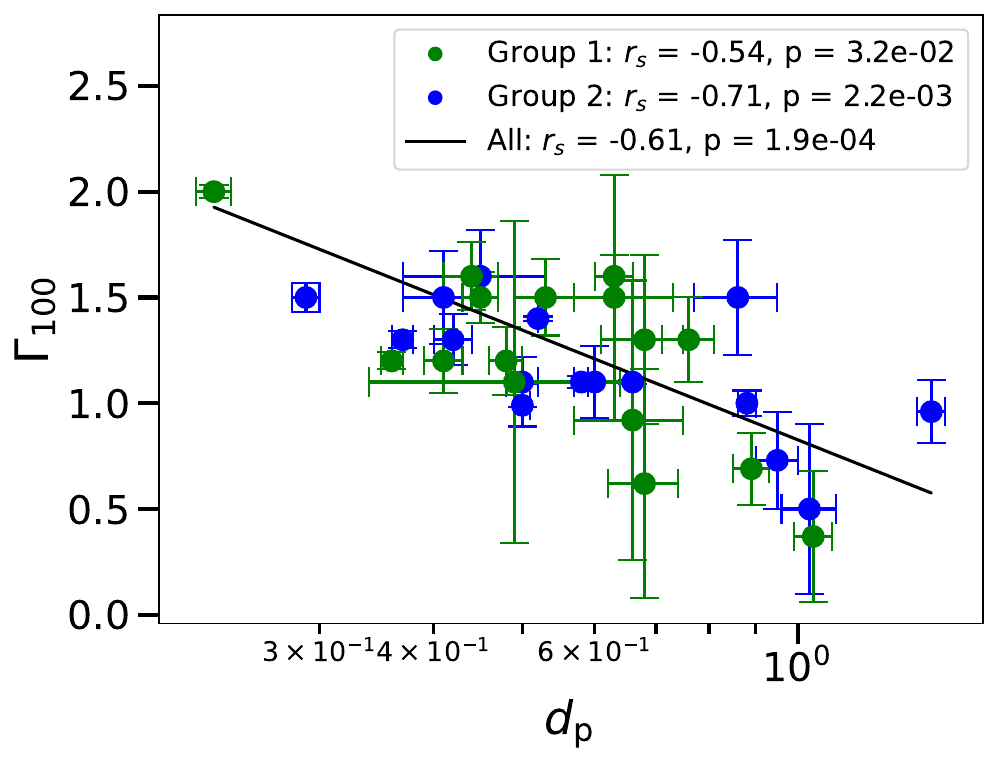}
    \end{subfigure}
    \hfill
    \begin{subfigure}{0.32\textwidth}
        \includegraphics[width=\linewidth]{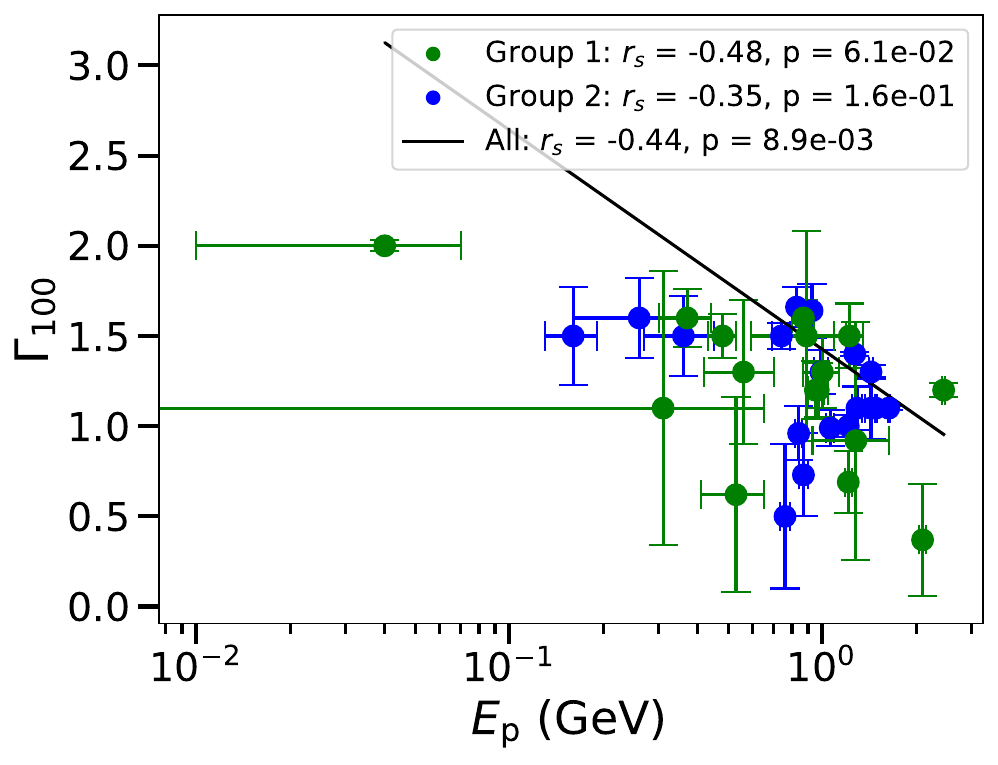}
    \end{subfigure}
    \hfill
    \begin{subfigure}{0.32\textwidth}
        \includegraphics[width=\linewidth]{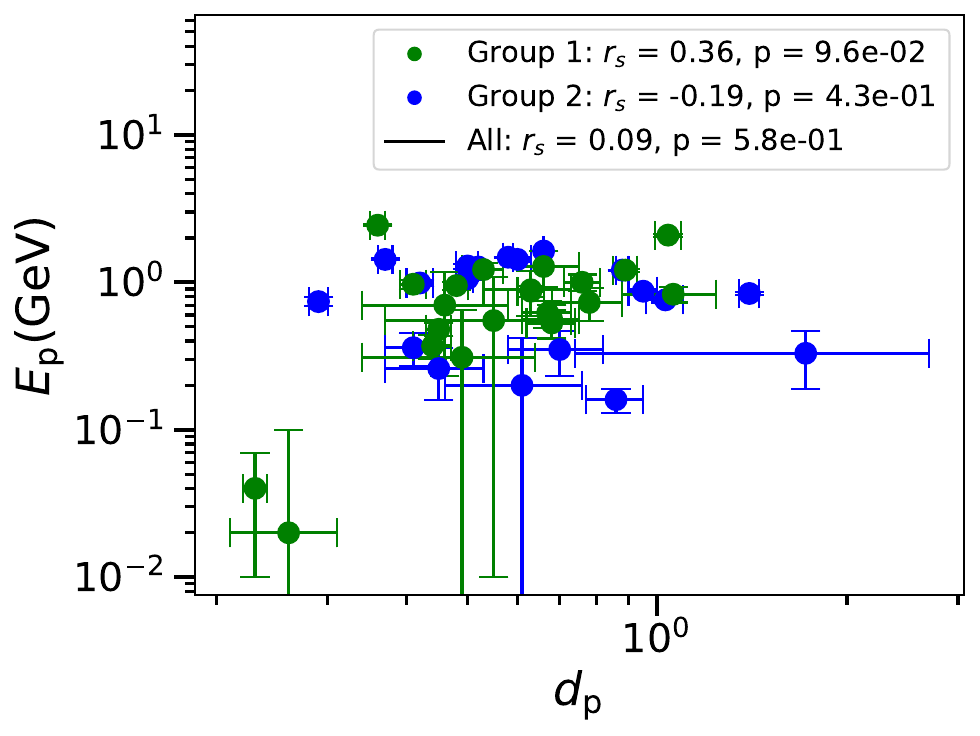}
    \end{subfigure}

    \caption{Correlations between Gamma-ray spectral parameters}
    \label{fig:gamma_spectral}
\end{figure*}

The correlations among the gamma-ray spectral parameters are shown in Fig.\ref{fig:gamma_spectral}. Except for $L_\gamma$ versus $d_{\rm p}$ and 
$\Gamma_{100}$ versus $d_{\rm p}$, no clear correlations are found among them.
There is no indication for Group 1 and Group 2 pulsars having different behavior either.
The best fit relationships for those having a more significant correlation are listed below:

\begin{equation}
    L_{\gamma} \propto d_{\rm p}^{-5.98\pm0.8} \:\: (\chi^2_{\nu}=6.68) 
\end{equation}
\begin{equation}
    \log L_{\gamma} =({2.35\pm0.36}) \Gamma_{100} + (31.74\pm0.53) \:\: (\chi^2_{\nu} =3.61)
\end{equation}
\begin{equation}
    \Gamma_{100} = (-1.72\pm0.22) \log d_{\rm p} + (0.83\pm0.06) \:\: (\chi^2_{\nu}=9.66)
\end{equation}

\section{Discussion}
Because of unknown beaming factors, viewing geometry and distance uncertainties, there is usually a large scatter in the correlation among spectral and timing properties
\citep{possenti2002,Li2008,Kalapotharakos2022ApJ...934...65K,HK2023}.
All the statistical descriptions of correlation function forms obtained from best fits are therefore only suggestive. 
These correlations are, however, still helpful in illustrating possible trends and dependencies among spectral and timing properties.
It may inspire or constrain theoretical models of pulsar emissions.

In the sample we studied, clear correlations are found for the gamma-ray luminosity with pulsars' age, spindown power and magnetic field strength at the light cylinder. It is worth noting that it does not correlate with the magnetic field strength at the stellar surface (see Fig.\ref{fig:Lgamma_vs_timing}), similar to the non-thermal X-ray luminosity \citep{HK2023}.
Although actual physical mechanisms need to be modeled, 
it seems that both the gamma-ray and non-thermal X-ray luminosities are more linked to outer magnetospheres or the wind zones of pulsars. Other gamma-ray spectral parameters, 
$\Gamma_{100}$, $E_{\rm p}$ and $d_{\rm p}$, do not show such a clear correlation.

\begin{figure*}
  \begin{subfigure}{\columnwidth}
  \includegraphics[width=\textwidth]{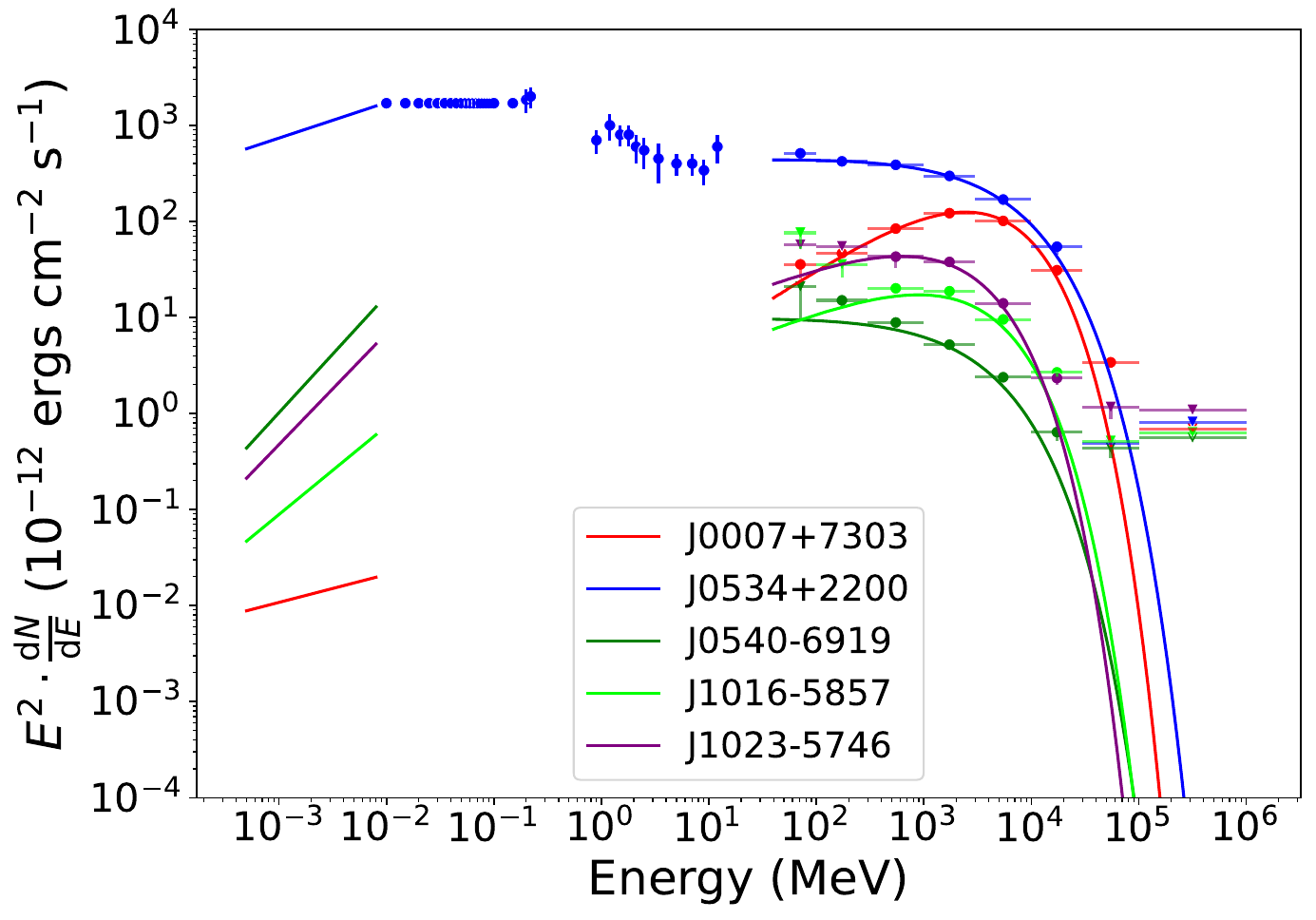}
  \end{subfigure}
  \hfill
  \begin{subfigure}{\columnwidth}
  \includegraphics[width=\textwidth]{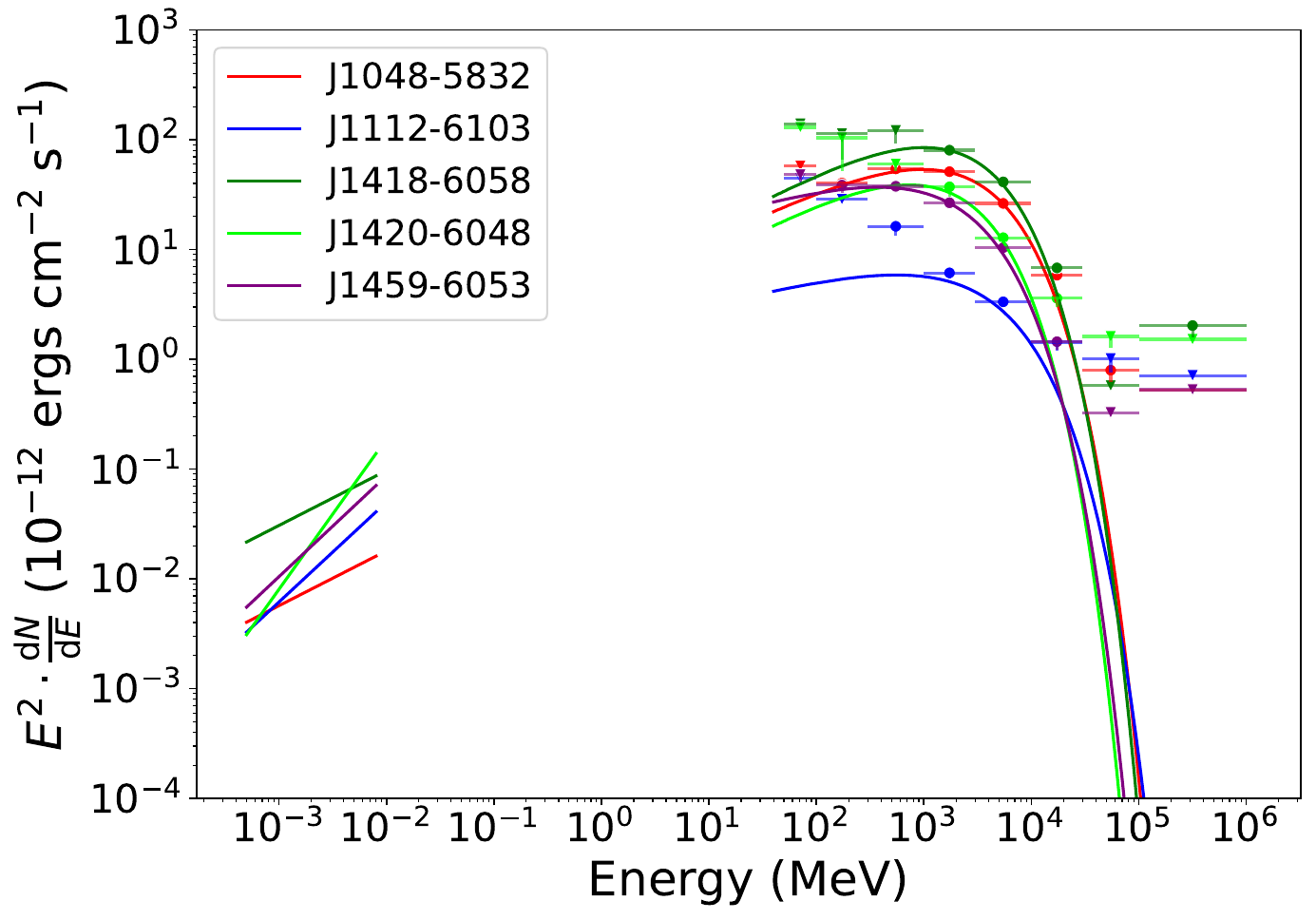} 
  \end{subfigure} 
  \begin{subfigure}{\columnwidth} 
  \includegraphics[width=\textwidth]{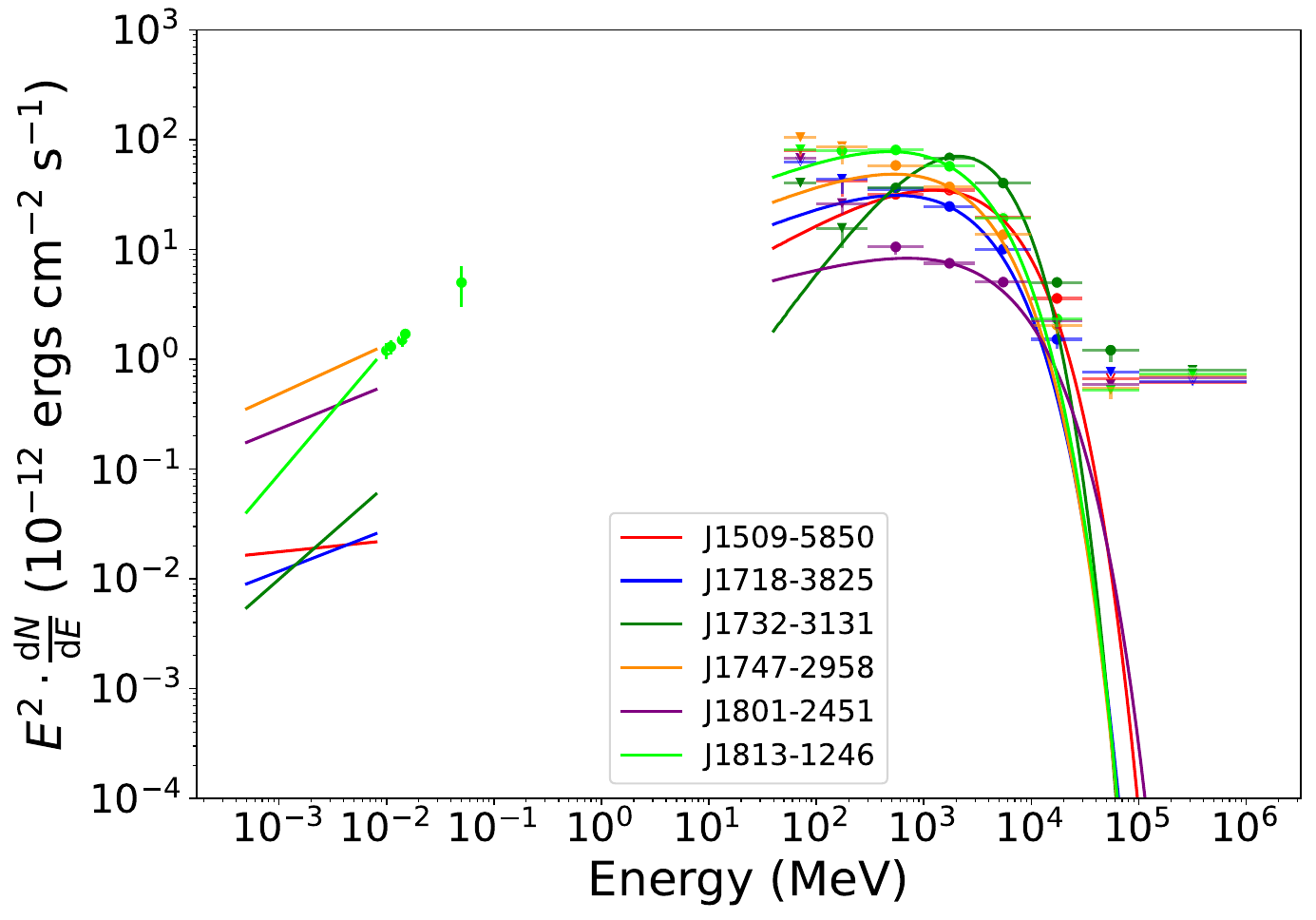} 
  \end{subfigure}  
  \hfill 
  \begin{subfigure}{\columnwidth} 
  \includegraphics[width=\textwidth]{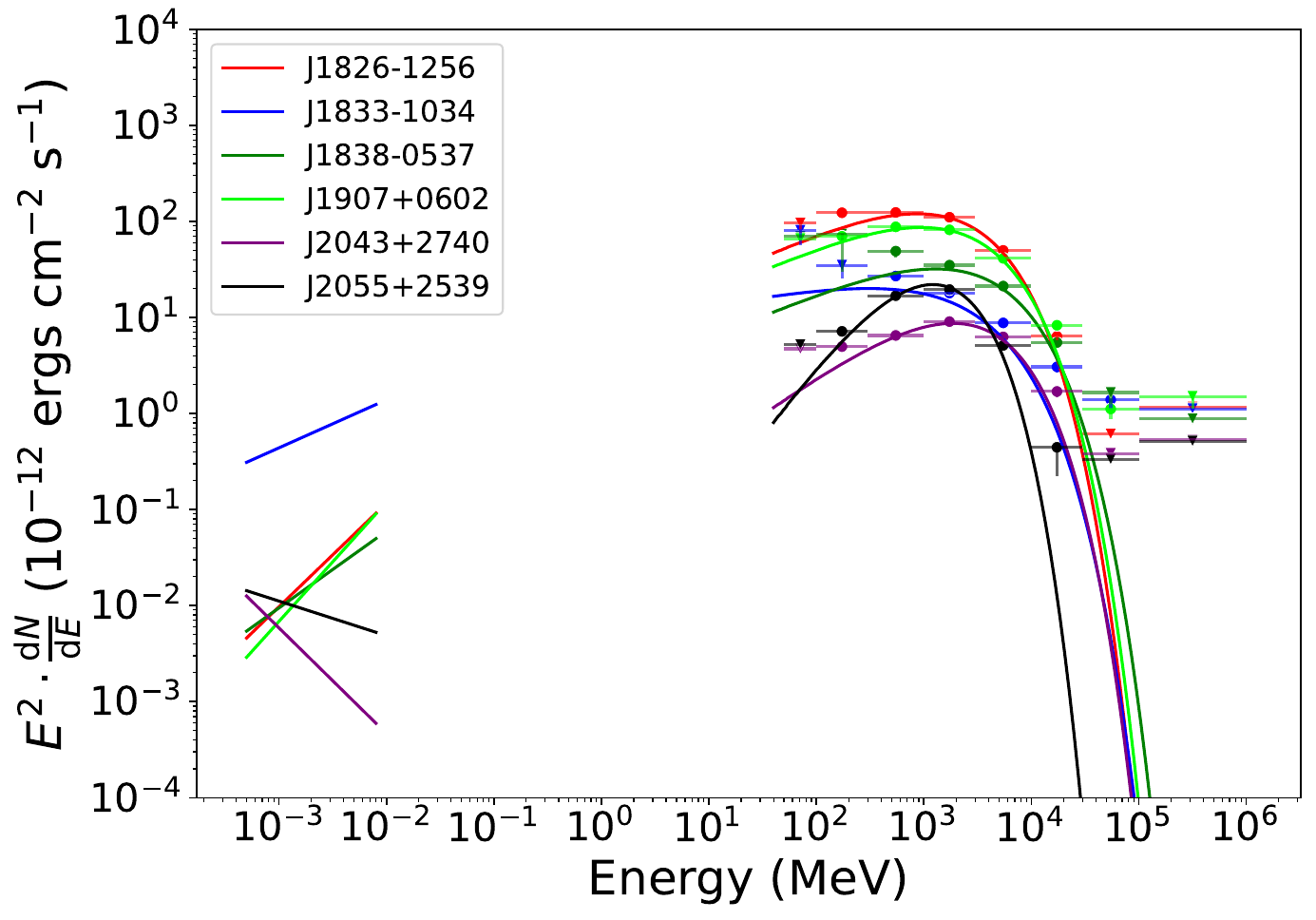}  
  \end{subfigure}
  \caption{Broad-band spectra of Group-1 pulsars.}
  \label{fig:spectrum_PL}
\end{figure*}

\begin{figure*}
  \begin{subfigure}{\columnwidth}
  \includegraphics[width=\textwidth]{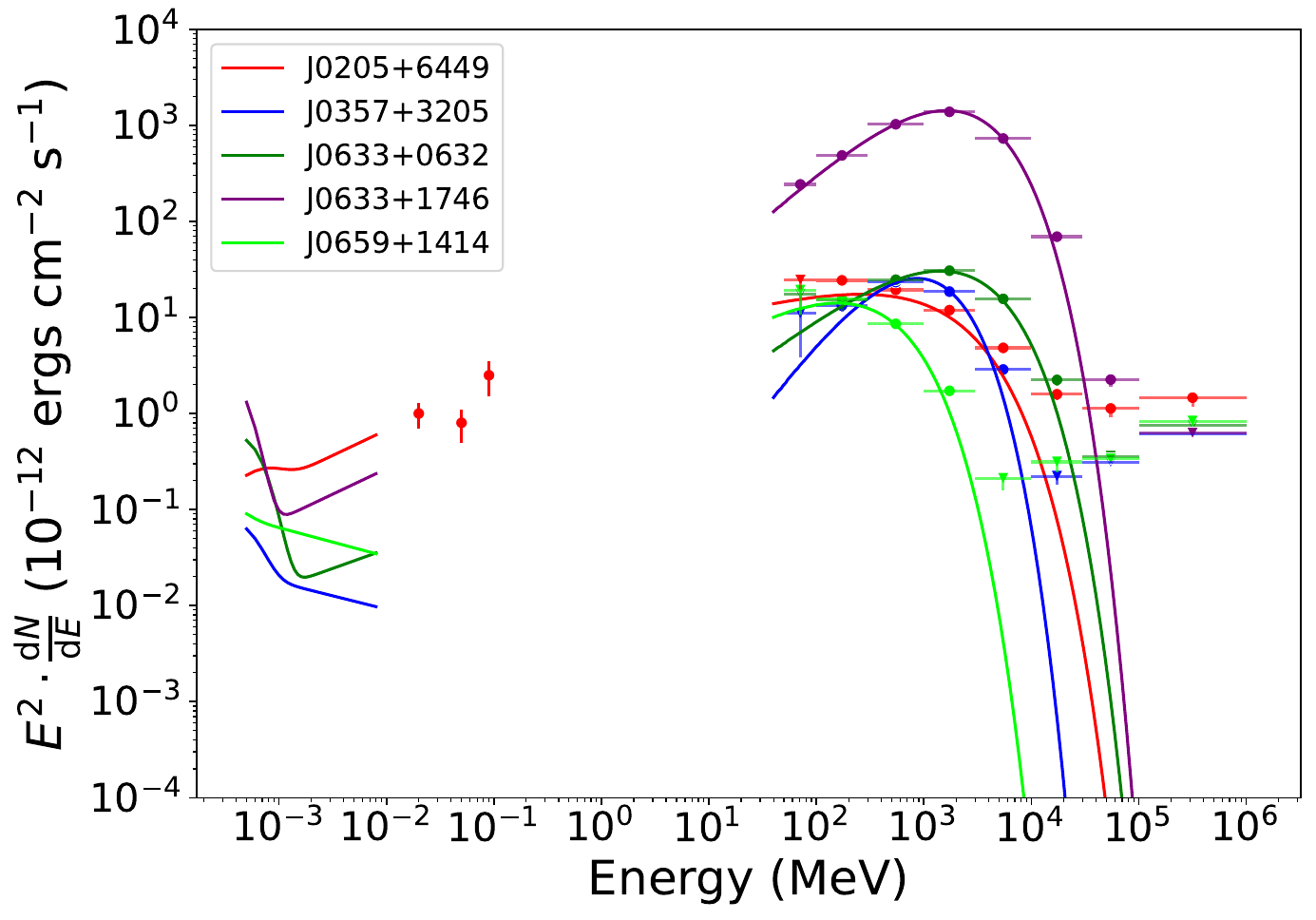}
  \end{subfigure}
  \hfill
  \begin{subfigure}{\columnwidth}
  \includegraphics[width=\textwidth]{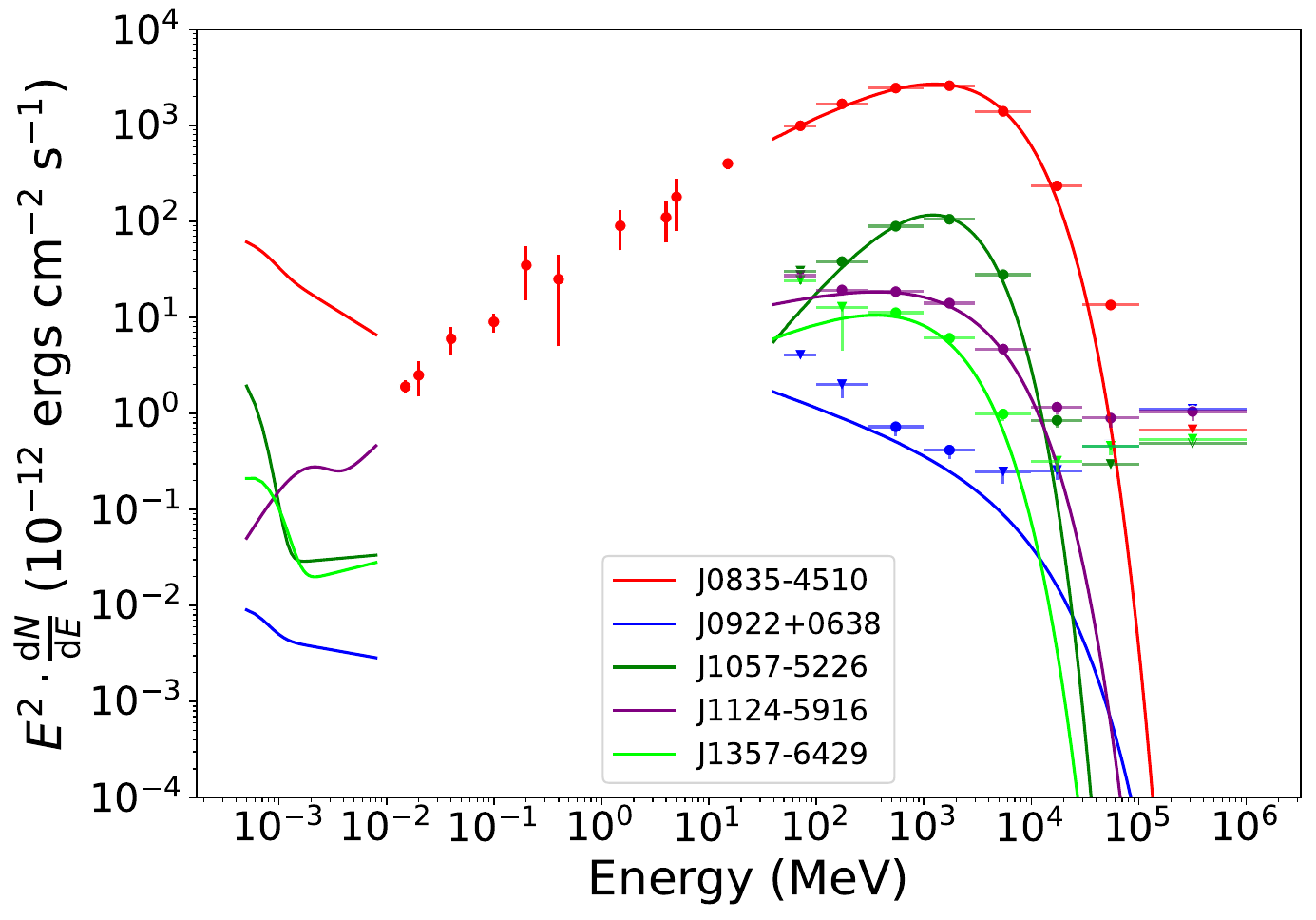} 
  \end{subfigure} 
  \begin{subfigure}{\columnwidth} 
  \includegraphics[width=\textwidth]{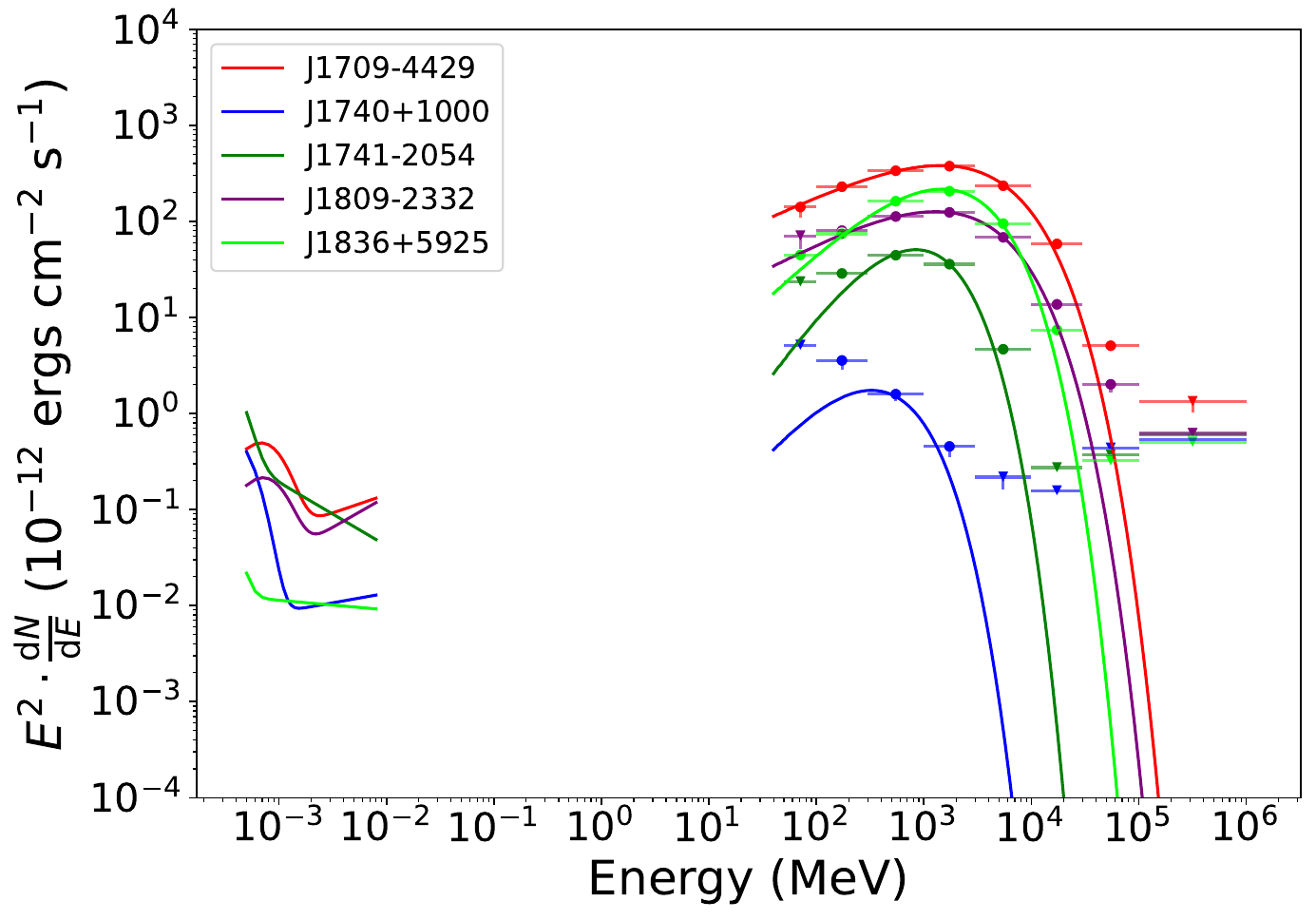} 
  \end{subfigure}  
  \hfill 
  \begin{subfigure}{\columnwidth} 
  \includegraphics[width=\textwidth]{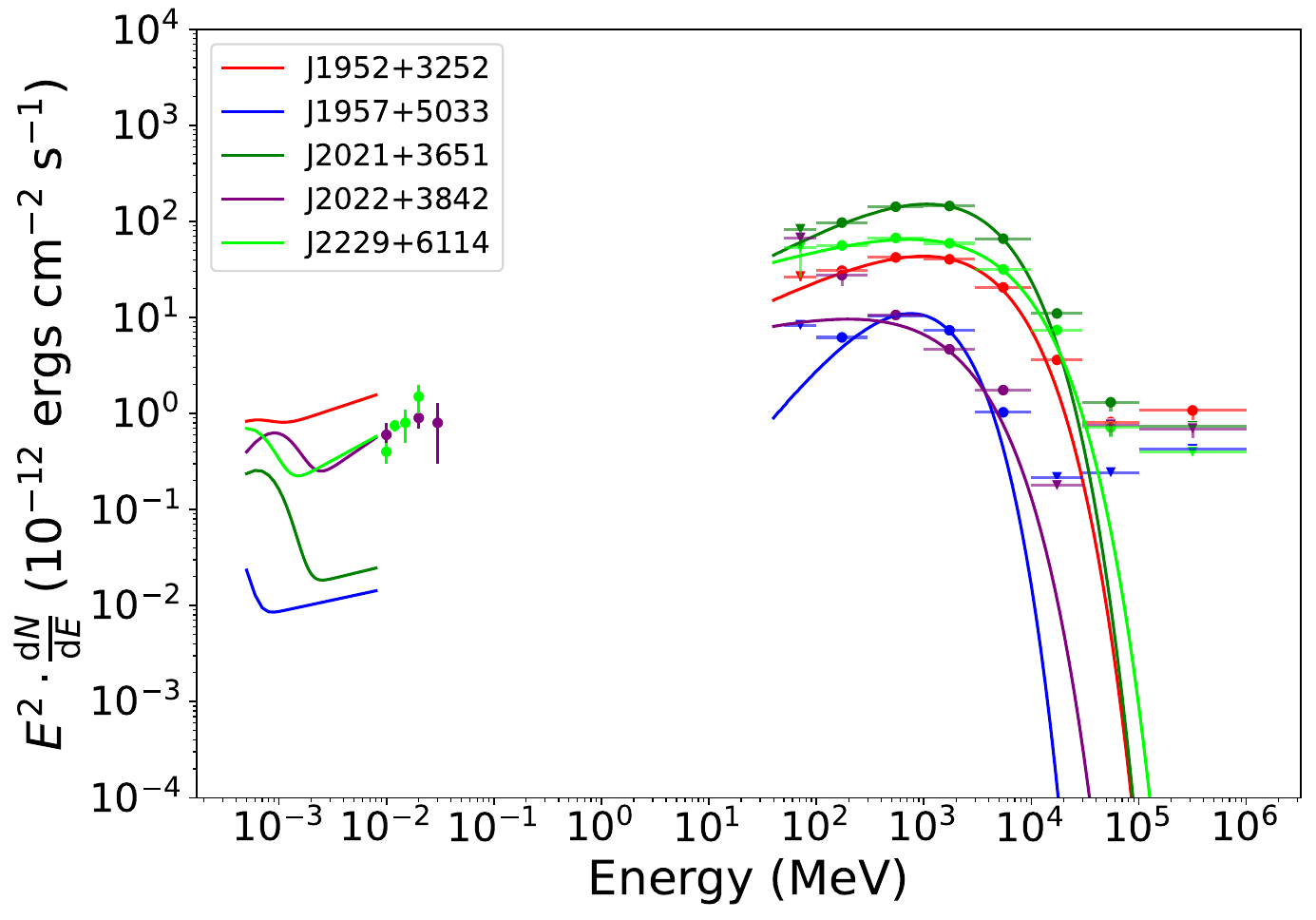} 
  \end{subfigure}
  \caption{Broad-band spectra of Group-2 pulsars.}
  \label{fig:spectrum_PL+BB}
\end{figure*}

The main purpose of this paper is to investigate possible correlations between the gamma-ray and X-ray spectral properties. 
We found that $L_\gamma$ has a strong correlation with the non-thermal X-ray luminosity $L_{\rm p}$ and it goes like $L_{\gamma} \propto L_{\rm{p}}^{0.49 \pm 0.05}$, or tentatively, $L_{\gamma} \propto L_{\rm{p}}^{0.5}$. While the non-thermal X-ray luminosity and gamma-ray luminosity are correlated, the gamma-ray energy flux shows no correlation with the radio flux density, as reported in the 3PC. This is one of the major results obtained in this study. Non-thermal X-rays from pulsars are usually attributed to synchrotron radiation of electron-positron pairs right after they are created somewhere in the magnetosphere of pulsars or to the low-energy continuum tail of a broad-band high-energy emission, which peaks around GeV.  
The X-ray power-law component of some pulsars obviously represent a separate component from the gamma-ray one in the broadband SED, but some others seem to have the non-thermal X-ray and gamma-ray emissions forming a single component in pulsar broadband spectra. 
The broadband spectra for the listed pulsars are shown in Figs. \ref{fig:spectrum_PL} and \ref{fig:spectrum_PL+BB}. 
The X-ray spectra in the 0.5–10 keV range are plotted using the spectral indices provided in Table~\ref{Spectral_table}. Fig. \ref{fig:spectrum_PL} is for Group-1 pulsars and Fig. \ref{fig:spectrum_PL+BB} is for Group 2. 
For energies above 100 MeV, the spectra are taken from the Fermi 3PC catalog, and for the intermediate energy range (0.01–100 MeV), the spectra are based on face values taken from figures in \citet{Coti2020}.
Although spectral fitting uncertainties hinder a clear conclusion, one can see from Figs.~\ref{fig:spectrum_PL} and \ref{fig:spectrum_PL+BB} that X-ray  emissions of some pulsars, such as J0007+7303, J2043+2740, J1509–5850, J1718–3825, J0357+3205, and J2021+3651, do not form a single spectral component with their gamma-ray emissions.
More observations in the soft gamma-ray regime will help identify if the other pulsars have a single spectral component from X-ray to gamma rays.
In any case, however, the relation that
$L_\gamma\propto L_{\rm p}^{0.5}$ seems to indicate
that the non-thermal X-ray emission is not simply the low-energy tail of the gamma-ray emission. They could be related to each other in a more complicated way through
radiating-particle initial energy distribution (right after their creation), acceleration mechanisms after/during synchrotron cooling and the gamma-ray radiation mechanims.

An interesting result is that pulsars without (Group 1) and with (Group 2) detected thermal emissions seem to distinguish themselves in their relationships of $L_\gamma$ versus $L_{\rm p}$;
see Fig. \ref{fig:Gamma_spectral_vs_Lp}, Eq. \ref{L_gamma_L_xray_group1} and \ref{L_gamma_L_xray_group2}.
If it is indeed the case, which requires a larger sample to verify, it strongly suggests that pulsars of Groups 1 and 2 are from different populations. 
It is clear that gamma-ray emission is related to X-ray emission and the relation changes when thermal emission is present. 
This confirms that thermal X-rays do play a role in the emission of non-thermal X-rays and gamma rays, since its presence makes a difference. Another interesting issue is why the surface thermal emission of Group-1 pulsars is not detected. These two groups do not show different distributions in timing parameters. There is no indication of data quality difference for these two groups either. The non-detection of a thermal component is not due to data quality.
In particular, Group-1 pulsars are not older. There is no reason for those pulsars to have a lower surface temperature and smaller thermal luminosity so that detection is more difficult, unless they have a cooling curve different from that of Group-2 pulsars. It is tempting to investigate whether they are actually quark stars rather than neutron stars or they cool down through a more exotic process; see, e.g., \citet{2002ApJ...571L..45S}, \citet{2020MNRAS.496.5052P} and \citet{2022A&A...663A..19Z}.

A strong anticorrelation is also found between $L_\gamma$ and the photon index $\Gamma_{\rm p}$ of the power-law component in X-rays (Fig.\ref{fig:Gamma_spectral_vs_gammap}). This is not surprising since a similar anti-correlation is also reported in \citet{HK2023} between
$L_{\rm p}$ and $\Gamma_{\rm p}$ and we have 
$L_\gamma\propto L_{\rm p}^{0.5}$ here.
A harder X-ray spectrum comes with a higher gamma-ray and non-thermal X-ray luminosities. This may have some implications on the physical conditions of their acceleration and emission mechanisms.
A best-fit function form of $L_\gamma(L_{\rm p}, \Gamma_{\rm p})$ is shown in Eq. \ref{eqn:Lgamma_Lp_gammap}. It may \textbf{provide some constraints} on future modeling works.
The aforementioned Group 1 and Group 2 pulsars do not show statistically significant distinctions in these relations involving $\Gamma_{\rm p}$. 

The strong correlation between $L_\gamma$ and the stellar surface temperature, represented by $kT$ throughout this paper, is noteworthy. 
Similarly to $L_{\rm p}$, studied in \citet{HK2023} with a larger sample, $L_\gamma$ shows a strong dependence on temperature $kT$ but not on the emission-region radius $R$ and a much better fit can be found when the dependence on both $kT$ and $R$ is taken into account at the same time. 
This indicates a possible connection between thermal photons and high-energy emissions. 
As discussed in \citet{HK2023} and references therein, ideas of pair production near the polar cap region and in the outer-gap region have been investigated for a long time. In either locations thermal photons both play a role in pair production via inverse Compton scattering or two-photon pair production. 
A clear dependence function form may help to model these processes. Unfortunately, a precise and more statistically reliable description of the fundamental planes in the spaces of \{$L_\gamma$, $kT$, $R$\} or \{$L_{\rm p}$, $kT$, $R$\} has yet to come with improved measurement uncertainties. 

\section*{Acknowledgements}
This work was supported by the National Science and Technology Council (NSTC) of the Republic of China (Taiwan) under grant NSTC 114-2112-M-007-042.

\section*{Data Availability}
The data employed by this article are available in the article and in
the quoted references.

\bibliographystyle{mnras}
\bibliography{example}

\appendix
\section{}
The relationships between gamma-ray spectral parameters $\Gamma_{100}$, $E_{\rm p}$, $d_{\rm p}$ and timing parameters are in weak or negligible, unlike $L_\gamma$, as described in the main text. These are shown in this Appendix. Those relationships with a p-value $\leq 10^{-2}$ are given a best fit description in the following:
\begin{equation}
    \Gamma_{100} =(0.095\pm 0.037) \log \dot{E} + (-1.96\pm 1.31) \:\: (\chi^2_{\nu}=61.9)
\end{equation}
\begin{equation}
    \Gamma_{100} =(0.11\pm0.05	) \log B_{\rm lc} + (1.00\pm0.20) \:\: (\chi^2_{\nu}=66.1)
\end{equation}

\begin{equation}
    E_{\rm p} \propto P^{-0.38\pm 0.13} \:\: (\chi^2_{\nu}=511.8)
\end{equation}
\begin{equation}
    E_{\rm p} \propto \dot{E}^{-0.019\pm 0.032} \:\: (\chi^2_{\nu}=613.0)
\end{equation}
\begin{equation}
    E_{\rm p} \propto B_{\rm lc}^{-0.022 \pm 0.049} \:\: (\chi^2_{\nu}=616.0)
\end{equation}

\begin{equation}
    d_{\rm p} \propto P^{0.245\pm0.002} \:\: (\chi^2_{\nu}=51.1)
\end{equation}
\begin{equation}
    d_{\rm p} \propto \dot{P}^{0.099\pm0.001} \:\: (\chi^2_{\nu}=61.5)
\end{equation}
\begin{equation}
    d_{\rm p} \propto \dot{E}^{0.045\pm0.001} \:\: (\chi^2_{\nu}=50.7)
\end{equation}
\begin{equation}
    d_{\rm p} \propto \tau^{0.070\pm0.001} \:\: (\chi^2_{\nu}=54.4)
\end{equation}
\begin{equation}
    d_{\rm p} \propto B_{\rm lc}^{-0.066 \pm 0.002} \:\: (\chi^2_{\nu}=50.0)
\end{equation}

\begin{figure}
    \centering

    \begin{subfigure}{0.48\columnwidth}
        \includegraphics[width=\linewidth]{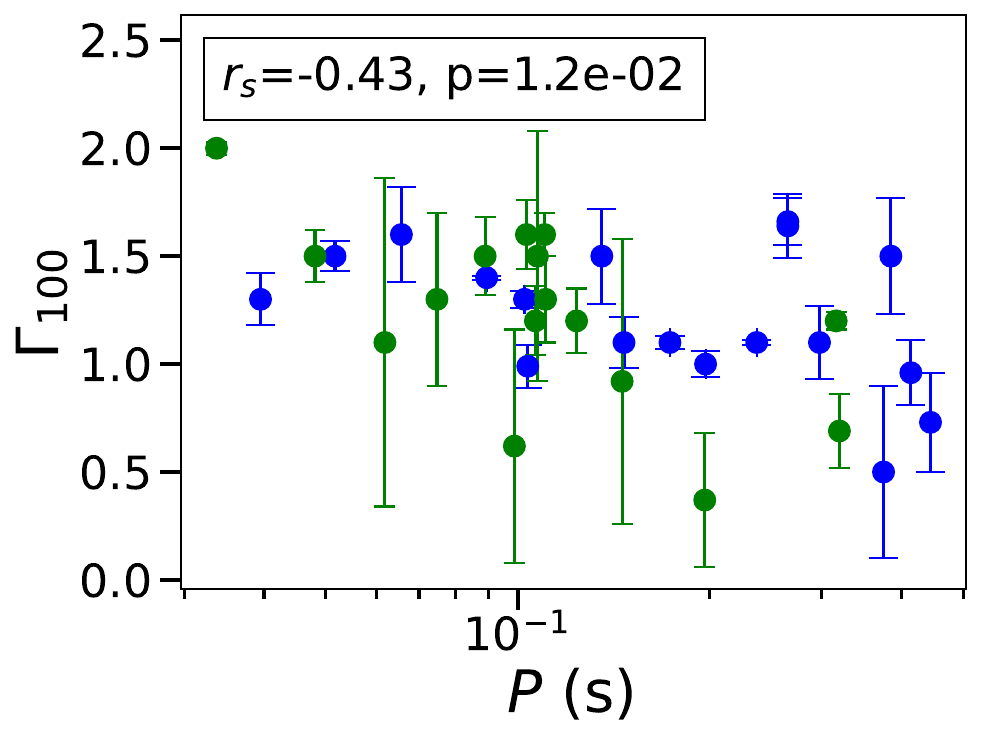}
    \end{subfigure}
    \hfill
    \begin{subfigure}{0.48\columnwidth}
        \includegraphics[width=\linewidth]{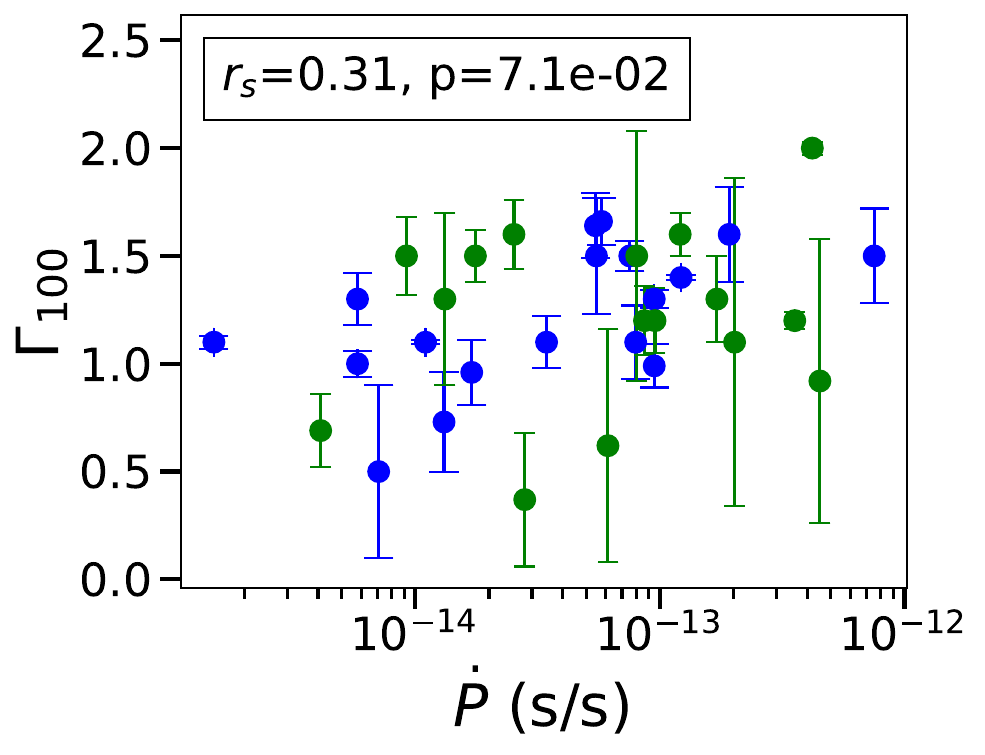}
    \end{subfigure}

    \vspace{0.6em}

    \begin{subfigure}{0.48\columnwidth}
        \includegraphics[width=\linewidth]{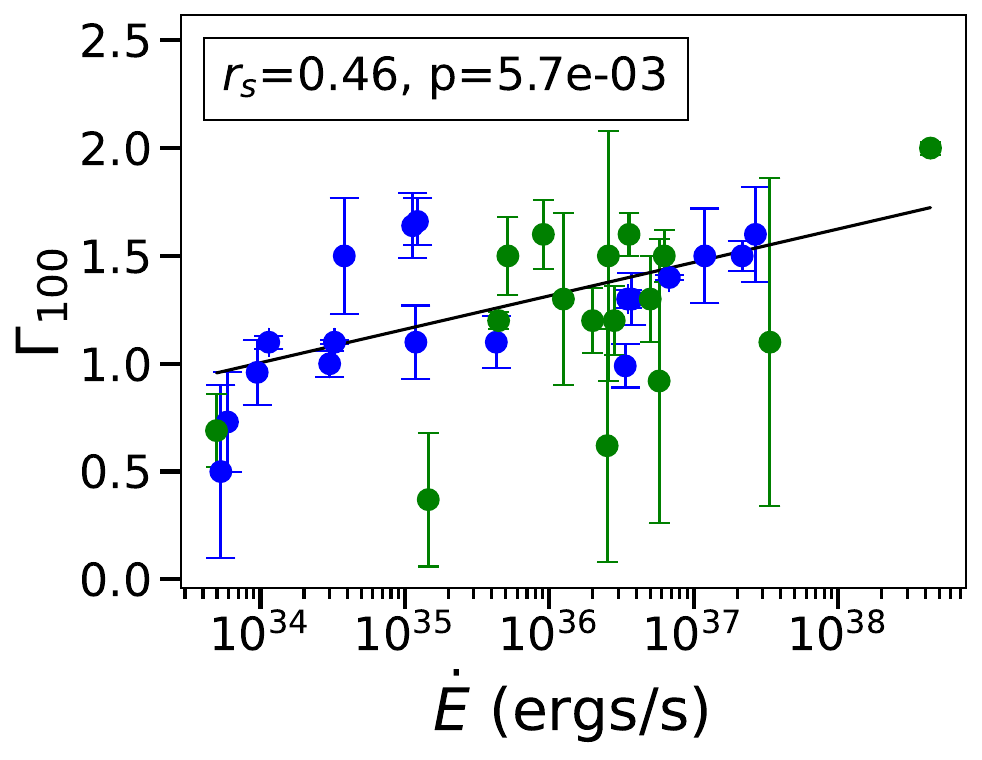}
    \end{subfigure}
    \hfill
    \begin{subfigure}{0.48\columnwidth}
        \includegraphics[width=\linewidth]{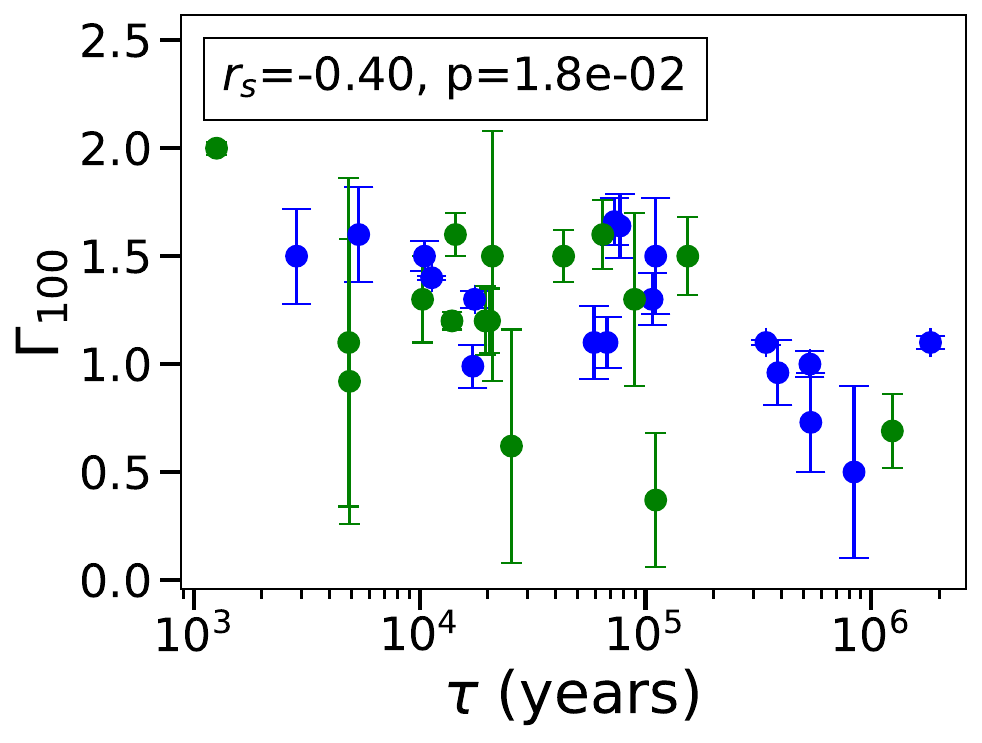}
    \end{subfigure}

    \vspace{0.6em}

    \begin{subfigure}{0.48\columnwidth}
        \includegraphics[width=\linewidth]{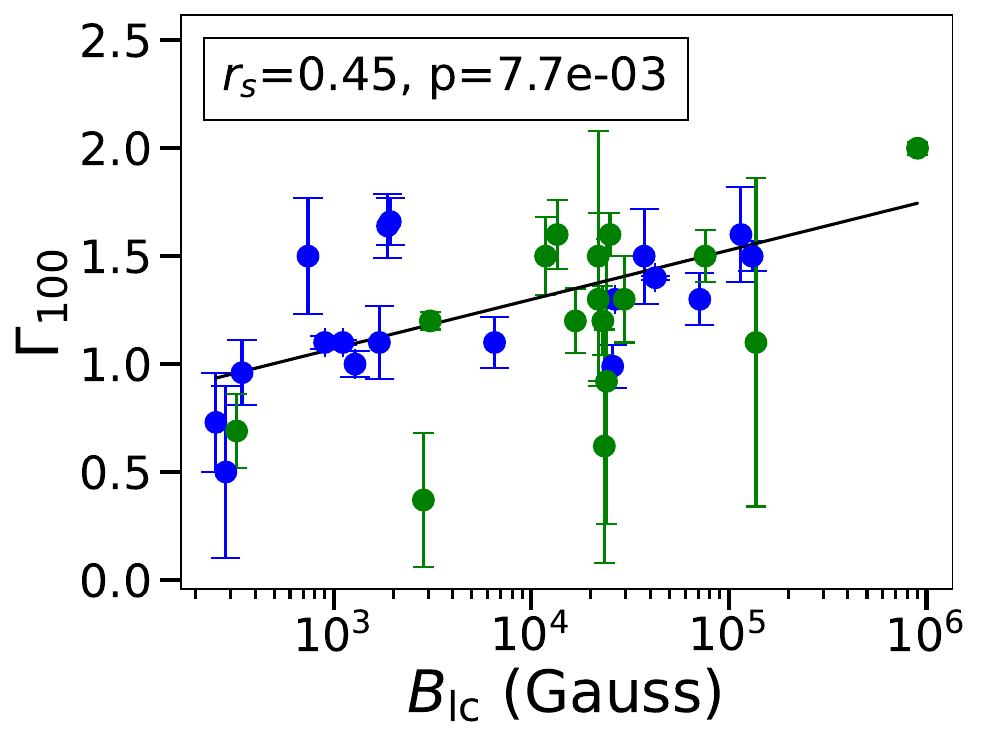}
    \end{subfigure}
    \hfill
    \begin{subfigure}{0.48\columnwidth}
        \includegraphics[width=\linewidth]{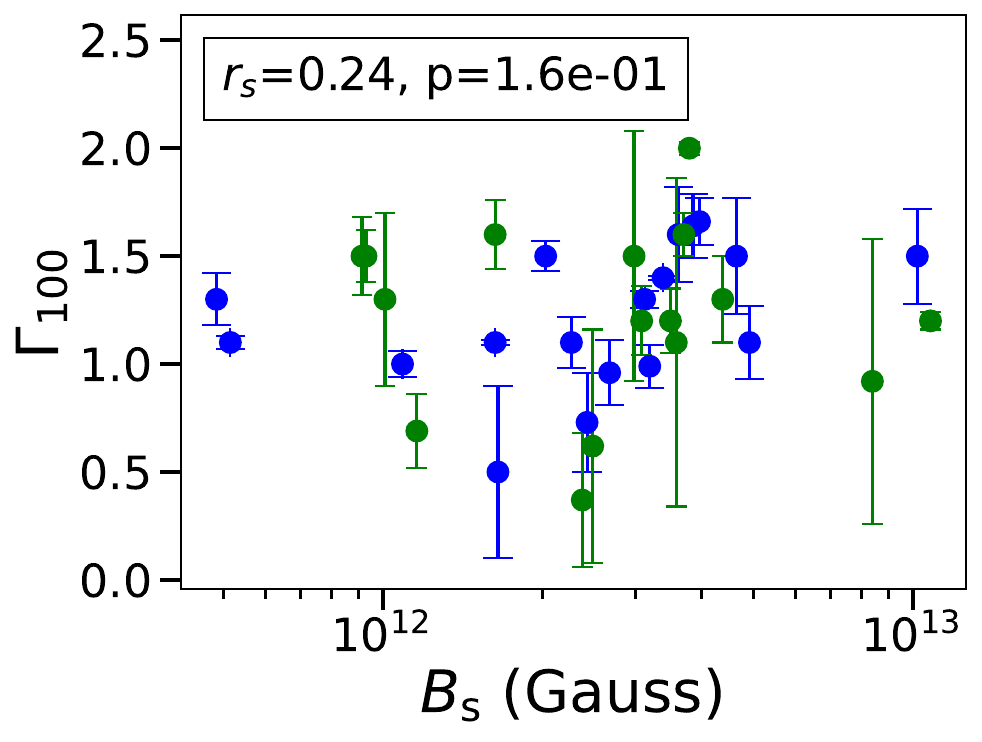}
    \end{subfigure}

    \caption{$\Gamma_{100}$ vs timing parameters}
    \label{fig:gamma_100_timing}
\end{figure}

\begin{figure}
    \centering

    \begin{subfigure}{0.48\columnwidth}
        \includegraphics[width=\linewidth]{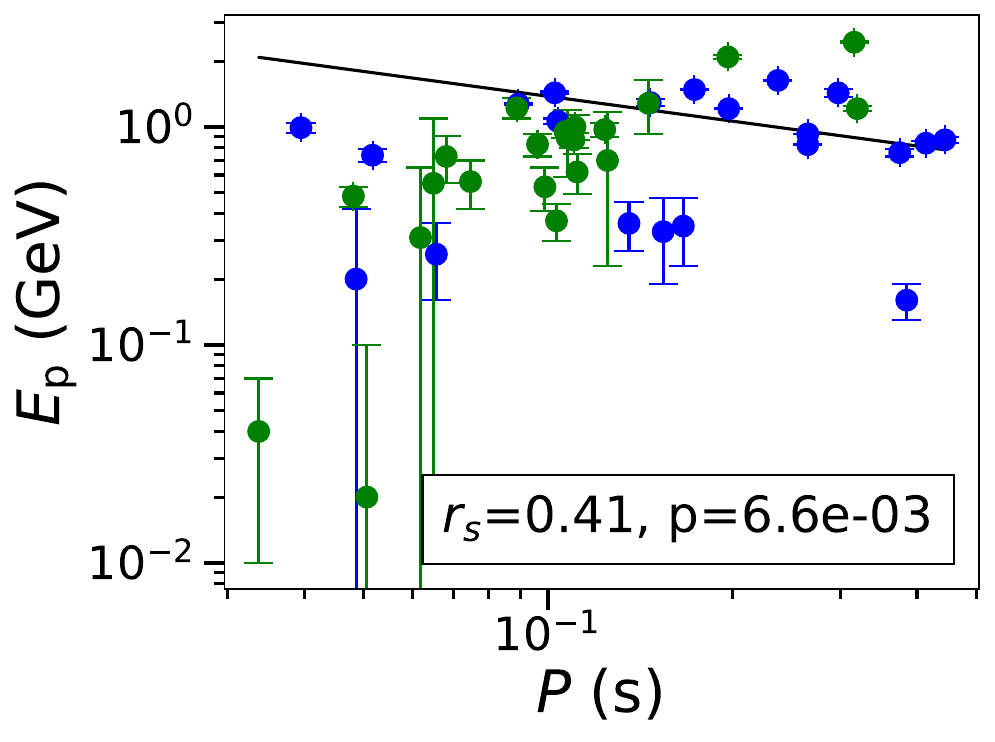}
    \end{subfigure}
    \hfill
    \begin{subfigure}{0.48\columnwidth}
        \includegraphics[width=\linewidth]{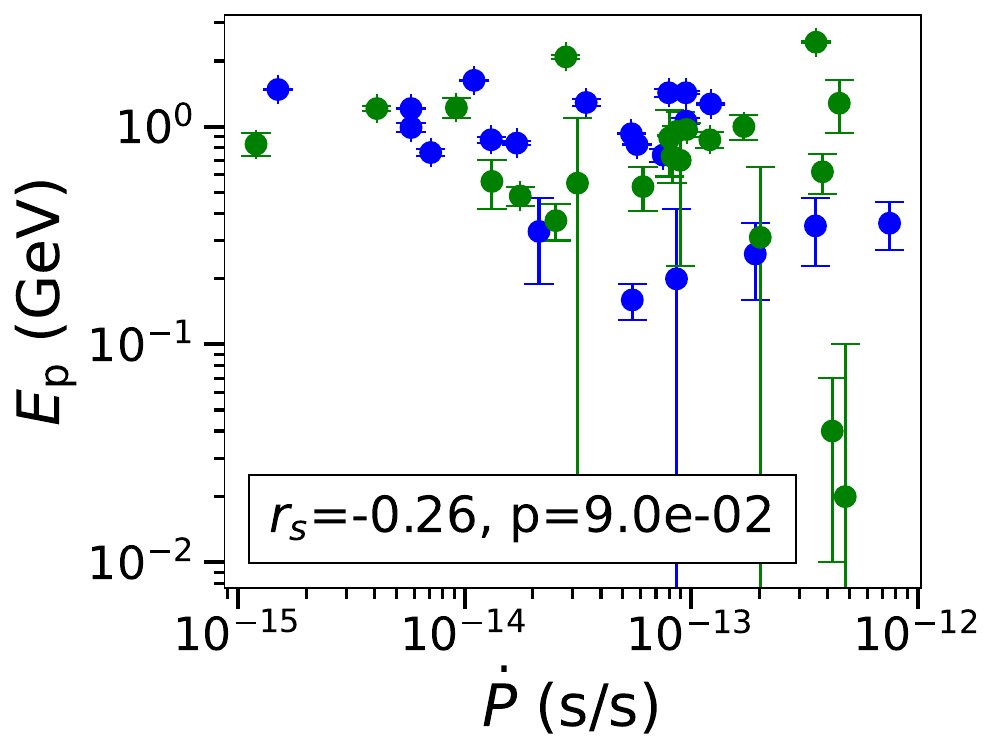}
    \end{subfigure}

    \vspace{0.6em}

    \begin{subfigure}{0.48\columnwidth}
        \includegraphics[width=\linewidth]{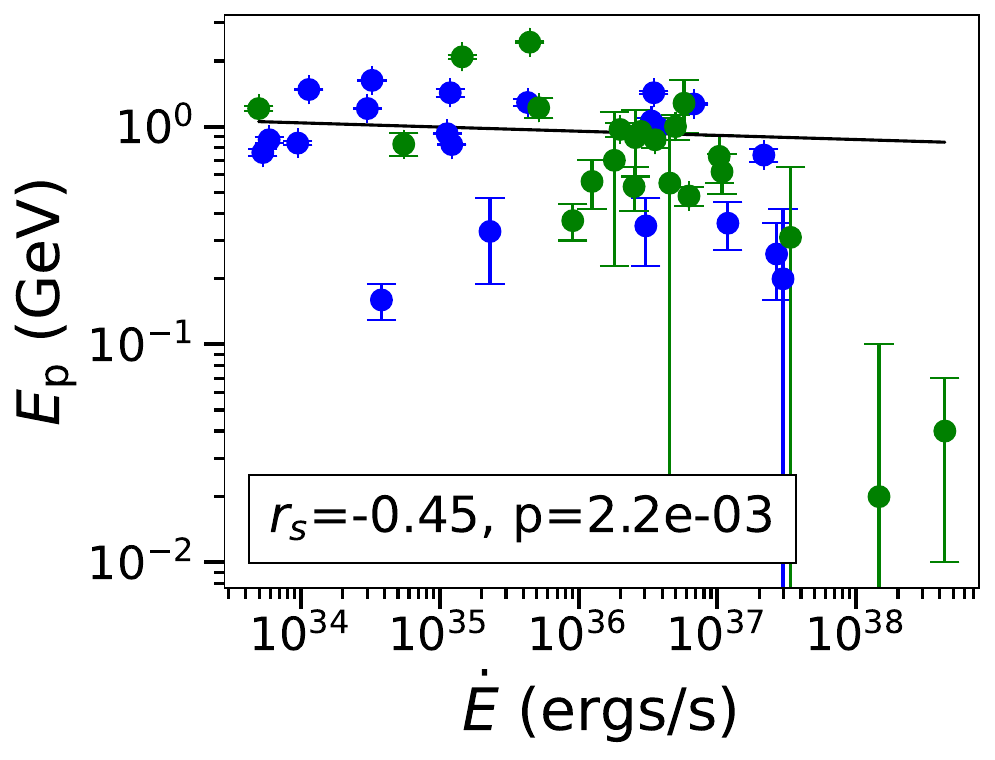}
    \end{subfigure}
    \hfill
    \begin{subfigure}{0.48\columnwidth}
        \includegraphics[width=\linewidth]{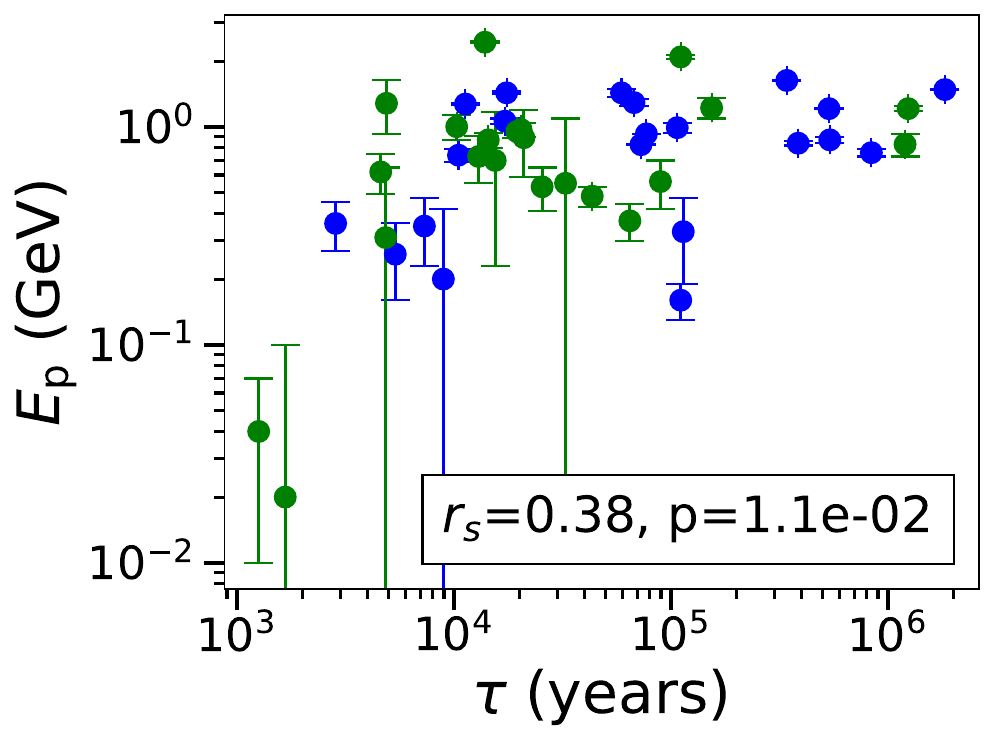}
    \end{subfigure}

    \vspace{0.6em}

    \begin{subfigure}{0.48\columnwidth}
        \includegraphics[width=\linewidth]{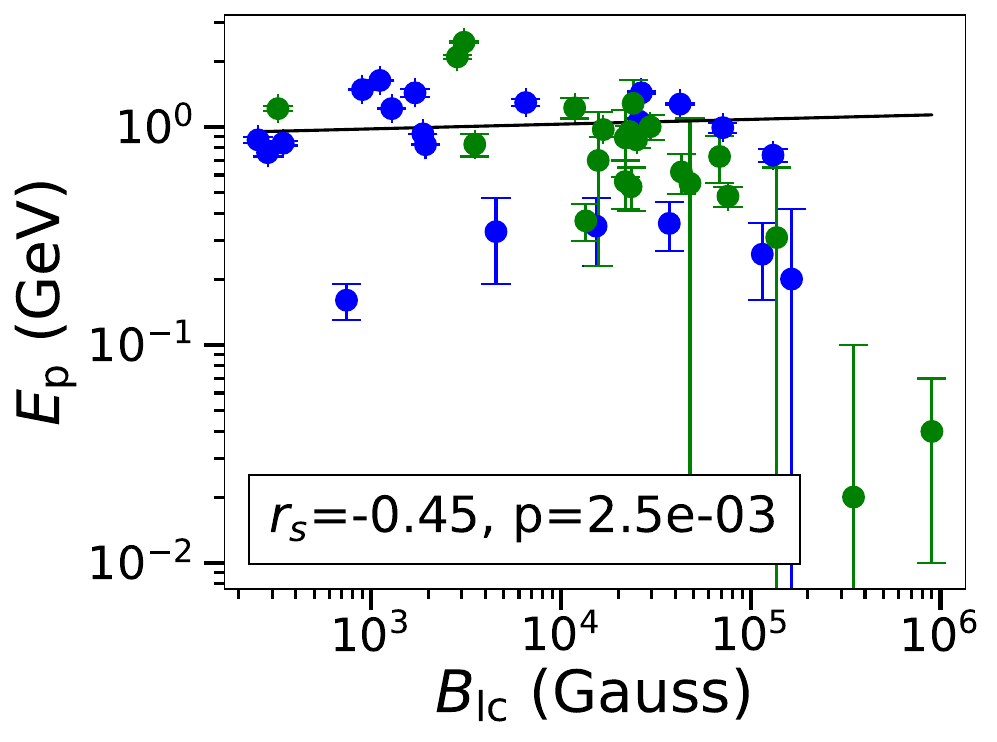}
    \end{subfigure}
    \hfill
    \begin{subfigure}{0.48\columnwidth}
        \includegraphics[width=\linewidth]{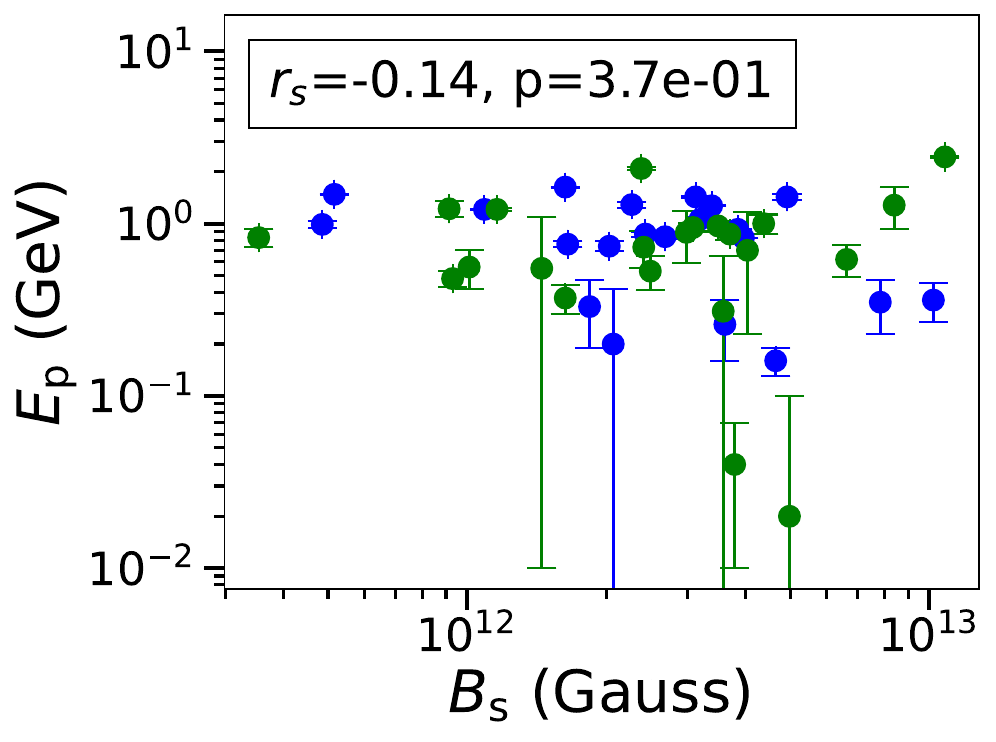}
    \end{subfigure}

    \caption{$E_{\rm p}$ vs timing parameters}
    \label{fig:Ep_timing}
\end{figure}

\begin{figure}
    \centering

    \begin{subfigure}{0.48\columnwidth}
        \includegraphics[width=\linewidth]{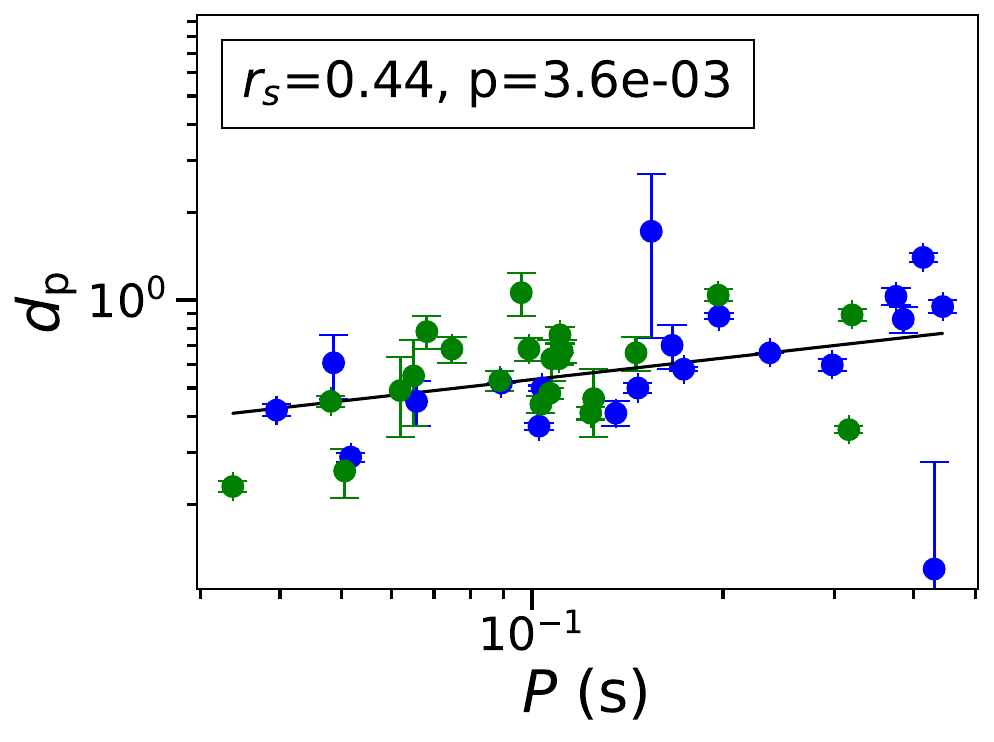}
    \end{subfigure}
    \hfill
    \begin{subfigure}{0.48\columnwidth}
        \includegraphics[width=\linewidth]{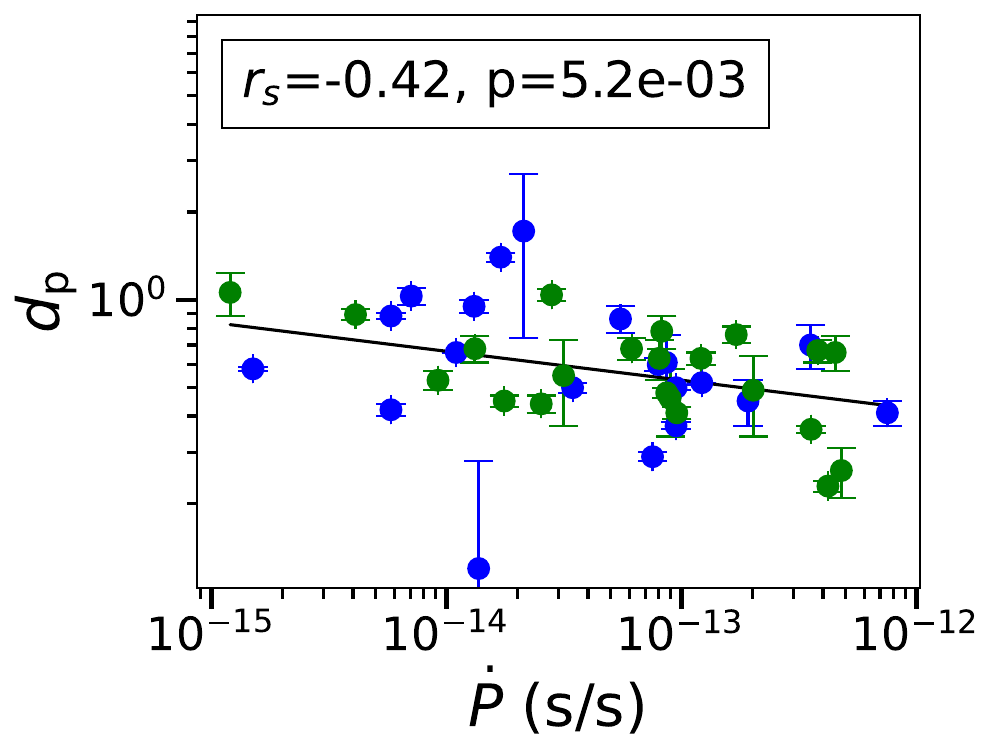}
    \end{subfigure}

    \vspace{0.6em}

    \begin{subfigure}{0.48\columnwidth}
        \includegraphics[width=\linewidth]{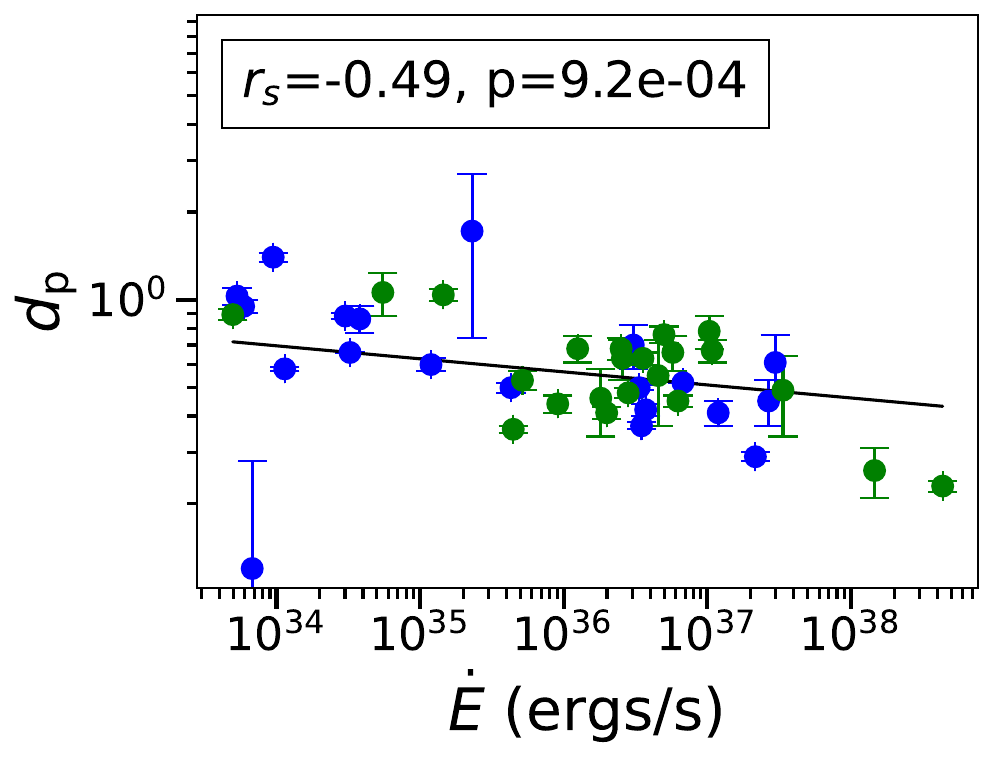}
    \end{subfigure}
    \hfill
    \begin{subfigure}{0.48\columnwidth}
        \includegraphics[width=\linewidth]{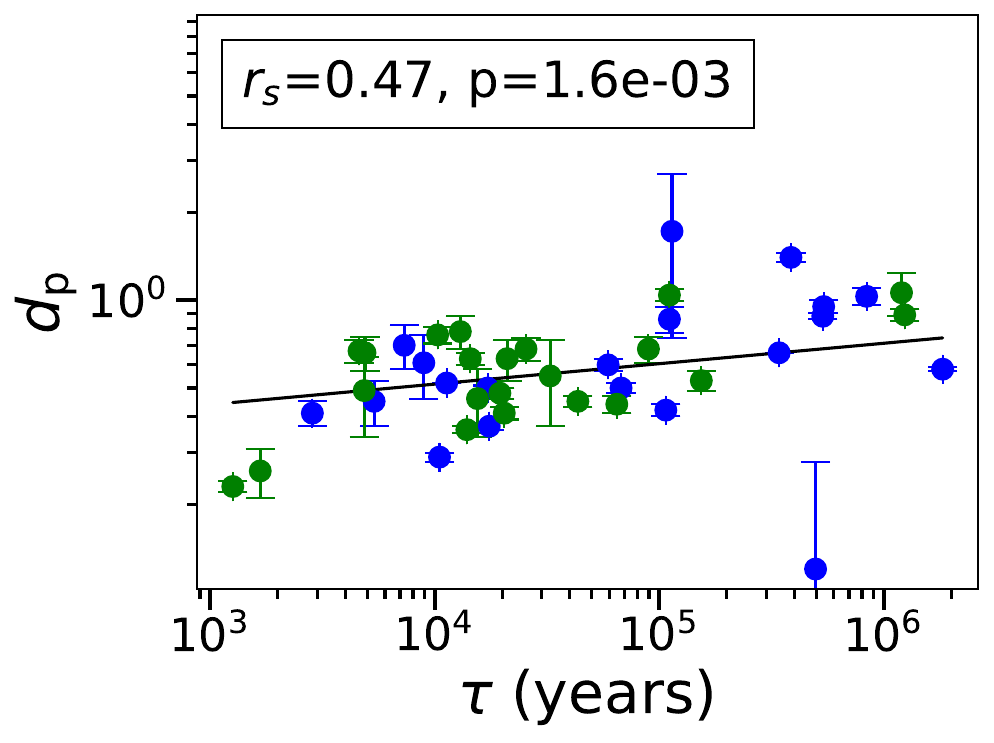}
    \end{subfigure}

    \vspace{0.6em}

    \begin{subfigure}{0.48\columnwidth}
        \includegraphics[width=\linewidth]{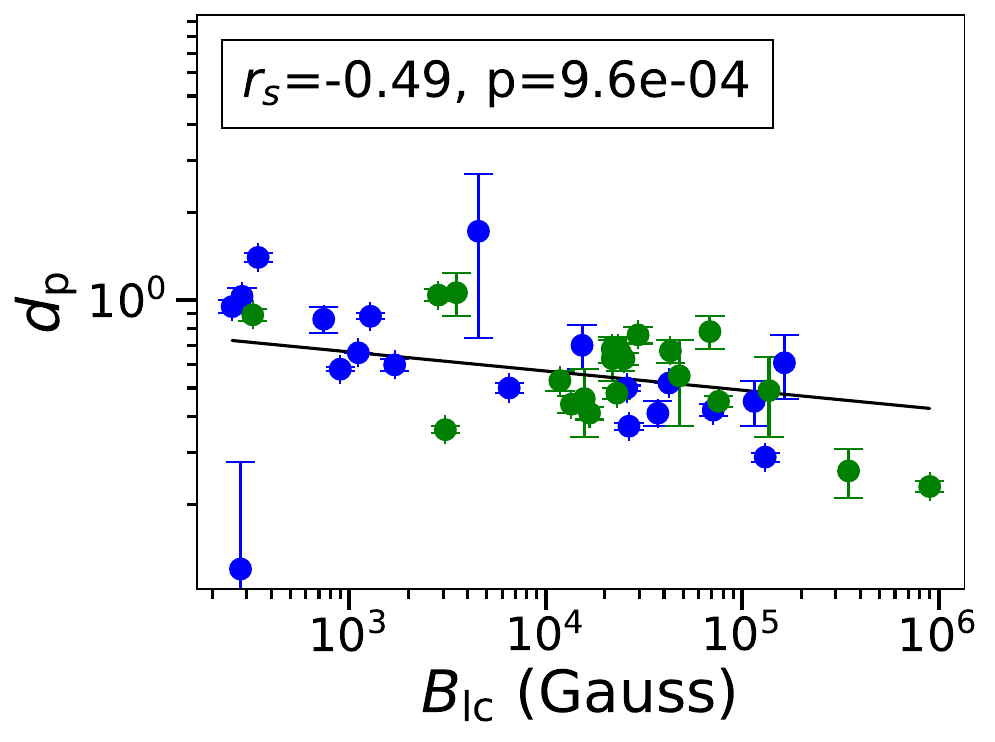}
    \end{subfigure}
    \hfill
    \begin{subfigure}{0.48\columnwidth}
        \includegraphics[width=\linewidth]{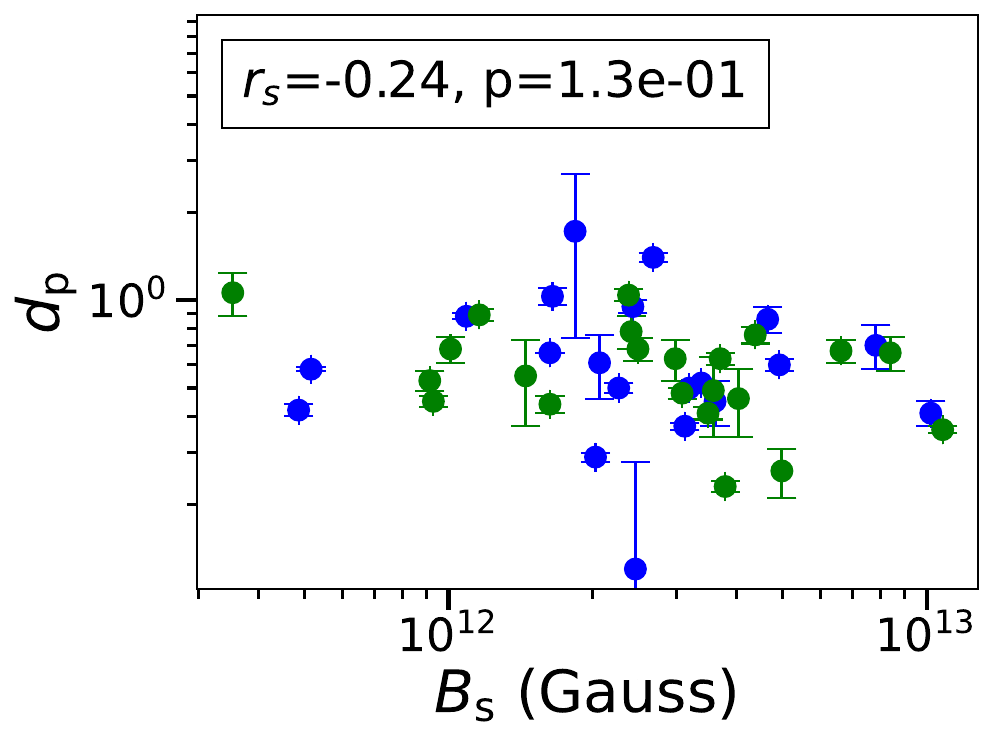}
    \end{subfigure}

    \caption{$d_{\rm p}$ vs timing parameters}
    \label{fig:dp_timing}
\end{figure}








\bsp	
\label{lastpage}
\end{document}